# Carbon nanomaterials for electronics, optoelectronics, photovoltaics, and sensing

Deep Jariwala,[a,‖] Vinod K. Sangwan,[a,‖] Lincoln J. Lauhon,[a] Tobin J. Marks,[a,b] and Mark C. Hersam[a,b,c,*]

[a] Department of Materials Science and Engineering, Northwestern University, Evanston, IL 60208, USA.

[b] Department of Chemistry, Northwestern University, Evanston, IL 60208, USA.

[c] Department of Medicine, Northwestern University, Evanston, IL 60208, USA.

[‖] These authors contributed equally to this manuscript.

[*] Corresponding author: Mark C. Hersam (m-hersam@northwestern.edu)

## Abstract

In the last three decades, zero-dimensional, one-dimensional, and two-dimensional carbon nanomaterials (i.e., fullerenes, carbon nanotubes, and graphene, respectively) have attracted significant attention from the scientific community due to their unique electronic, optical, thermal, mechanical, and chemical properties. While early work showed that these properties could enable high performance in selected applications, issues surrounding structural inhomogeneity and imprecise assembly have impeded robust and reliable implementation of carbon nanomaterials in widespread technologies. However, with recent advances in synthesis, sorting, and assembly techniques, carbon nanomaterials are experiencing renewed interest as the basis of numerous scalable technologies. Here, we present an extensive review of carbon nanomaterials in electronic, optoelectronic, photovoltaic, and sensing devices with a particular focus on the latest examples based on the highest purity samples. Specific attention is devoted to each class of carbon nanomaterial, thereby allowing comparative analysis of the suitability of fullerenes, carbon nanotubes, and graphene for each application area. In this manner, this article will provide guidance to future application developers and also articulate the remaining research challenges confronting this field.

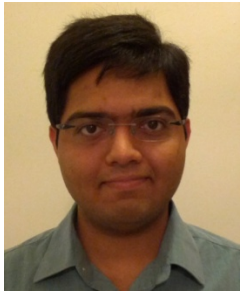

Deep Jariwala is currently working as a graduate student under supervision of Prof. Mark Hersam and Prof. Tobin Marks in Department of Materials Science and Engineering, Northwestern University. He received his B.Tech in Metallurgical Engineering from the Indian Institute of Technology-Banaras Hindu University (IIT-BHU) in 2010 where he worked on high temperature synthesis, doping, and shaping of carbon nanomaterials; namely carbon nanotubes and graphene. His current research interests include fabrication and characterization of novel electronic devices based on one and two dimensional nanomaterials.

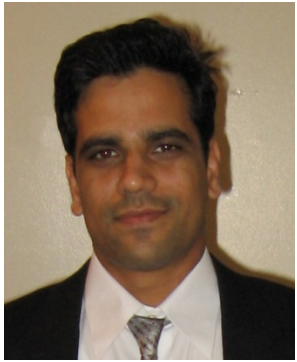

Vinod K. Sangwan is currently pursuing post-doctoral training with Prof. Mark C. Hersam, Prof. Lincoln J. Lauhon, and Prof. Tobin J. Marks in the Department of Materials Science and Engineering at Northwestern University. He received a B.Tech. in Engineering Physics from the Indian Institute of Technology (IIT), Mumbai in 2002. He earned a Ph.D. in Physics from University of Maryland, College Park in 2009. His doctoral work focused on large-area flexible electronics based on organic semiconductors and carbon nanotube thin films. His current research interests include electronic and optoelectronic applications based on carbon nanotubes and graphene.

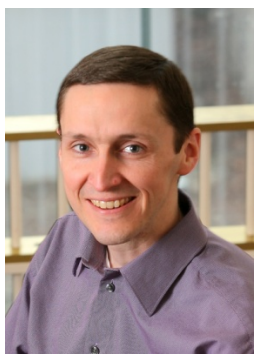

Lincoln J. Lauhon is an Associate Professor and Associate Chair of the Department of Materials Science and Engineering at Northwestern University. He received a Ph.D. in Physics from Cornell in 2000 and a B.S. in Physics from the University of Michigan in 1993. Prior to joining

Northwestern in 2003, he was a postdoctoral researcher at Harvard University with Charles Lieber. At Northwestern, the Lauhon group investigates novel structure-property relationships in nanostructured materials with an emphasis on spatially resolved measurements. His work has been recognized with an NSF CAREER Award, a Sloan Fellowship in Chemistry, and a Camille Dreyfus Teacher Scholar Award.

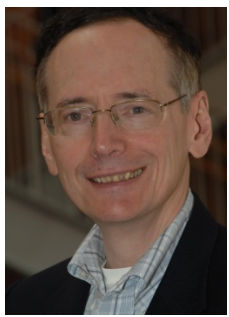

Tobin Marks is the Vladimir N. Ipatieff Professor of Catalytic Chemistry and Professor of Materials Science and Engineering at Northwestern University. He received a B.S. degree from the University of Maryland (1966) and a Ph.D. degree from MIT (1971). Among his recognitions, he received the 2006 U.S. National Medal of Science, the 2008 Principe de Asturias Prize in Science and Technology, the 2009 the MRS Von Hippel Award, the 2011 Dreyfus Prize in the Chemical Sciences, and the 2012 U.S. National Academy of Sciences Award in the Chemical Sciences. He is an elected member of the U.S., German, and Indian National Academies of Sciences, an elected member of the U.S. National Academy of Engineering, and a Fellow of the Royal Society of Chemistry, and of the American Academy of Arts and Sciences. He has 1050 peer-reviewed publications and 207 issued U.S. patents.

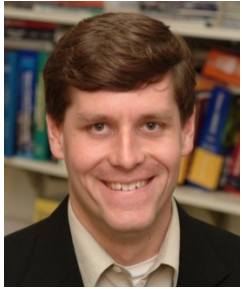

Mark C. Hersam is a Professor of Materials Science and Engineering, Chemistry, and Medicine at Northwestern University. He earned a B.S. in Electrical Engineering from the University of Illinois at Urbana-Champaign (UIUC) in 1996, M.Phil. in Physics from the University of Cambridge in 1997, and a Ph.D. in Electrical Engineering from UIUC in 2000. His research interests include nanofabrication, scanning probe microscopy, semiconductor surfaces, and carbon nanomaterials. Dr. Hersam co-founded NanoIntegris, which is a commercial supplier of high performance carbon nanotubes and graphene. Dr. Hersam is a Fellow of MRS, AVS, and SPIE, and serves as an Associate Editor of *ACS Nano*.

## 1. Introduction

Recently, the emerging need for high-speed electronics and renewable energy has motivated researchers to discover, develop, and assemble new classes of nanomaterials in unconventional device architectures. Among these materials, carbon-based nanomaterials have attracted particular attention due to their unique structural and physical properties. Carbon nanomaterials, composed entirely of $sp^2$ bonded graphitic carbon, are found in all reduced dimensionalities including zero-dimensional fullerenes, one-dimensional carbon nanotubes (CNTs), and two-dimensional graphene. With nanometer-scale dimensions, the properties of carbon nanomaterials are strongly dependent on their atomic structures and interactions with other materials. Consequently, significant recent effort has been devoted to the mass production of structurally homogeneous samples and their large-scale assembly into device architectures with well-controlled surfaces and interfaces. Although developments in the growth and post-synthetic purification of monodisperse carbon nanomaterials have been reviewed elsewhere,[1-5] this topic is briefly summarized below in Section 1.

Advances in producing highly monodisperse carbon nanotube and graphene samples have renewed interest in employing them as the basis of electronic, optoelectronic, photovoltaic, and sensing applications, thus forming the central theme of this review. In Section 2, we discuss digital electronics, analog electronics, and optoelectronic devices based on CNTs (individual CNTs as well as CNT thin films) and graphene. Section 3 explores photovoltaic applications of fullerenes, CNTs, and graphene, while Section 4 focuses on chemical and biological sensing enabled by carbon nanomaterials. The review concludes with a summary of the most salient points and a perspective on the future prospects and challenges for this field.

Although this review summarizes much of the historically significant work on carbon nanomaterials, the most recent developments are emphasized. Moreover, no comprehensive literature review exits on the impact created by sorted carbon nanomaterials in device applications. Consequently, readers who are interested in more thorough coverage of the early literature are referred to previously published review articles since this review highlights major breakthroughs following the advent of sorting techniques. For example, the fundamental properties of carbon nanotubes relevant to electronic and optoelectronic applications have been reviewed extensively in Ref.[6-9], while large-area electronics based on carbon nanotube thin films are discussed in Ref.[10, 11]. Electronic applications of graphene with a particular emphasis on fundamental physics are covered in Ref.,[12-15] while research on graphene-based optoelectronic and photovoltaic applications is covered in Ref[16]. Fullerene-based photovoltaics are summarized in Ref.,[17, 18] and sensing applications using carbon nanotubes and graphene are overviewed in Ref.[19-22]. In addition to these review articles, several pedagogical books have been written on carbon nanotubes,[23-25] and similar treatises on the fundamental properties and applications of graphene are currently in press.[26]

## 1.1. Carbon allotropes

Carbon is well known to form distinct solid state allotropes with diverse structures and properties ranging from $sp^3$ hybridized diamond to $sp^2$ hybridized graphite. Mixed states are also possible and form the basis of amorphous carbon, diamond-like carbon, and nanocrystalline diamond. Diamond is a metastable form of carbon that possesses a three-dimensional cubic lattice with a lattice constant of 3.57 Å and C-C bond length of 1.54 Å. In contrast, graphite is the most thermodynamically stable form of carbon at room temperature and consists of a layered two-dimensional structure where each layer possesses a hexagonal honeycomb structure of $sp^2$ bonded carbon atoms with a C-C bond length of 1.42 Å. These single atom thick layers (i.e., graphene layers) interact via noncovalent van der Waals forces with an interlayer spacing of 3.35 Å. The weak interlayer bonding in graphite implies that single graphene layers can be exfoliated via mechanical or chemical methods as will be outlined in detail below. Graphene is often viewed as the two-dimensional building block of other $sp^2$ hybridized carbon nanomaterials in that it can be conceptually rolled or distorted to form carbon nanotubes and fullerenes.

Fullerenes are the zero-dimensional form of graphitic carbon that can be visualized as an irregular sheet of graphene being curled up into a sphere by incorporating pentagons in its structure. Fullerenes come in various forms and sizes ranging from 30 to 3000 carbon atoms. As a fullerene is elongated in one dimension, it approaches the structure of a carbon nanotube (CNT). Conceptually, CNTs are seamless cylinders of single or few layered graphene with a high aspect ratio (i.e,. length to diameter ratio) that ranges from $10^2$ to $10^7$. The structure, diameter, and electronic type of a single-walled carbon nanotube (SWCNT) are determined by the chiral vector (i.e., roll-up vector) that defines the circumference of the SWCNT with respect to the graphene lattice (Fig. 1a). Multi-walled carbon nanotubes (MWCNTs) consist of nested, concentric shells of SWCNTs with a spacing between individual walls of 3.4 Å.

## 1.2. Synthesis and growth of carbon nanomaterials

The first carbon nanomaterial to be successfully isolated was $C_{60}$ (i.e., buckminsterfullerene) using laser ablation of graphite in a high flow of helium by Kroto, et al.[27] Although reports of even numbered carbon clusters[28] existed prior to the landmark $C_{60}$ paper, these clusters were produced in large size distributions and thus were unsuitable for characterization. Fullerenes have since been synthesized by a large number of groups using a variety of processes which include electric arc discharge, electron beam ablation, and sputtering.[29, 30] Most of these processes use graphite electrodes or targets as the carbon source. In some cases, composites of graphite and metal oxides are employed as targets to generate endohedral fullerenes where a metal atom is encapsulated inside the fullerene carbon cage.[31, 32] Fullerenes have been detected in common combustion flame soot[33-35] and have also been synthesized using bottom-up chemical methods.[36]

CNTs were subsequently isolated as an offshoot of fullerene synthesis since the initial techniques that resulted in CNT synthesis were either intended to produce fullerenes[37] or derived from existing fullerene production techniques such as the Kratschmer-Huffman method.[38, 39] The first observations of CNTs and their subsequent large-scale synthesis using arc discharge techniques were reported by Iijima and coworkers.[38, 40, 41] Laser ablation was later demonstrated as an alternative method for growing CNTs by Smalley et al.[42-44] Since it was observed that transition metals embedded in graphite electrodes/targets produced carbon nanotubes with higher yield and reproducibility,[37, 44] chemical vapor deposition (CVD) using transition metal nanoparticle catalysts was then developed to produce high quality single-walled and multi-walled CNTs in vertically aligned arrays.[45, 46] Vertically aligned arrays can also be grown on metallic[47] and quasicrystalline[48] substrates. Extending the metal nanoparticle catalyst concept, CNTs were later synthesized by pyrolysing metal carbonyls in the presence of other hydrocarbons.[49, 50] When optimized in a high pressure carbon monoxide (HiPco) environment, the carbonyl pyrolysis process led to high yield production of SWCNTs.[51] The synthesis of both SWCNTs and MWCNTs have since been thoroughly studied and reviewed by many.[1, 52-55] CNTs synthesized by arc discharge, HiPco, and CVD (using Co-Mo catalysts) are now commercially available in kilogram quantities.

Graphene, although often referred to as the mother of all graphitic carbon, was the final carbon nanomaterial to be isolated on an insulating substrate and electrically characterized by Geim and coworkers in 2004.[56] While graphene had been theoretically discussed and some attempts at mechanical isolation were made prior to 2004,[57, 58] the most definitive evidence of monolayer graphene and its electrical properties was provided by the Manchester group using mechanically exfoliated graphene. Although historically significant, this so-called "Scotch tape" technique for producing graphene lacks sufficient scalability for most applications. To address this issue, epitaxial graphene has been realized by graphitization of both doped and undoped silicon carbide (SiC) single crystal wafers at high temperatures.[59, 60] Although this recent work has triggered substantial interest in epitaxial graphene, it should be noted that reports of SiC graphitization date back several decades.[61] Furthermore, claims of epitaxial monolayer graphite (MG) on metal carbides at high temperatures using hydrocarbon precursors also exist from the early 1990's as have been comprehensively reviewed elsewhere.[62] Nevertheless, only the recent work on SiC has demonstrated graphene growth with sufficient quality for electronic applications. While this approach to epitaxial graphene offers wafer-scale growth, it is difficult to achieve uniform monolayer graphene coverage, and the product suffers from inferior electronic properties compared to mechanically exfoliated graphene.[60]

Solution-processing is another important technique to synthesize graphene at low cost in a scalable manner. The earliest reports of 'graphite oxide' synthesis trace back to the work of Hummers,[63] Brodie,[64] and Staudenmeier.[65] Similar methods were employed by the Ruoff group in 2006 to create graphene oxide (GO) that was mostly single layered.[66, 67] All of these methods result in an aqueous dispersion/colloid of thin GO flakes by subjecting graphite to highly oxidizing conditions that functionalizes the basal plane of graphene with hydrophilic functional groups.[66, 67] The resulting GO can be partially reduced to form reduced graphene oxide (r-GO) via chemical methods,[68] annealing in reducing environments,[69] or laser irradiation.[70] Several variants of these oxidation and reduction processes have been developed and are summarized in recent reviews.[71-74] Although this method is promising for large-scale solution processing of graphene-based materials, the harsh oxidizing conditions irreversibly damage the basal plane of graphene, leading to deterioration of its properties. This problem can be partially circumvented by directly exfoliating graphene from graphite using ultrasonication with suitable choice of surfactants and solvents.[75-77] In all cases, solution-based methods for preparing single-layer graphene result in relatively small flakes that are sub-optimal for wafer-scale applications.

CVD growth on metallic substrates such as nickel[78, 79] and copper[80, 81] provide an alternative pathway to large-area graphene. The concept of CVD synthesis of thin graphitic layers on transition metal surfaces has been discussed for several decades with the first demonstration of few layer growth on nickel substrates by exposure to a gaseous hydrocarbon source.[82] The earliest confirmed monolayer growth on Ni (111) was published in 1979.[83] Many reports followed on Pt, Ru, Pd, Re and Ir, which have been reviewed elsewhere.[62] Only recently the process has been optimized and extended to polycrystalline films/foils as well as other metals such as copper. CVD-grown graphene is continuous with uniform thickness over large areas, thus making it promising for electronic applications. CVD graphene on copper/nickel and solution-processed graphene are currently produced in bulk quantities and are commercially available. Readers are referred to recent reviews[5, 84] for more details on graphene synthesis.

## 1.3. Sorting and purification of carbon nanomaterials

As outlined above, synthetic methods for carbon nanomaterials tend to lack control over all structural parameters, resulting in raw materials that possess considerable polydispersity in their physical and electronic structure. Since most applications require uniformity and reproducibility, methods for sorting and purifying carbon nanomaterials to improve their monodispersity are of critical importance. For example, fullerene production methods commonly yield mixtures of $C_{60}$, $C_{70}$, and many higher homologues, necessitating the use of high performance liquid chromatography (HPLC),[85] column chromatography,[86] or selective chemistry[87, 88] to isolate monodisperse fullerene populations.

For CNTs, the polydispersity problem has been even more crucial since the electronic structure of CNTs is highly dependent on diameter and chiral vector. Although synthetic procedures have been developed to selectively grow SWCNTs in high yields, the as-grown material is a mixture of both metallic and semiconducting SWCNTs having a wide range of diameters and chiralities. Even advanced growth procedures that produce a narrow diameter distribution of SWCNTs[89] do not provide electronic type selectivity because many metallic and semiconducting chiralities can have nearly the same diameter. Several attempts to selectively remove semiconducting and/or metallic SWCNTs have been accomplished using controlled breakdown,[90] selective chemical reactions,[91, 92] electrophoresis,[93, 94] and chromatography[95-97] techniques as reviewed by Hersam[4, 98] and Zhang et al.[99] Perhaps the most flexible and commercially successfully method for sorting SWCNTs is centrifugal sorting in density gradients. This method, commonly known as density gradient ultracentrifugation (DGU), enables separation by diameter and/or electronic type by varying the surfactant concentration.[100, 101] DGU also allows sorting of double-walled carbon nanotubes,[102, 103] individual SWCNT chiralities,[104] and SWCNT enantiomers,[105] using both ionic and non-ionic[106] surfactants.

Since the properties of few-layer graphene are strongly dependent on the number of layers, similar polydispersity issues exist for graphene when the thickness is not carefully controlled during synthesis. To address this issue, centrifugal sorting techniques have been developed to separate graphene by thickness.[107, 108] Similarly, centrifugation has also been employed for lateral size sorting of solution-processed graphene.[109, 110] However, unlike carbon nanotubes where single chirality growth has proven to be elusive, growth of single-layer graphene over large areas has been achieved. Specifically, in CVD synthesis of graphene, the number of graphene layers in the resulting film is tunable by proper choice of metal and growth conditions. For example, the growth process on copper is self limiting to one layer due to the low solubility of carbon in copper. Furthermore, by tuning the growth conditions, CVD on copper can be extended to grow bilayers over large areas.[111] Multilayer graphene can also be grown using metals that have higher solubility for carbon such as nickel. The implications of these highly monodisperse carbon nanomaterials for various applications are discussed below.

2. Carbon Nanomaterials for Electronics and Optoelectronics

The highly delocalized electronic structure of $sp^2$ hybridized carbon nanomaterials suggests their utility as high mobility electronic materials. Furthermore, the ability to tune the band gap of semiconducting CNTs via control of diameter provides unique opportunities for customizing optical and optoelectronic properties. For these reasons, carbon nanomaterials are often cited as a potential successor to conventional semiconducting materials such as silicon[6] in electronic and optoelectronic applications. This section provides an overview of recent efforts to employ carbon nanomaterials for this class of devices.

## 2.1. Carbon nanotubes for electronic and optoelectronic applications

The diverse range of CNT electronic properties as a function of their chiral vector coupled with their quasi-one dimensional structure presents a number of attractive opportunities for electronic applications. For example, semiconducting CNTs are promising channel materials in field-effect transistors (FETs), whereas metallic CNT thin films are potentially useful as transparent conductors. A CNT FET is a three-terminal switch where current is passed through the CNT connected to two electrodes (source and drain; Fig. 1b).[112] Switching is achieved by modulating the carrier density in the CNT by a third electrically isolated electrode (gate). Here, we first briefly review the fundamental aspects of CNT charge transport that make them attractive materials for electronic applications.

First, the small capacitance of CNTs (< 0.05 aF/nm) enables low switching energies, efficient gate coupling, and minimal parasitic capacitance for low-power, high-speed electronics. Secondly, the atomically smooth surface of CNTs with no dangling bonds results in decreased carrier scattering and therefore increased carrier mobility. In addition, the surface structure of CNTs minimizes issues related to surface states and interface roughness that are prevalent in conventional semiconductor technology. Thirdly, the one-dimensional structure of CNTs eliminates small angle scattering of carriers, resulting only in forward scattering and back scattering. Since the momentum transfer required for back scattering is high and can only be provided by sharp defects and high energy optical phonons, carrier back scattering is suppressed, especially under low-field conditions.[113] Furthermore, long-range Coulomb scattering is also relatively ineffective, ultimately implying that the elastic mean free path in CNTs can be up to a few microns.[114] On the other hand, significant inelastic scattering can be induced by low energy acoustic phonons [115, 116] and radial breathing mode (RBM) phonons.[116] Scattering phase-space restrictions in one-dimensional CNTs result in an inverse relationship between carrier mobility and temperature (1/T) as opposed to the $1/T^5$ behavior in three-dimensional metals.[117] Therefore, CNTs can have unusually high low-field mobility at room temperature in contrast to other high mobility semiconductors such as InSb that can have ultrahigh mobility at low temperature but significantly reduced mobility at room temperature.[118] Specifically, CNTs have shown field-effect mobilities exceeding 100,000 $cm^2$/Vs [114] at room temperature and current densities up to $10^8$ $A/cm^2$ without electromigration.[119, 90] At high bias, energetic electrons can interact with optical phonons resulting in current saturation in metallic CNTs[119, 120] and velocity saturation in semiconducting CNTs.[121] At even higher energies, strong electron-electron interactions can induce impact excitation as discussed in detail below.

The one-dimensionality of CNTs also imposes limitations for nanoelectronic applications. In particular, the contact between a one-dimensional CNT with a three-dimensional metal electrode gives rise to a fundamental lower limit of contact resistance (~ 6.45 k$\Omega$) even in the ballistic regime.[122] Furthermore, unlike Ohmic contacts in Si FETs, Schottky contacts between most metals and CNTs further increase the contact resistance.[123, 124] Since CNT FETs are intrinsically ambipolar,[125] the difference in metal work function and CNT Fermi level can be tuned to enable both p-type (hole conducting) and n-type (electron conducting) FETs.[126] In the absence of intentional doping or electrode work function tuning, CNT FETs are generally p-type in ambient conditions due to atmospheric adsorbates such as oxygen.[127, 128] In addition to quantum resistance, the intrinsic band structure of CNTs also results in a carrier density dependent quantum capacitance of $10^{-16}$ F/μm.[129] Therefore, the performance of a CNT FET can be limited by this quantum capacitance rather than geometrical capacitance when integrated with ultra-thin high-κ gate dielectrics.

The unique optical properties of CNTs also present opportunities for novel optoelectronic devices. Semiconducting CNTs are direct band gap materials that possess free electron-hole pair excitations as well as strongly bound electron-hole pair states called excitons.[130, 131] The one-dimensional nature of CNTs produces van Hove singularities in the density of states that result in strong optical absorption and emission with energies determined by the CNT chirality.[132] The exciton binding energies in CNTs are higher (few hundred meV)[130] than conventional bulk semiconductors such as GaAs (< 10 meV)[133] due to strong Coulombic interactions. The large binding energy of one-dimensional excitons results in large radiative lifetimes (up to 100 ns)[134] and fluorescence lifetimes (up to 100 ps)[135] at room temperature, thereby enabling straightforward study of exciton dynamics at room temperature (in contrast to fabrication-intensive coupled-quantum well heterostructures of III-V semiconductors).[136] Excitons in CNTs can be created optically as well as electrically, and the corresponding radiative recombination results in photoluminescence and electroluminescence in CNTs. Electroluminescence in CNTs has been demonstrated at room temperature in ambipolar FETs[137] as well as through impact excitation in unipolar FETs.[138]

## 2.1.1. Single carbon nanotube transistors for digital electronics

The first electrical characterization of a metallic CNT was reported in 1997,[139] and the first semiconducting CNT FET was demonstrated by the same group in 1998 (Fig. 1c).[140] These early studies sparked significant interest in a host of novel transport phenomena and electronic devices based on CNTs such as Luttinger liquid behavior,[141] quantum wires and single electron transistors,[139, 142, 143] ballistic transistors,[144] and ambipolar FETs.[125] The importance of metal contacts was quickly realized for efficient charge injection in CNTs, and high work function metals such as Pd were found to provide Ohmic contacts to p-type CNTs in ambient conditions.[144] Along with high mobility, CNT FETs also show high on/off ratios ($10^4 - 10^6$) and low off-currents, both of which are desirable for low-power digital electronics. Additional metrics of device performance include transconductance and sub-threshold swing. Transconductance ($d(I_d)/d(V_g)$) is the variation in drain current ($I_d$) with respect to gate bias ($V_g$) at a constant drain bias ($V_d$), while sub-threshold swing is given by ($d(V_g)/d(Log(I_d))$), which indicates the variation in gate bias required for an order of magnitude change in drain current. A high transconductance and a low sub-threshold swing are highly desirable for high-speed electronic circuits. Transconductance as high as 30 µS has been achieved in FETs based on single CNTs.[145] Although Schottky contacts in CNT FETs result in a sub-threshold swing of 100-150 mV/decade, which is larger than the quantum limit for themionic emission from the contacts (60 mV/decade), a sub-threshold swing lower than the thermal limit (40 mV/decade) has been achieved in CNTs via band-to-band tunneling in dual-gated CNT FETs.[146] Early research on CNT FETs focused primarily on thermally grown thick oxides as gate dielectrics; however, it was soon realized that ultra-thin high-κ gate dielectrics, such as $ZrO_2$ and $HfO_2$, achieve far better gate coupling in small channel devices.[145, 147] The scaling of channel length remains an active area of research as CNT FETs have recently been demonstrated at the sub-10 nm scale (Fig. 1c).[112]

Progress in CNT electronics has moved beyond individual transistors. In particular, a complementary metal-oxide-semiconductor (CMOS) architecture is desired for energy efficient circuits. CMOS circuits consist of pairs of p-type and n-type transistors, and in steady state, one of the transistors is always in the off state, resulting in a low standby current dissipation.[148] Stable n-type doping in CNTs remains a significant challenge. Early demonstrations of logic gates, such as inverters and NOR gates, was achieved n-type CNTs by using a low work function metal electrode such as Al[149] or by annealing the devices in vacuum.[150, 151] Ambipolar CNT FETs (Fig. 1d)[6] can also been used in logic gates; however, the threshold voltages for the n-type and p-type branches have to be precisely controlled by using different work function gate electrodes. This strategy has been employed to fabricate 5-stage ring oscillators on single CNTs with operating frequencies up to 52 MHz (Fig. 1e).[152] The difficulty in precisely controlling threshold voltage in CNT-based devices has inspired the exploration of alternative circuit architectures beyond CMOS. In this regard, pass-transistor logic was recently demonstrated in complex circuits such as adders from 6 transistors instead of the 28 transistors required in CMOS-based architectures.[153] Despite significant progress, single CNT FETs have not yet advanced into commercial technology mainly due to two technical bottlenecks – heterogeneity in CNT-based device performance and large-scale assembly of CNTs. As outlined above, significant advances have been made in addressing CNT structural polydispersity with highly purified semiconducting CNTs now commercially available. In addition, directed assembly strategies such as dielectrophoresis are being actively pursued to address the second issue.[94, 154-161]

### 2.1.2. Carbon nanotube thin-film transistors for digital electronics

In recent years, the proliferating popularity of hand-held, portable consumer electronics has motivated researchers to develop semiconductor materials that can be incorporated in large-area, flexible macroelectronics. Amorphous Si and emerging organic and inorganic semiconductors have already found widespread usage in commercially available electronic devices, sensors, and flexible displays.[162, 163] CNT TFTs (Fig. 2a)[164] have shown equal or higher field-effect mobility than most of the organic and inorganic semiconductors that are being investigated for these applications.[162] In addition, CNT thin films are chemically inert in ambient, and possess attractive mechanical and optical properties that make them well-suited for flexible, stretchable, and transparent electronics.[10, 11] However, the integration of CNT thin films with flexible substrates also presents unique fabrication and processing challenges that must be overcome. In this section, we present an overview of large-area CNT TFT electronics with a focus on device characteristics and performance metrics required for practical applications.

The earliest reports of CNT TFTs employed as-grown random CNT networks on oxide gate dielectrics with a bottom-gate geometry.[165] These CNT TFTs exhibited a field-effect mobility of 10 $cm^2$/Vs (300 $cm^2$/Vs) at an on/off ratio of $10^5$ (10), thereby revealing the underlying trade-offs between the different device parameters such as field-effect mobility and on/off ratio. Charge transport in these CNT TFTs is found to be dominated by percolation effects as confirmed by simulations[164, 166] and experiments.[167, 168] Percolation effects show a power law behavior of channel resistance with channel length (Fig. 2b), [164, 167, 169] channel width, CNT network density, and CNT alignment.[166] The field-effect mobility of CNT TFTs (< 100 $cm^2$/Vs) is significantly less than that for a single CNT (> 10,000 $cm^2$/Vs) mainly for the following two reasons. First, the assumption of a parallel plate capacitor geometry overestimates the capacitance of a random network of CNTs and therefore underestimates the field-effect mobility. Analytical models have been developed for more accurate calculations of the capacitance of random CNT thin films that take into account the effects of sub-monolayer coverage of CNTs, nanotube-nanotube capacitive coupling, and the quantum capacitance of CNTs.[170] However, even these more realistic descriptions cannot account for the two orders of magnitude lower field-effect mobility in CNT thin films. Therefore, the second reason for reduced mobility can be attributed to the CNT-CNT contact resistance in the percolating CNT network.[171, 172] The contact between a metallic CNT and a semiconducting CNT is approximately 1000 times more resistive than that between two metallic or two semiconducting CNTs.[171] Overall, realistic modeling of a CNT TFT has to take into account this complex set of variables within a percolation model, including variable contact resistance due to CNT heterogeneity, metal-CNT contact resistance, and capacitive coupling between CNTs and the gate electrode. The physics is further complicated by large drain current hysteresis in ambient conditions due to atmospheric adsorbates and trapped charges in the oxide gate dielectric.[127, 128, 173] While significant advances have been made in developing phenomenological models that accurately predict scaling behaviors, quantitative predictions of basic device parameters such as on/off ratio are still lacking. From a technology perspective, it is often sufficient to gain an intuitive understanding of trade-offs in device performance parameters. In particular, for as-grown CNTs, a trade-off exists between on-state current (and thus field-effect mobility) and on/off ratio (Fig. 2c)[167] due to increased percolation of metallic CNTs in thicker films.

In addition to as-grown random CNT networks, as-grown aligned CNTs have been considered for TFTs. Almost perfectly aligned (alignment angle < 0.01°) CNTs with nanotube densities up to 50 CNTs/μm have been grown by CVD on miscut quartz substrates (Fig. 2d).[174-176] Printing methods have also been developed to transfer these aligned CNT arrays onto plastic substrates.[174] Although these aligned CNT films result in large device currents and high field-effect mobilities, the on/off ratio is compromised due to directly bridging metallic CNTs.[175] To overcome such low on/off ratios for digital circuits, multiple approaches have been utilized to minimize the effect of metallic CNTs, including selective removal of metallic CNTs in as-grown CNT TFTs and reducing the population of metallic CNTs in solution-processed CNT TFTs. The earliest report of selective removal of metallic CNTs utilized controlled electrical breakdown of metallic CNTs by Joule heating, while the semiconducting CNTs were electrostatically depleted with an appropriate gate bias.[177] This method is particularly successful for CVD-grown aligned CNTs where all of the CNTs directly bridge the source-drain electrode gap.[175, 178, 179] However, this process causes collateral damage to nearby semiconducting nanotubes that leads to reductions in on-state currents. This process of correlated breakdown has been systematically investigated in dielectrophoretically aligned nanotube array transistors.[180, 181]

Metallic CNTs from random CNT thin films have also been selectively etched by methane plasma etching.[182] Another method for minimizing the effect of metallic CNTs is to define a parallel array of narrow CNT strips in the channel.[183] This strategy has allowed the demonstration of large-area, flexible integrated circuits consisting of more than 100 transistors (Fig. 2e, f).[183] Optimum device performance can also be realized by fine-tuning the density of CNTs above the percolation threshold of all CNTs and below the percolation threshold of just the metallic CNTs (Fig. 2c).[167] A recent effort using a selective cycloaddition reaction of metallic nanotubes in an as-grown film has been particularly successful in achieving field-effect transistors with high mobility (>100 $cm^2$/Vs) and on/off ratios of $10^5$.[91] Further improvement in as-grown CNT TFTs has been achieved by realizing covalent bonding between CNTs grown by CVD (Fig. 2g).[184] In this case, reduced CNT-CNT junction resistance yields a high field-effect mobility (~70 $cm^2$/Vs) with a high on/off ratio, allowing flexible circuits consisting of 21-stage ring oscillators to be realized.[184] Nevertheless, despite this significant progress, the presence of metallic species in as-grown CNTs continues to limit the full potential of CNT thin films in electronic applications.

Post-growth solution processing to produce monodisperse CNTs has not only significantly reduced the issues related to CNT heterogeneity but has also allowed CNTs to be integrated with flexible substrates using low-temperature, solution-based assembly methods. As discussed above, density gradient ultracentrifugation (DGU) has enabled the scalable production of semiconducting and metallic CNTs.[4, 100, 101, 103, 104, 106, 185-186] By reducing the fraction (< 1%) of metallic species in sorted CNTs, substantially thicker CNT films can be incorporated into TFTs than is possible with as-grown CNTs. Consequently, DGU-sorted semiconducting CNT inks result in high performance devices with concurrently high field-effect mobilities, current densities, and on/off ratios.[187, 188] These CNT inks are compatible with facile assembly methods such as drop-casting, dip-coating, and transfer printing, thus allowing wafer-scale fabrication of logic gates (Fig. 3a).[188] DGU-sorted CNT TFTs have also been used for light-emitting diode (LED) control circuits.[189] Furthermore, 99% single-chirality (6,5) SWCNTs have enabled high-current (up to 0.1 A) CNT TFTs.[104] Solutions of semiconducting CNTs have also been incorporated into aligned arrays of CNTs by evaporation driven methods (Fig. 3b,c)[190] and dielectrophoresis[191] to obtain further improvements in on-state current. The marked improvements of monodisperse semiconducting CNT inks compared to as-grown CNTs can be attributed to a fundamental shift in the trade-off relationships between competing device parameters. While metallic CNTs begin to dominate at low densities of as-grown CNTs (~1 CNT/$\mu m^2$, average CNT length ~5 $\mu m$, Fig. 2c),[167] the density of DGU-sorted semiconducting CNTs can be increased (~25 CNT/$\mu m^2$, average CNT length ~1.3 $\mu m$) without compromising the on/off ratio.[192] Besides DGU, sorted CNTs from gel-based techniques have also resulted in high performance TFTs with on/off ratios exceeding $10^4$.[97, 193] Recent developments in self-sorting and alignment of nanotubes using amine terminated surfaces have also resulted in TFT devices with high on/off ratios (~$10^5$).[92, 194]

These significant developments have advanced the field of CNT TFTs into a new regime where concurrent improvement in other device components such as gate dielectrics is required to fully realize the potential offered by purified CNT inks (Fig. 3,a,b,c).[189, 190, 195] Ultra-thin, high-κ gate dielectrics are desirable for low-power, hysteresis-free operation of TFTs. Large-area printed electronics using aerosol jet printing of semiconducting CNTs with ultra-high capacitance ion-gel dielectrics (~10 μF/cm$^2$) enables sub-3 V operation logic gates and 5-stage ring oscillators with operating frequencies up to 5 kHz (Fig. 3d,e,f).[187] In addition, a new class of ultra-thin, self-assembled nanodielectrics (SANDs) has been successfully integrated with purified CNTs to achieve significant improvements in all device performance metrics, including hysteresis-free operation in ambient conditions (Fig. 3g,h).[192, 196-199] These SAND gate dielectrics consist of a multilayer structure of vapor-deposited organic chromophore and inorganic oxide grown by atomic layer deposition (Fig. 3g).[192] The advantages of SANDs over commonly used high-κ inorganic oxides can be attributed to their reduced trapped charge densities and leakage currents. A recent systematic study achieved an intrinsic field-effect mobility of ~150 cm$^2$/Vs and a sub-threshold swing of ~150 mV/decade with an on/off ratio exceeding $10^5$.[192] Importantly, large-area processability and compatibility with plastics makes SANDs ideal for printed electronics.[198]

Although monodisperse semiconducting CNT inks appear to be the most viable route toward CNT TFT applications, these sorted CNTs also introduce new challenges that require further research. For example, the surfactants used in dispersing CNTs in aqueous solution must be efficiently removed from the device channel for reduced contact resistance and CNT-CNT junction resistance. In addition, the selective preparation of semiconducting CNTs with lengths that are longer than currently available (~1-2 µm, compared to the as-grown CNT length > 10 µm) would also improve device performance. Finally, an assembly strategy that achieves dense arrays of aligned, individual CNTs (as opposed to bundles of CNTs[190]) would likely allow CNT TFT performance to approach the performance limit of single CNT FETs. Strategies based on DNA linkers show promise for addressing this challenge.[200, 201]

### 2.1.3. Carbon nanotubes for radio frequency analog circuits

While CNT-based digital electronic applications still face significant challenges as outlined above, radio frequency (RF) analog circuits may prove to be a more realistic short-term goal.[202] A high on/off ratio, which is essential for low power digital electronics, is less important for analog electronic applications where one of the main goals is power amplification at high frequencies. With these relaxed performance metrics, new materials and device concepts have advanced more quickly in RF applications than in CMOS-based digital electronics. The large current density[119] and low capacitance of CNTs are highly desired properties for RF circuits. Field-effect mobility (determined by the drift velocity of carriers) is considered an important metric for digital electronics, but is of reduced significance in short gate length ($L_{gate}$) RF devices where the speed of operation is determined by the saturation velocity ($v_{sat}$) of carriers.[203] Instead, the small signal equivalent circuit of an RF device is characterized by the current gain (power gain) defined as the ratio of output current (power) to the input current (power). The gain rolls off at high frequency, and an important figure of merit for RF circuits is the cutoff frequency ($f_T$), which is defined as the highest frequency at which the current gain is unity. The cutoff frequency is determined by $f_T \sim g_m/F(C_g, C_p)$, where $g_m$ is transconductance and $F(C_g, C_p)$ is a function of the gate capacitance ($C_g$) and the parasitic capacitance ($C_p$) between the gate and source-drain electrodes.[202] Alternatively, the cutoff frequency in small channel devices can also be expressed as $f_T \sim v_{sat}/L_{gate}$. Calculations suggest that CNTs have a higher $v_{sat}$ of 4 x 10$^7$ cm/s [121] compared to that in state-of-the-art GaAs (2 x 10$^7$ cm/s) and Si (10$^7$ cm/s) devices, resulting in an intrinsic cutoff frequency of ~1 THz in 10 nm channel devices.[202]

Early studies of single CNT-based ring-oscillators reported $f_T$ up to 52 MHz,[152] while more recent CNT-based RF devices have yielded $f_T$ in the range of 100 MHz – 2.6 GHz.[204, 205] Furthermore, the nonlinear transfer characteristics of CNT FETs and CNT-metal Schottky diodes have been used as mixers and rectifiers at frequencies up to 50 GHz.[206-208] The large discrepancy between intrinsic and experimental $f_T$ results from the two orders of magnitude higher parasitic capacitance of the electrodes compared to the gate capacitance of a CNT. As discussed above, the contact resistance of CNT FETs is higher than the fundamental quantum limit (6.45 kΩ) for each of the contacts. However, the input impedance of a RF circuit must match the 50 Ω impedance in the external circuitry, which has been achieved in isolated cases via capacitive coupling between the contacts and CNTs at 4 K.[205] Both of these issues can be partially addressed by utilizing parallel arrays of CNTs where the transconductance can be significantly increased without increasing the limiting parasitic capacitance from the electrodes (Fig. 4a).[202] One approach to achieve this goal is to use CVD-grown parallel arrays of CNTs on quartz substrates where unity power gains have been achieved up to 9 GHz.[209, 210] Another approach is to align unsorted solution-processed CNTs using dielectrophoresis, leading to an input impedance of 50 Ω up to 20 GHz.[211] However, the presence of metallic CNTs in these studies significantly degrades the performance by decreasing the output impedance (and thus the gain) of the devices. Therefore, monodisperse semiconducting CNT thin films have also been considered for RF electronics, allowing the extrinsic $f_T$ to be pushed beyond 5 GHz with linearity up to 1 GHz.[212] Recently, DGU-sorted 99% semiconducting CNT thin films were assembled by dielectrophoresis to achieve the highest reported $f_T$ of 80 GHz to date (Fig. 4b).[213] Finally, Fig. 4c compares the highest performing CNT RF devices with graphene FETs (discussed in detail below), GaAs high electron mobility transistors (HEMTs), InP HEMTs, and Si FETs as a function of gate length.[214] In spite of the high intrinsic limit for CNTs (~1 THz), the extrinsic $f_T$ of CNT devices is still one order of magnitude lower than the highest $f_T$ of 660 GHz for 20 nm GaAs HEMTs.

CNT-based RF devices have recently been incorporated into fully functional proof-of-concept circuits. At least two groups have reported CNT circuits being used as a demodulator for RF signals.[209, 215] However, replacing only one part of the radio with a CNT device does not constitute a major technological advance since the size and performance of the radio remains limited by other components such as the batteries and antenna.[215] Therefore, one group has utilized CNT FETs as both the RF mixer and audio amplifier in a fully working AM (amplitude modulation) radio system that was able to detect signals from a local radio station.[209] In addition, fully operational roll-to-roll printed RFID tags and ring oscillators using CNT TFTs have been demonstrated at 13.56 MHz.[216]

### 2.1.4. Carbon nanotubes for optoelectronics

Semiconducting CNTs are direct band gap materials that have been incorporated in a variety of optoelectronic devices such as light detectors, light emitters, and transparent conductors.[6, 9, 217] As previously discussed, van Hove singularities in the one-dimensional density of states and strongly bound excitons make CNTs interesting candidates for optoelectronics (Fig 5a).[132] The exciton binding energy depends on the diameter of CNTs[218, 219] as well as on the dielectric constant of the surrounding environment.[219] The earliest experiments on CNT optical excitations were conducted in dispersed aqueous solutions containing individual CNTs coated with surfactants.[131, 132] In most cases, photoexcitation generates excitons in the second sub-band of CNTs ($E_{22}$), followed by radiative decay to the first sub-band ($E_{11}$) (Fig. 5a).[132] Consequently, a two-dimensional plot of photoluminescence as a function of excitation and emission energies (Fig 5b) provides peaks that uniquely identify the CNT chirality.[132] Early experiments showed the fluorescence quantum efficiency ($Q_F$) of dispersed CNTs to be $10^{-3}$ -$10^{-4}$ with an effective radiative lifetime of 1-10 ns at room temperature.[132, 220] Recent experiments on as-grown suspended CNTs have yielded $Q_F$ up to 10%.[221] The limited $Q_F$ in CNTs can be ascribed to multiple non-radiative processes such as exciton-exciton annihilation,[134] presence of low energy dark excitons that cannot relax to the ground state radiatively,[219] and efficient non-radiative, phonon-mediated decay of excitons from $E_{11}$.[222] The wide range of $Q_F$ values reported in literature also suggests sensitivity of the radiative decay rate to the quality of the CNTs and the nature of the surrounding environment. The diameter-dependent CNT excitonic binding energy allows tunability of the photoresponse, especially in the near-infrared (900 nm – 2000 nm).

In addition to photoluminescent applications, semiconducting CNTs are suitable candidates for photocurrent and electroluminescent devices. For example, CNTs have been utilized as the basis of light-emitting transistors and photodetectors.[190] The first study of electroluminescence in ambipolar CNT FETs revealed a number of interesting observations: (1) Unlike conventional p-n junctions, electroluminescence in ambipolar CNT FETs does not require extrinsic doping; (2) The maximum electroluminescence efficiency is observed in the off-state; (3) Polarization of the emitted light was parallel to the CNT axis.[137] In addition, the position of light emission from the CNT channel can be tuned by controlling the recombination site via biasing conditions, as demonstrated later by the same group.[223] However, the emitted light power (~100 pW) and $Q_F$ ($10^{-6}$) are rather low for typical light-emitting applications.[224] In addition, spectral broadening was observed in small channel devices due to hot carrier recombination mediated by optical/zone boundary phonons. Improvement in emission power per unit area has been attempted by utilizing large arrays of electrolyte gated aligned CNTs[225] and top-gated aligned 99% semiconducting CNTs[190] in ambipolar device operation. However, these devices show decreased $Q_F$ ($10^{-9}$) compared to individual CNT FETs.[225] Increasing the electroluminescence intensity was also attempted in random CNT TFTs;[226] however, red-shifting and spectral broadening was observed due to exciton transfer from large band gap CNTs (narrow diameter) to small band gap CNTs (large diameter) in the heterogeneous mixture of as-grown CNTs. The exciton energy transfer mechanism was later confirmed by a spatially and spectrally resolved photoluminescence experiment on a crossed junction of CNTs with different chiralities sorted by DGU.[227] Spectral red-shifting of the electroluminescence spectrum from aligned bundles of monodisperse CNTs was also attributed to the same mechanism.[190]

Electroluminescence can also be achieved in unipolar CNT FETs by impact excitation processes from hot carriers.[138, 228-230] At high electric fields, the electrons in CNTs can gain sufficient kinetic energy to drive electronic excitations across the band gap upon scattering. The impact excitation process in CNTs is found to be at least 3 orders of magnitude more efficient than in conventional bulk semiconductors.[9, 138] Impact excitation can also be induced by high local electric fields from inhomogeneities such as defects, trapped charges, and CNT-metal contacts,[231] while localized unipolar CNT electroluminescence is achieved in artificially constructed regions of high electric field in the channel. One approach is based on the fabrication of CNT FETs where a portion of the CNT is supported by a dielectric and the remainder is suspended over a trench etched in the channel (Fig. 5c).[138] The abrupt discontinuity in the dielectric constant results in band bending that accelerates electrons, thus generating excitons via impact excitation that can recombine radiatively (Fig. 5d).[229] The $Q_F$ of this device was reported to be 1000x higher than that from ambipolar CNT FETs. Another efficient approach for achieving sufficiently large electric fields for impact excitation is by creating p-n junctions between electrostatically doped p-type and n-type regions in the same CNT (Fig. 5 e,f).[232, 233] Additionally, impact excitation processes in CNTs are not constrained by the same selection rules as optical excitations and can lead to the creation of excitons as well as free electron hole pairs.[228] Electrically driven thermal light emission corresponding to interband transitions, in contrast to featureless blackbody radiation, has also been observed in suspended quasi-metallic CNTs.[234]

The application of semiconducting CNTs to photodetectors has also been studied extensively.[6, 9] Photodetection can be thought of as the inverse of electroluminescence where optically generated excitons are separated into free electrons and holes to produce a photocurrent with an applied field[235] or an open circuit photovoltage in an asymmetric field configuration.[236] Early photoconductivity measurements on individual CNT ambipolar FETs showed internal $Q_F$ up to 10% with expected resonances at optical energies corresponding to CNT excitonic states.[235] Sufficiently large fields to separate excitons can also be generated locally by asymmetric Schottky contacts,[236] p-n junctions,[237] and local charge defects.[238] An electrostatically gated p-n junction in a CNT has shown efficient generation of multiple electron-hole pairs per absorbed photon through a process similar to impact excitation in electroluminescence.[239] While the mechanism of photodetection in individual CNT devices is mostly based on generation and dissociation of excitons, the photoconductivity in CNT networks is often dominated by the thermal effects of nonradiative recombination. In particular, bolometric increases in CNT resistance due to increased local temperature have been shown to be a viable method for photodetection.[240]

Since the highest $Q_F$ for light emission in CNTs has not exceeded $10^{-3}$, there is currently a limited scope of practical applications for CNTs in light-emitting optoelectronic applications. Consequently, researchers have attempted to incorporate CNTs into alternative technologies such as light-emitting diodes based on conjugated polymers to obtain enhanced electroluminescence efficiencies.[241] Since CNT thin films are flexible, optically transparent, and highly conductive, they have been considered as an alternative to indium tin oxide (ITO) in organic light-emitting diodes and organic TFTs.[242] In particular, CNTs hold promise for overcoming the limitations of ITO for large-area flexible electronics such as brittleness and patterning-related issues, while maintaining desirable properties such as high optical transparency and low sheet resistance.[242] DGU-sorted 99% pure metallic CNTs have been shown to enhance sheet resistance by over 5 times compared to as-grown CNTs.[185] Additional sorting by diameter can also produce conductive films with tunable optical transmittance (Fig. 5g).[185] CNT thin film electrodes provide up to 3 times lower contact resistance than commonly used Au electrodes in organic semiconductor TFTs (Fig. 5h).[243, 244] Furthermore, CNT-based transparent conductors have application in photovoltaics as will be described below.

CNTs have also been considered as components of non-linear optics.[245] For example, dispersed CNTs can serve as saturable absorbers with up to 40% attenuation efficiency and passive mode-lockers for femtosecond lasers.[246] The advantage of CNTs in this application is that they offer facile fabrication with widely tunable wavelengths based on CNT diameter. In this regard, the isolation of monodisperse CNTs with $E_{11}$ excitation near the fiber optic communication wavelength (1550 nm) is particularly useful.

## 2.2. Graphene for electronics and optoelectronics

While the explosion of research interest in graphene was triggered by the seminal paper on single layer graphene by Geim and Novoselov in 2004,[56] the unique electronic properties of graphene were known from theoretical predictions for more than half a century. Specifically, the first calculations of the electronic structure of graphene were reported in 1947,[247] and the physical structure of graphene (Fig. 6a)[12] is often discussed in condensed matter physics textbooks. The existence of single layer graphene was also known as an undesirable coating on metal surfaces since the electrical properties in this form had been difficult to characterize.[62] The two breakthrough results of the 2004 paper[56] were the successful isolation of a truly two-dimensional material and the carrier concentration dependent conductivity of graphene that was reminiscent of a FET (Fig. 6b).[12] Both of these results were considered counterintuitive at the time. In particular, isolated two-dimensional materials were previously thought to be impossible to isolate due to thermodynamics arguments.[248, 249] In fact, other examples of known monolayer materials only existed when tightly bound to other bulk materials.[250] In contrast, single layer graphene is stable on an arbitrary substrate, and can even be freely suspended in space or dispersed in solution, thus implying that it can be considered as an intrinsically stable two-dimensional material. From the perspective of conductivity, early calculations performed on short-range defect scattering had predicted a constant electrical conductivity in graphene at a finite carrier density.[251] Consequently, the linear dependence of conductivity on carrier concentration in graphene forced researchers to revisit electronic transport models.[252-255] In this section, we will outline the electronic structure of graphene and then critically assess its viability for applications in digital electronics, RF analog electronics, and optoelectronics.

The unique electronic structure of graphene stems from its honeycomb lattice in which a carbon atom is bonded to a neighboring three carbon atoms through $sp^2$ hybridized bonds (Fig. 6a).[12] The unit cell of graphene consists of a pair of two neighboring carbon atoms. The π-orbitals of these carbon atoms delocalize to form bonding and anti-bonding bands that cross each other at the corners of the Brillouin zone (K or Dirac points) (Fig. 6c).[12] Near the K points, the bands have a linear dispersion relation, $E = \hbar v_F |k|$, where $E$ is energy, $k$ is the wave vector, $\hbar$ is Planck's constant divided by 2π, and $v_F$ is the Fermi velocity in graphene (~$10^6$ m/s) (Fig. 6c).[12-14] This linear dispersion relation has similarities with that for massless photons (light), in contrast to the parabolic dispersion relation of electrons in conventional semiconductors. In addition, at low energies near the K points, the electron states have contributions from two sub-lattices (two carbon atoms in a unit cell) that can be represented by spinors, resulting in a Hamiltonian that is reminiscent of the Dirac Hamiltonian in quantum electrodynamics (QED). However, the electron spin in the Dirac-like equation is represented by a new quantum number that is often called a pseudospin (in addition to the real spin of the electrons). The QED-like spectrum in graphene allows the observation of novel phenomena such as Klein tunneling[256] and led to the measurement of two new kinds of anomalous quantum Hall effects (Fig. 6d).[257-259] Therefore, the linear dispersion and pseudospin are two central features of graphene and have led to the phrase "massless Dirac fermions" as a descriptor for carriers in graphene.

Another relevant feature of the graphene band structure is a zero band gap at the Fermi level, which limits the ability of a gate voltage to modulate current flow in FETs. From the perspective of charge transport, intervalley backscattering is nearly forbidden in graphene because of the required large momentum transfer that can only be supplied by small crystal defects (which are relatively rare in clean samples) or high energy optical phonons (that are only relevant for high-field transport).[113] Intravalley backscattering is also rare due to the conservation of chirality (or helicity) of Dirac fermions.[13] Consequently, the elastic mean free path in graphene has been found to be as long as a few microns, resulting in field-effect mobilities up to 200,000 cm$^2$/Vs in suspended graphene.[260] Moreover, since its two-dimensional structure is completely exposed to the environment, high carrier concentrations (up to 4 x 10$^{14}$ cm$^{-2}$) can be realized by electrostatic gating.[261] In this manner, graphene can concurrently attain high mobilities at high carrier concentrations. However, the exposed two-dimensional structure of graphene also implies that it is susceptible to interactions with substrate/environmental impurities and adsorbates. In particular, Coulomb scattering from charged impurities has been found to be the dominant factor in limiting the field-effect mobility of substrate-mounted graphene to 20,000 – 40,000 cm$^2$/Vs.[255, 262, 263] Similar to CNTs, inelastic scattering in graphene by acoustic phonons[263, 264] is rather weak, which implies that the field-effect mobility at room temperature remains dominated by impurity scattering. Remote interface phonon scattering in graphene from polar oxide substrates (such as SiO$_2$) has been shown to further reduce the mobility to 10,000-20,000 cm$^2$/Vs[263] at room temperature, thus inspiring efforts to integrate graphene with alternative gate dielectrics. For example, graphene on ultra-flat boron nitride (BN) has shown intrinsic mobility approaching 500,000 cm$^2$/Vs.[265, 266]

Fundamental charge transport experiments are often performed on graphene obtained by mechanical exfoliation. On the other hand, large-area production methods for graphene are needed for wafer-scale electronic circuits. As outlined above, two common methods for large-area production are epitaxial graphitization of SiC substrates and CVD growth of graphene on Cu and Ni films. Thus, epitaxial graphene and CVD-grown graphene exhibit field-effect mobilities in the range of 1000-1500 cm$^2$/Vs (highest 30,000 cm$^2$/Vs[267]) and 2000-5000 cm$^2$/Vs (highest 37,000 cm$^2$/Vs[268]), respectively.

**2.2.1. Graphene for digital electronics**

The high field-effect mobility of graphene has inspired significant efforts to explore its utility for digital electronics. In principle, the high mobility allows faster switching circuits, and the ideal two-dimensional structure enables ultimate scaling of the device channel.[214] The low contact resistance of graphene, in contrast to CNTs, also enables high conductance devices. However, the lack of a band gap and the resulting low on/off ratio (5-10) seriously compromises the prospects of graphene for digital electronics where an on/off ratio of $10^4 - 10^6$ is desired. Graphene shows a minimum conductivity of 4-8 e$^2$/h even at zero carrier concentration (i.e., unbiased gate) and thus it cannot be turned off completely. Although the origin of the minimum conductivity is not completely understood,[12, 14] it is widely accepted that graphene-based digital electronics will require a method to open a band gap in graphene.

Here, we discuss band gap engineering in graphene, focusing on the two most successful approaches to date. First, a band gap can be opened in graphene through quantum confinement in narrow graphene nanoribbons (GNRs) (Fig. 7a)[214], as discussed even before the realization of graphene FETs.[269] Unlike CNTs that can be found in metallic and semiconducting forms, GNRs are predicted to only exist in semiconducting forms. The confinement-induced band gap can be written as $\Delta E \approx \alpha \hbar v_F / d(nm)$, where $\alpha$ is the fine-structure constant of graphene and $d$ is the width of the GNR.[270, 271] Due to large $v_F$, a large confinement gap is expected in graphene compared to conventional semiconductors. This inverse relationship between band gap and GNR width has been verified experimentally by several groups. The band gap also depends on the type of GNR edge (i.e., zig-zag or armchair).[270, 272] In particular, a 2-3 nanometer wide perfect armchair GNRs can achieve a band gap of 500 meV. Transport in zig-zag edge GNRs is less understood, although the existence of ferromagnetic metallic edge states have been predicted.[270] The smoothness of the GNR edge is also critical since disordered edges can contribute to scattering and charge localization, thus reducing mobility.[273]

Various methods have been explored to fabricate GNRs[5] including patterning by nanolithography,[274] unzipping CNTs by plasma[275] or chemical etching,[276] chemical exfoliation of graphene (Fig. 7b),[277] nanocutting graphene and nanotubes via catalytic nanoparticles,[278, 279] epitaxial growth of graphene on templated SiC,[280] and directly synthesizing GNRs by self-assembly of polycyclic aromatic hydrocarbons.[281] Each of these methods has advantages and limitations. For example, nanolithography methods do not provide edge selectivity or smoothness, and have not produced GNRs with widths smaller than 20 nm.[274] In contrast, chemical exfoliation of graphene has produced sub-10 nm GNR FETs with an on/off ratio up to $10^6$ and mobility of 100-200 $cm^2/Vs$ (Fig. 7b).[277] Although graphene FETs are usually p-type in ambient conditions, chemically reactive GNR edges can be functionalized with nitrogen to obtain n-type GNR FETs.[282] The band gap energy versus GNR width is plotted in Fig. 7c from a compilation of experimental and theoretical data.[214, 274, 277, 283-285]

Quantum confinement in graphene can also be achieved by creating nanomeshes in micron-scale graphene FETs, leading to on/off ratios up to 100.[286] Similarly, a lateral confining potential can be achieved through covalent chemical modification of graphene. For example, chemisorption of aryl diazonium salts on graphene has provided evidence of band gap opening in scanning probe microscopy and spectroscopy.[287] One advantage of this approach is that graphene can potentially be patterned in arbitrary geometries containing chemically functionalized semiconducting regions with non-functionalized regions for metal contacts. Covalent modification also offers possibilities for additional functionalization in the exposed regions of graphene.[288] Although graphene is considered chemically inert, stoichiometric derivatives have been obtained by reacting with atomic hydrogen, fluorine, and oxygen. Fully hydrogenated graphene (graphane) is obtained by reacting both sides of graphene with atomic hydrogen,[289] and graphane has been shown to be a wide gap semiconductor with a band gap of 3.5 eV. Fully fluorinated graphene (fluorographene) obtained through reaction with xenon difluoride also has a band gap of ~3 eV.[290] Oxygenation of graphene can also be achieved by reacting graphene with atomic oxygen.[291] The resulting chemically uniform and reversible epoxide bonds in atomic oxygen functionalized graphene should be contrasted with insulating graphene oxide obtained from Hummer's method[63] that contains a heterogeneous mixture of carboxylic, hydroxyl, and carbonyl groups.[292]

The second approach to create a band gap in graphene is to break the pseudospin symmetry of the K and K' carbon atoms in graphene.[271] Since the two carbon atoms are only 1.4 Å apart, a sublattice selective interaction or chemical modification are viable routes to break the symmetry. For example, superlattice interactions in a boron nitride-single layer graphene stack[293, 294] or boron nitride-bilayer graphene stack[295] could in principle induce a band gap, although the requisite atomically precise alignment of graphene on boron nitride has not yet been achieved. A band gap can also be opened by creating a lateral superlattice potential in epitaxial graphene grown on SiC.[59, 296, 297] However, SiC-induced n-type doping pushes the Fermi level of graphene into the conduction band, so this approach has not yielded useful electronic devices.

Alternatively, band gaps have been achieved in bilayer graphene (BLG) and trilayer graphene. BLG consists of Bernal stacking of two graphene layers where half of the carbon atoms in one layer align with carbon atoms in other layer, and the other half of the atoms occupy the centers of the hexagons (Fig. 7d).[298] Intrinsic BLG is also gapless with parabolic valence and conductance bands touching each other at the K point (Fig. 7a).[257] However, a vertically applied electric field has been predicted to modify the valence and conduction bands, which would ultimately lead to a finite gap (Fig. 7a,d,e).[299, 300] An electric field induced band gap of more than 100 meV was observed in electrical (Fig. 7e)[298] as well as optical measurements.[301, 302] Since the band gap depends on the strength of the vertical field, ultra-thin high-κ gate dielectrics are desired for a large band gap. Since the deposition of a top-gate dielectric on graphene is also important for RF electronic applications, this topic will be discussed in detail below. Thus far, the highest on/off ratio reported is 100 at room temperature (band gap ~ 130 meV) for a 10 nm thick $HfO_2$ top-gate dielectric.[298] Similarly, vertically applied electric fields in trilayer graphene have produced a band gap of more than 100 meV.[303]

In addition to a band gap, controlled n-type and p-type doping are desired for CMOS-based graphene digital electronics. Graphene FETs are intrinsically ambipolar but, unlike CNTs, graphene doping cannot be controlled by varying the work function of the contact metal. Decorating graphene with charged species shifts the charge neutrality point to either positive or negative gate bias but the transfer characteristics around the Dirac point remain ambipolar.[262, 304] In contrast, chemical modification of graphene by substitutional doping has been predicted to dope graphene in a manner that results in asymmetric electron and hole conduction.[305] Molecular simulations suggest that replacing a $sp^2$ hybridized carbon atom in graphene with boron and nitrogen could dope graphene n-type and p-type, respectively.[305-307] Carbon-boron-nitrogen hybrids have also been grown, but it was found that boron and nitrogen tend to segregate to form domains of hexagonal boron nitride within the graphene lattice.[308] Thus far, substitutional doping has led to reduced mobilities.[282, 306, 309] Moreover, no experimental evidence of band gap opening has been reported yet by substitutional doping. Therefore, many fundamental materials issues remain to be addressed before graphene can be realistically considered for CMOS-based digital electronics.

### 2.2.2. Graphene for radio frequency analog circuits

The materials and device architecture requirements for high performance RF electronics were introduced above when discussing CNT-based RF analog circuits. Here, we highlight the properties of graphene that are best suited for RF electronics and summarize recent progress towards graphene-based RF analog devices. Since switching a device 'off' is not essential for analog circuits (e.g., amplifiers), the lack of a band gap in graphene does not necessarily impede its potential for RF electronics.[214] A large cut-off frequency ($f_T$) requires a large transconductance and small parasitic capacitance of the device channel. For a small signal RF amplifier, an AC input signal is superimposed on the DC gate-source voltage, and the output amplified signal is measured at the drain-source terminal while the device is in the 'on' state.[148] One critical aspect of this amplification process is that the drain conductance is desired to be minimal (i.e., large output impedance) for high $f_T$.[148] In other words, current saturation in the output characteristics is an essential feature of high speed RF electronics. We discuss this issue in detail below because graphene FETs, unlike CNT FETs, show rather unusual current saturation. Overall, the following four features are desired for high speed RF electronics: (1)

Short gate length; (2) Large transconductance; (3) Small contact resistance; (4) Drain current saturation.

Graphene is an excellent candidate for the first three RF device requirements. As a true two-dimensional material, graphene should be ideal in the limit of short gate length devices. Furthermore, high carrier mobility in graphene at room temperature (typically 10,000-20,000 cm$^2$/Vs on SiO$_2$)[56, 263] results in a normalized transconductance parameter of up to 7 mS,[310] which is higher than the transconductance in state-of-the-art Si MOSFETs and GaAs HEMTs.[214] The contact resistance of graphene (500-1000 $\Omega$cm[311]) is much smaller than that for a single CNT (12.9 k$\Omega$), but it is still an order of magnitude higher than in Si MOSFETs and GaAs HEMTs.[214] On the other hand, the saturation velocity in graphene (4 x 10$^7$ cm/s) exceeds that of GaAs HEMT and Si MOSFETs by approximately a factor of two.[312]

A potential issue for graphene RF devices is the absence of strong drain current saturation (Fig. 8a).[313] This weak saturation behavior in graphene can be explained as follows. At low bias, graphene shows a linear $I_d$-$V_d$ characteristic where transport is dominated by one carrier type throughout the channel (Fig. 8a). Current then begins to saturate at a $V_d$ such that $V_d - V_g$ is equal to the charge neutrality point gate voltage such that the Dirac point enters the channel near the drain electrode. At even higher bias, the channel region near the drain is doped with opposite charge carrier types, and the $I_d$-$V_d$ curve enters a second linear region (Fig. 8a).[313]

Bottom-gate graphene FETs impose a large parasitic capacitance, and thus they are not compatible with the fabrication of short channel devices. Consequently, top-gate FETs with high-$\kappa$ gate dielectrics are a more relevant geometry for high speed RF electronics.[214] Although the inert basal plane of graphene is not amenable to the growth of most dielectric materials, several methods have been developed to grow top-gate dielectrics on graphene since the first report of a top-gate graphene FET in 2007.[314] Two of the most common methods are oxidation of evaporated metals (e.g., Al) on graphene[315, 316] and atomic layer deposition (ALD) of dielectrics seeded by a spin-coated polymer on graphene.[298, 317] The first method has limited applicability since it limits the choice of dielectric materials to oxides of reactive metals, and it involves evaporation of hot metal atoms directly on graphene. The second approach involves a low-$\kappa$ 5-10 nm thick polymer film that reduces the effective gate capacitance. Other approaches based on seeding ALD through ozone functionalization[318] also degrade the underlying graphene. In contrast, self-assembled monolayers of perylene-3,4,9,10-tetracarboxylic dianhydride (PTCDA) have shown to be an effective seeding layer for oxide growth by ALD.[319] This non-covalently bound molecular layer also effects minimal perturbation of the electronic structure of the underlying graphene.[320] ALD-grown $Al_2O_3$ and $HfO_2$ gate dielectric stacks have demonstrated a high gate capacitance of 700 nF/cm$^2$ at a low gate leakage current density of 5 x 10$^{-9}$ A/cm$^2$.[319]

Early studies of graphene-based RF devices using top-gated, mechanically exfoliated graphene on highly resistive Si substrates demonstrated a cut-off frequency ($f_T$) of 26 GHz.[321] However, a completely insulating underlying substrate is necessary for further increases in speed. Therefore, epitaxial graphene grown on insulating SiC has been a natural choice for RF device research. A breakthrough study reported wafer-scale arrays of graphene RF devices on SiC (gate length of 240 nm) operating at $f_T$ = 100 GHz (Fig. 8b).[322] This high $f_T$ was achieved despite the modest carrier mobility (1500 cm$^2$/Vs) in epitaxial graphene. Similar devices later enabled wafer-scale integrated circuits containing broadband RF mixers operating at 10 GHz (Fig. 8c).[323] Recently, CVD-grown graphene transferred onto diamond-like carbon substrates achieved $f_T$ = 155 GHz at a gate length of 40 nm.[324] Unlike Si MOSFETs, these devices show no carrier freeze-out at low temperatures (4.3 K). Self-assembled core-shell nanowires have also been employed as high performance top-gates for graphene RF devices.[325] Nanowires enable fabrication of self-aligned source-drain electrodes, resulting in reduced parasitic capacitance. Graphene FETs with a channel length 140 nm (defined by the nanowire diameter) show a transconductance of 1.27 mS/μm and $f_T$ of 100-300 GHz. It is also possible to achieve reasonable RF performance using solution-processed single layer graphene, with devices fabricated on flexible substrates showing unity current gain at 2.2 GHz (Fig. 8d).[326]

Recently, graphene has also shown promise for frequency multiplication. The v-shaped transfer curve of graphene FETs has been exploited for frequency doubling at frequencies up to 100 kHz.[327] Similarly, a circuit consisting of two graphene FETs in series produces a w-shaped transfer curve that has been utilized as a frequency tripler operating at 1 kHz.[328] Overall, considering the impressive advances achieved over the relatively short span of graphene-based RF research, this application appears considerably more promising than graphene-based digital electronics. However, the absence of strong current saturation is a clear weakness that will need to be addressed for graphene to realize its full potential for commercial RF applications.

### 2.2.3. Graphene for optoelectronics

Graphene has several attributes that makes it well-suited for optoelectronic applications.[16] For example, its linear dispersion with zero band gap suggests the possibility of widely tunable optical excitations (Fig. 9a)[13], leading to an optical absorption spectrum that is featureless over wavelengths ranging from 300 nm to 2500 nm.[329] A single layer of graphene absorbs 2.3% of the incident light with minimum reflection (<0.1 %) over this wavelength range.[330] The transmittance of graphene (1-πα ~ 97.7 %) is given by the effective fine structure constant (α) of graphene that depends on the dielectric constant of the environment.[330] The absorption peak at 250 nm results from a saddle-point singularity near the *M* point in the Brillouin zone of graphene.[329, 331] Graphene is not luminescent but its chemical derivatives such as graphene oxide (GO) exhibit photoluminescence over a broad range.[332, 333] This light emission has been speculated to occur in islands of $sp^2$ carbon within GO or at oxygen-induced defect sites. Recently, electroluminescence from pristine graphene was also reported,[334] although the underlying light emission mechanism was found to be different from that in GO. Specifically, optical phonon-assisted radiative recombination of carriers results in light emission in graphene (similar to metallic CNTs).[334]

While graphene itself may have limited potential as a light-emitting material, it is effective as a transparent conductor in flexible organic light-emitting diodes (OLEDs), organic TFTs, and organic photovoltaics (OPVs). Graphene-based transparent conductors show performance metrics close to that of ITO in addition to superior mechanical flexibility.[16] However, the minimum finite conductivity of undoped graphene (4-8 $e^2/h$) results in a sheet resistance ($R_s$) of 4-7 kΩ for high-quality single-layer graphene, thus necessitating chemical doping methods for graphene-based transparent conductors, as will be described in detail below. Here, we compare the performance metrics of CNT thin films and graphene-based transparent conductors with ITO in Fig. 9b.[16] The theoretical curves of graphene at two different doping levels and mobilities are also plotted. The best CNT thin films show slightly larger $R_s$ than ITO at the same transparency. On the other hand, graphene outperforms ITO for transparencies above 90%, and it is expected to improve further with the availability of higher quality large-area graphene (Fig. 9b).

Graphene has also been used in photodetectors where optically generated electron-hole pairs can be separated by an externally applied bias. Unlike CNTs and conventional bulk semiconductors, the linear dispersion of graphene provides a uniform photoresponse from the THz to the ultraviolet range.[16] In addition, the high mobility of graphene has been reported to yield ultra-fast photodetection up to 40 GHz.[335] Since the operation speed in this early report was limited by parasitic capacitance, it is anticipated that the intrinsic photoresponse speed of graphene could be as high as 500 GHz. An internal built-in electric field at metal-graphene contacts can also provide an efficient way to achieve photodetection. Metal-graphene contacts have shown internal photocurrent efficiencies of 15-30% and external photocurrent yield of 6 mA/W (Fig. 9c,d).[336] Recently, photoresponse was also observed in a p-n junction created via top-gate architecture on graphene FETs.[337, 338] This device exploits hot carriers in graphene, the non-local transport of which contributes to the photoresponse in additional to the expected photovoltaic effect.

Another related optical application of graphene-derived materials is in bioimaging. In particular, photoluminescence in nanoscale graphene oxide has been used for live cell imaging in the visible and near infrared.[110] For in vivo applications, issues surrounding the potential toxicity of graphene will need to be addressed, although a recent report suggests that encapsulation of graphene in the biocompatible block copolymer Pluronic can reduce toxicity and inflammation in the lungs of mice.[339] These biocompatible dispersion methods could lead to additional opportunities for graphene-based biomedical applications including imaging contrast agents and drug delivery.

### 2.2.4. Emerging graphene device concepts

While the absence of a band gap poses a serious issue for graphene-based digital electronics, the unique electronic properties of graphene hold promise for fundamentally different device architectures. One such device is the bilayer pseudospin field-effect transistor (BiSFET).[340] In a BiSFET, two single layers of graphene are separated by an ultra-thin dielectric. Under certain conditions, the electrons in one graphene layer and holes in the other graphene layer can form bound excitons. These bosonic quasiparticles then reduce the tunneling resistance through the ultra-thin dielectric, leading to significant current flow at low bias. However, at elevated bias, the bosonic state is disrupted, inducing increases in the tunneling resistance and corresponding reductions in current. The net result is a decreasing current with increasing voltage (i.e., negative differential resistance). While simulations predict that the BiSFET should be operational at room temperature, this device has not yet been demonstrated experimentally, presumably due to challenges in fabrication. Nevertheless, a related device architecture has led to a successful switching ratio of up to $10^4$ by utilizing an alternative operating principle.[341] These vertical graphene heterojunctions were fabricated by separating two graphene layers with atomically thin boron nitride and $MoS_2$ on a $SiO_2$/Si substrate. Biasing the Si substrate (gate) changes the Fermi level on the two graphene layers by different amounts due to differences in screening. The resulting difference in carrier density then induces tunneling which can be further controlled by applying a vertical bias via the top layer of graphene (drain), ultimately yielding an output characteristic that resembles a MOSFET. Recently, a novel triode-like device, termed a barrister, was also realized using a Schottky junction between the graphene and a hydrogenated Si substrate (Fig. 9e,f).[342] This device relies on the lack of Fermi level pinning due to the absence of dangling bonds in graphene and hydrogen-passivated Si. In this case, the gate-controlled Schottky barrier allows current modulation by up to 5 orders of magnitude (Fig. 9f). Inverters and half-adder circuits were also demonstrated using graphene barristers, thus providing a pathway for graphene-based digital logic.

**3. Carbon Nanomaterials for Photovoltaic Applications**

In addition to electronics, an emerging applications area for carbon nanomaterials is phovoltaics. The mechanical flexibility, chemical stability, and elemental abundance of carbon nanomaterials present unique opportunities for solar technology. For example, fullerenes are effective electron transport materials in organic photovoltaics (OPVs), while CNTs and graphene hold promise as transparent conductors. Recent work has also shown progress in the utilization of CNTs and graphene as the photoactive components in solar cells. This section will explore recent progress and future prospects of fullerenes, CNTs, and graphene for solar energy applications.

**3.1. Fullerenes in photovoltaics**

Although fullerenes were the first carbon nanomaterials to be experimentally isolated, they are relatively underutilized in electronics in comparison to CNTs or graphene. However, in solar technology, the situation is the reverse as fullerenes and their derivative are among the most important electron acceptor materials in OPVs. OPVs are solar cells fabricated from conjugated organic small molecules or polymers. The low temperature, solution-based processing of organic materials makes OPVs a promising option for large-area, mechanically flexible solar technology. Fullerenes were recognized as strong electron acceptor materials for OPVs as early as 1992 with the first report of charge transfer from poly*[2-methoxy-5-(2'-ethylhexyloxy)-p-phenylene vinylene]* (MEH-PPV) to $C_{60}$ by Sariciftci et al.[343] Thereafter, the same group reported the first photovoltaic cells from bilayer heterojunctions of MEH-PPV (donor) and $C_{60}$ (acceptor).[344] Other donor polymers were subsequently reported in a follow up study.[345] However, the efficiency values of bilayer devices were limited to about 0.05% due to the limited surface area of the bilayer heterojunction and the short diffusion length of photogenerated excitons.

In an effort to address the limitations of bilayer OPVs, the bulk heterojunction concept was introduced in which the donor and acceptor were incorporated into an interpenetrating network. Due to the poor miscibility and solubility of $C_{60}$, it is challenging to form bulk heterojunctions from pristine fullerenes. Consequently, the Wudl group developed a chemically modified form of $C_{60}$ in 1995 ([6,6]-phenyl-$C_{61}$-butyric acid methyl ester (PCBM) (Fig. 10 (a)) that improved solubility in many organic solvents.[346] PCBM was subsequently combined with MEH-PPV to form a phase-segregated, interconnected bulk heterojunction with large junction area. These bulk heterojunction OPVs had efficiencies that were approximately 60x higher (2.9%) than bilayer heterojunction devices.[347] PCBM has since become one of the best performing fullerene derivatives and is often cited as a benchmark for new acceptor materials. Recent PCBM analogs include 1-(3-methoxycarbonyl)propyl-1-thienyl-[6,6]-methano-fullerene (ThCBM)[348] and [6,6]-phenyl-$C_{71}$-butyric acid methyl ester (PC$_{71}$BM)[349] as shown in Fig. 10b[350] and 10c[350]. Since PC$_{71}$BM absorbs a significantly larger proportion of incident solar light, it is a promising candidate for higher efficiency OPVs, although significantly more expensive.[351] Similarly, metalloendohedral fullerenes have shown improved efficiency values.[352] Additional recent developments in fullerene-based acceptor materials for OPV devices are well summarized in recent reviews by He et al.[353] and Delgado et al.[350]

## 3.2. Carbon nanotubes in photovoltaics

The unique optical, electrical, chemical, and mechanical properties of CNTs make them enticing materials for photovoltaic applications. Furthermore, the optical excitation of strongly bound excitons in semiconducting SWCNTs at room temperature shows similarities to the conjugated organic molecules and polymers that are commonly employed in OPVs.[131, 354] In addition, CNTs show significantly higher carrier mobility and reduced trap density compared to organic electronic materials. Consequently, CNTs are gaining popularity as components of photovoltaic devices.[355-358]

One of the earliest attempts focused on the incorporation of CNTs with the conjugated polymer poly 3-octyl thiophene (P3OT) as the active layer in OPVs. While this CNT-P3OT device showed a large open circuit voltage ($V_{oc}$) of 0.75 V, the power conversion efficiency (PCE) was low at 0.062%.[359] Annealing the CNT-P3OT system for 10 min at 120ºC led to modest improvements in PCE to 0.22 %.[360] Later, Berson et al.[361] incorporated both single-walled and multi-walled CNTs (in separate devices) as additives to a poly 3-hexathiophene (P3HT):PCBM active layer mixture, resulting in a 100% increase in short circuit current density ($J_{SC}$). However, a tradeoff was observed between the $J_{SC}$ and the fill factor (FF) which is the ratio of maximum obtainable power to the product of $V_{OC}$ and $J_{SC}$, as well as between $J_{SC}$ and $V_{OC}$, ultimately limiting PCE values to 1.3% and 2% for SWCNTs and MWCNTs, respectively. This tradeoff was attributed to increased short circuit contacts by CNTs with high aspect ratios.[361] Recently, CNT-induced crystallinity and improved ordering of the active layer have been predicted by molecular dynamics simulations[362] and observed experimentally.[363, 364] CNTs have also been incorporated hierarchically between various layers of an OPV device, namely the anode, hole transport layer, bulk heterojunction active layer, and cathode.[365] It was observed that the presence of CNTs at the anode or hole transport layer were beneficial to the cell performance, improving PCE to 4.9% versus 4% for the control device. This improvement resulted from a reduction in the series resistance due to conductive pathways from interpenetrating CNTs in the hole transport layer.[365]

All of the above studies utilized as-grown material that contains both metallic and semiconducting CNTs, suggesting that further improvements may be possible using electronically and/or optically monodisperse CNTs. For example, a comparative analysis of enriched semiconducting and metallic CNTs in time-resolved microwave conductivity measurements revealed the negative impact of metallic CNTs in OPVs. In particular, 90% semiconducting SWCNT-P3HT composites showed a 45% increase in carrier decay time compared to 88% metallic SWCNT-P3HT composites.[366] This effect was recently exploited in functional OPV devices where purified semiconducting SWCNT:P3HT blends were used (see Fig. 11a (inset)) to get PCE values of 0.72%.[364] In this case, the SWCNT is apparently acting as the electron acceptor. However, photosensitive capacitor measurements by the Arnold group have shown that semiconducting SWCNTs can either act as acceptors with conjugated poly thiophenes or as donors in blends with fullerenes.[367] Subsequent work illustrated the use of enriched semiconducting SWCNTs as donor materials in SWCNT/$C_{60}$ bilayer heterojunction devices, achieving PCE values of 0.6 % with an internal quantum efficiency approaching 100% (Fig. 11b).[368]

From these reports, it is clear that CNTs can be successfully incorporated into the active layer of OPVs. However, the debates continues whether semiconducting SWCNTs are appropriate acceptor materials to completely replace fullerene derivatives or are better suited as additives to improve morphology, absorption, charge separation, and charge collection. Recent studies on planar heterojunctions[369] do suggest that semiconducting SWCNTs can be as effective as fullerenes for electron acceptors, and that their performance in the bulk heterojunction geometry[370] is primarily limited by bundling and CNT-CNT contacts.[369, 370] The detailed control of CNT morphology in bulk heterojunction OPVs remains an active area of investigation.

In addition to the active layer of photovoltaic devices, CNTs show promise as transparent conductive electrodes. As discussed above, indium tin oxide (ITO) is the most common transparent conductive electrode in organic light-emitting diodes and OPVs.[371] However, due to their greater mechanical flexibility and the earth abundance of carbon, CNTs hold promise for overcoming the brittleness that can compromise device lifetime with ITO, especially on flexible substrates, and the scarcity of indium.[372,373] The earliest reports on CNT electrodes for OPVs focused primarily on MWCNTs,[374, 375] the high surface roughness and inhomogeneity of which led to suboptimal performance. Subsequent work addressed these issues, leading to CNT-based electrodes with OPV device performance comparable to that of ITO.[376,377] Although similar performance metrics were reproduced in later studies,[378] the anisotropic nature of CNT films showed a strong dependence of transparency and cell performance on incident light angle.[379] The polydispersity of as-grown CNT films also compromised performance since the contact resistance between CNTs of different electronic types is about 100x higher than that CNTs of the same electronic character.[171, 380] Thus, CNT films of monodisperse metallic or semiconducting CNTs are expected to have a lower $R_s$ compared to a heterogeneous mixture of as-grown CNT films. Recently, the effect of SWCNT electronic type purity was systematically studied in the transparent anode for OPVs.[381] Electrodes based on 99.9% pure metallic SWCNTs were found to perform 50x better than 99.9% semiconducting SWCNTs and showed comparable performance to that of control ITO-based devices.[381] The improvement was primarily due to a rise in $J_{SC}$ as seen in Fig. 11c.[381] The SWCNT films used in this study were doped with nitric acid to reduce roughness and sheet resistance.[373] Since nitric acid treated semiconducting SWCNTs exhibit lower sheet resistance than comparably treated metallic SWCNTs,[382, 383] the dramatically reduced PCE for semiconducting SWCNTs appears counterintuitive. However, it was shown that the films were dedoped by the hole transport layer PEDOT:PSS in the fabricated OPVs,[381] thus compromising the sheet resistance of the semiconducting SWCNTs to a much greater degree than the metallic SWCNTs. Further improvements in sheet resistance and thus OPV power conversion efficiency can be achieved with longer CNTs as is evident in Fig. 11d.[373]

Beyond OPVs, CNTs have been explored in a variety of other photovoltaic devices, especially photoelectrochemical or dye sensitized solar cells (DSSCs).[384] The earliest studies incorporated as-grown CNTs in the titania ($TiO_2$) nanoparticle matrix to improve cell efficiency.[385] PCE values of ~4.5 % were achieved, which was comparable to that of control cells.[386] Refluxing acid treatment of CNTs introduces carboxyl groups that improves the binding of $TiO_2$ nanoparticles and dye molecules, thereby raising $J_{SC}$ by 25% without compromising $V_{OC}$.[387,388, 389] Vertically aligned MWCNT forests and $TiO_2$ composites also improve $J_{SC}$ by 5x versus bare $TiO_2$.[390] In one of the seminal papers in this field, CNT electrodes were incorporated with $TiO_2$ nanoparticles to fabricated DSSCs having internal power conversion efficiencies of 16% compared to 7.3% for bare $TiO_2$ at 350 nm.[391] This improvement was mainly attributed to better dispersion of the $TiO_2$ nanoparticles and enhanced charge transport to the electrode by the CNTs.[391] When this assembly was packed in a full cell and sensitized with a dye, the IPCE values were maintained. Furthermore, the $J_{SC}$ value increased by 45%, which corroborates the high IPCE values, but the $V_{OC}$ fell somewhat due to charge equilibration between the $TiO_2$ and the CNTs. The net result was an external PCE value of ~0.18%.[392] PCE values greater than 5% have also been recently achieved using vertically aligned CNT forests as counter electrodes in iodine free DSSCs.[393] In terms of monodisperse CNTs, Belcher and coworkers dispersed semiconducting SWCNTs in TiO$_2$ nanoparticles via virus-templated self-assembly.[394] The TiO$_2$ nanoparticles were directly biomineralized on top of the CNTs to achieve superior contact and charge separation. A TEM image of the resulting biomineralized TiO$_2$-CNT structure is shown in Fig. 11e.[394] For semiconducting CNT contents as low as 0.2 wt% in TiO$_2$, the PCE value was 10.3% compared with 8.3% for no CNTs and 6.2% for 0.2 wt% metallic CNTs.[394]

Another emerging class of photovoltaic devices is based on semiconducting quantum dots.[395] CNTs have been utilized in such systems as acceptor materials, analogous to the role of fullerenes in OPVs. Early attempts at using CNTs as acceptors with porphyrins resulted in monochromatic (435 nm) IPCE values of 8.5%.[396] Subsequently, cadmium sulphide (CdS) nanoparticles were covalently attached to CNTs, and the photoelectrochemical behavior of the resulting system showed an internal quantum efficiency of 25%. The photocurrent improved with increasing CNT content, which led to the conclusion that the CNTs were acting as efficient charge separation and transport sites.[397] Soon studies using CdTe[398] and CdSe[399] followed with similar results. A schematic of the CNT-quantum dot charge separating junction is shown in Fig. 11f.[399] Stacked-cup CNTs were also used with CdSe, leading to a 10-fold increase in photocurrent compared to the control sample.[400] It has been hypothesized that both metallic and semiconducting CNTs would play a similar role in charge separation and transport for such devices,[399] however, no studies using electronic type sorted CNTs have been reported to date. Although both metallic and semiconducting CNTs may act as effective charge carriers, semiconducting CNTs with a finite band gap could also act as charge separation sites, whereas metallic CNTs could act as recombination centers in addition to increasing the short circuit current, thus suggesting the attraction of monodisperse semiconducting CNTs in this application.

Since semiconducting SWCNTs are direct band gap materials, they can potentially be used alone as the active layer in photovoltaic devices. Moreover, multiple exciton generation in CNTs[239] and photovoltage multiplication in CNT arrays[401] provide further motivation for CNT-based solar cells. However, the difficulty in achieving stable n-type doping of CNTs in ambient conditions has hindered efforts to realize large-area CNT-based p-n junctions. To overcome this issue, researchers have begun integrating CNTs with conventional n-type semiconductors to realize a photovoltaic effect. A pioneering approach in this direction was taken by the Wu group who integrated double-walled carbon nanotubes (DWCNTs) with a single crystal n-type silicon wafer to create a p-n junction thin-film solar cell with a PCE value greater than 1%.[402] A schematic of this device structure is provided in Fig. 11g.[402] In this heterojunction geometry, the CNTs participate in charge separation as well as charge collection. Further optimization of the Si-CNT interface improved the efficiency to 7%. Here, the CNT networks contained both metallic and semiconducting species that created both metal-semiconductor Schottky junctions and semiconductor-semiconductor heterojunctions with Si, respectively.[403] These results were then reproduced by another group who used air-brushed CNT networks. These CNT network films were additionally doped with thionyl chloride ($SOCl_2$), leading to cell efficiencies of 4.5%.[404] A similar doping strategy was employed by the Wu group using nitric acid ($HNO_3$) as the dopant, leading to improvements in efficiency to 11% versus 6% for undoped CNTs.[405] The current-voltage characteristics of their cells before and after doping are shown in Fig. 11h.[405] CNTs have also been integrated with CdTe nanobelts[406] and Si nanowires[407] to form nanostructured versions of the CNT-semiconductor solar cell design. Future work will likely explore the integration of semiconducting p-type CNTs with other inorganic n-type semiconductors with the goal of further improving the performance and processability of CNT-semiconductor solar cells.

### 3.3. Graphene in photovoltaics

Due to the lack of a band gap in graphene, it is not well-suited as an active layer material for efficient carrier generation and separation in photovoltaic devices. However, modern solar cells consist of an assembly of several layers such as transparent conducting electrodes and charge blocking interfacial layers. In this regard, the large conductivity and high transparency of single layer graphene make it a promising material for transparent conducting electrodes in OPVs. In addition, graphene and its chemically modified derivatives are potentially candidates for charge blocking interfacial layers. This section reviews recent efforts to incorporate graphene in photovoltaic devices.[16, 408, 409]

The first demonstration of a graphene-based electrode in a solar cell was in a DSSC device in 2007. Here, reduced graphene oxide (r-GO) was used as the electrode[410] without degrading cell efficiency compared to a control device with an ITO electrode. Similar r-GO electrodes were employed in small molecule[411] and polymer[412] OPVs as well as hybrid solar cells.[413] The cell performance in all of these reports[411-413] was equal or inferior to control ITO devices mainly because of the large sheet resistance of graphene (1kΩ/sq to 1MΩ/sq). In particular, the discontinuous boundaries between interconnected r-GO flakes, as seen Fig. 12a,[414] contribute to the large sheet resistance. Recently, an r-GO/CNT composite significantly reduced the sheet resistance ($R_s$ = 240 Ω/sq) at 86% transparency, albeit with a PCE of only 0.85%.[415]

To overcome the limitations imposed by percolating networks of small r-GO flakes, researchers have turned to large-area continuous graphene grown by CVD. Wang et al. were the first authors to report CVD-grown graphene as anodes for polymer (P3HT:PCBM) OPVs.[416] They reported PCE values of 0.21% for as-grown graphene and 1.71% for graphene modified with pyrene buanoic acid succidymidyl ester (PBASE). Nevertheless, the PCE values of the chemically treated graphene remained inferior to the control ITO device (3.10%) due to the lower $J_{sc}$ value in the graphene anode device. The lowest reported sheet resistance value of CVD graphene is 210 Ω/sq at a 72% transparency,[416] which is still significantly larger than that of ITO. For CuPc/$C_{60}$ based OPVs,[417] the PCE values observed for CVD graphene and control ITO devices were nearly equal (1.18% and 1.27%, respectively) even though the $R_s$ value for ITO (25 Ω/sq) was more than two orders of magnitude lower than that of graphene (3.5 kΩ/sq). In addition, the graphene-based devices were more robust to repeated bending versus the ITO devices which failed at a bending angle of 60%, as seen in Fig. 12b.[417] CVD-grown graphene films have been used subsequently in CdTe/CdS-based quantum dot solar cells with PCE values of 4.17 %.[418]

Doping strategies have been attempted by many groups to achieve higher carrier concentrations and thus lower $R_s$ values in graphene. Following early reports on CNTs,[381, 382] doping of graphene has been achieved through chemical treatments with nitric acid and gold chloride.[329, 419-422] However, even these doped CVD graphene anode devices fail to surpass the performance of control ITO anode devices.[420] A recent effort to dope mechanically exfoliated few layer graphene intercalated with $FeCl_3$ resulted in a record low $R_s$ value 8.8 Ω/sq with a transparency of 84% in the visible range. For CVD graphene, the lowest $R_s$ value reported to date is 30 Ω/sq with a transparency of 90%.[329] These values are well within the minimum industry standard of 100 Ω/sq at 90% transparency.[423] However, these reports did not integrate these high-performance graphene transparent conductors into fully fabricated photovoltaic devices. Stable chemical doping following device integration thus remains an important area for future investigation.

Beyond transparent anodes, graphene-based materials have been explored as charge blocking interfacial layers for OPVs. The conventionally used hole transport layer (HTL), PEDOT:PSS, suffers from photo-induced degradation, ITO corrosion, and sensitivity to ambient conditions that compromises the lifetime of OPV devices.[424-426] Lithium fluoride is the canonical electron transport layer (ETL), but is typically deposited in vacuum, thus conflicting with the solution processing methods that are often cited as the advantage of OPVs compared to competing photovoltaic technologies. While pristine graphene is not a promising choice for charge blocking layers due to its zero band gap, chemically functionalized graphene (e.g., graphene oxide (GO)) can possess a tunable band gap and work function based on processing conditions.[427, 428] The work function of GO is ~4.7-4.9 eV, thus presenting a small hole injection barrier from P3HT, which has a work function ~4.3 eV. In addition, the band gap of GO, which can vary over the range of 2.8 eV to 4.2 eV, is sufficiently large to prevent electron transport from PCBM. GO was first demonstrated as a HTL in OPVs by the Chhowalla group in 2010.[429] In particular, OPV devices with P3HT:PCBM as the active layer were fabricated with varying thicknesses of spin-coated GO as HTLs in addition to two kinds of control devices using PEDOT:PSS as the HTL and no HTL. The highest PCE value (3.5%) was observed for the lowest GO HTL thickness of ~2 nm, and the PCE steadily fell with increasing GO thickness due to increasing series resistance and decreasing optical transmittance. The best GO PCE value was nearly equal to the control device with 30 nm thick PEDOT:PSS as the HTL (3.6%).[429] Similar results were obtained by Yun et al. with GO partially reduced by p-toluenesulfonyl hydrazide.[430]

Further advances in the use of GO as a HTL were demonstrated by Murray et al.[431] In this case, the device active layer consisted of poly[[4,8-bis[(2-ethylhexyl)oxy]benzo[1,2-b:4,5-b0]dithiophene-2,6-diyl][3-fluoro-2-[(2-ethylhexyl)carbonyl]-thieno[3,4-b]thiophenediyl]] (PTB7) and the fullerene electron acceptor $PC_{71}BM$, which achieves significantly higher PCEs than do P3HT:PCBM OPVs.[432, 433] The reported PCE values for the GO (7.39%) and control PEDOT:PSS devices (7.46%) were nearly equal as seen in Fig. 12c.[431] More importantly, the GO-based devices were far more durable and robust under high humidity (80%) and high temperature (80ºC) conditions. In addition, it was found that the PTB7 stacks in a more ordered manner compared to PEDOT:PSS, thus leading to better electronic coupling and charge transfer.[431]

Recently, cesiated GO (GO-Cs) was also used as an electron transport/hole blocking layer between PCBM and OPV cathodes. The cesium neutralizes the peripheral carboxyl groups (-COOH) in GO to –COOCs groups, yielding a work function of ~4 eV (see band diagram Fig. 12d) that provides selective electron transport. Both GO and GO-Cs were used together in normal and inverted OPV geometries with P3HT-PCBM as the active layer, and achieved equivalent or superior performance versus control devices using PEDOT:PSS and lithium fluoride (LiF) as the HTL and ETL, respectively.[434] GO-CNT composites have also been used as effective HTLs in P3HT:PCBM devices by the Huang group. It was found that a non-percolating amount of CNTs (1:0.2 GO:CNTs) in the GO film improved the carrier transport and afforded PCEs comparable in efficiency to PEDOT:PSS.[435] GO-CNT composites have since been used in tandem cells in both regular and inverted geometries.[436] All of the above examples used solution-processed GO or functionalized graphene to coat an ultra-thin films. However, solution-processed graphene films are often discontinuous and electronically inhomogeneous, thus leading to unwanted recombination currents.[431] Efforts to covalently modify large-area CVD graphene may overcome this limitation, and future work will likely explore CVD graphene that is covalently modified with diazonium salts[287] and/or fluorine[437] to achieve improved graphene-based HTLs.

Although pristine graphene lacks a finite band gap, various forms of chemically functionalized graphene materials have been used in the active layers of OPV devices. The first attempt was to use GO as an acceptor material in a simple mixture with donor polymers such as P3HT and P3OT.[438, 439] The reported PCE values were low (1.1-1.4%) but comparable to other non-fullerene acceptor devices.[438, 439] Subsequent attempts involved covalently grafting $C_{60}$ directly onto the basal plane of r-GO via a nucleophilic addition reaction.[440] Although cells with the grafted composite (see Fig. 12e[440]) had somewhat improved efficiencies (1.22%) compared to non-grafted mixtures (0.44%) in a bilayer geometry, they remain inferior to conventional P3HT:PCBM BHJ devices.[440] A similar approach was adopted on the donor side by grafting P3HT onto r-GO. However, no significant improvements in device performance over direct mixtures or control devices were observed, and the PCE values remained less than 1%.[441]

Graphene has also been explored as an additive to the active layer of DSSCs. In this case, graphene provides efficient charge transport and reduced recombination in addition to higher optical absorption due to increased light scattering. A schematic illustration of this system is provided in Fig. 12f.[442] The PCE value of a standard $TiO_2$-ruthenium dye-based DSSC with the graphene additive was reported to be 6.97% compared to 5.01% for the control sample.[442]

In addition to providing enhanced charge transport and reduced recombination, graphene has served other unique roles in emerging variants of organic solar cells. The Huang group has recently demonstrated the use of GO as a surfactant to assemble CNTs and $C_{60}$ into an all-carbon photoconductive composite with a PCE value of 0.21%.[443] In another approach, GO was mixed with PEDOT:PSS to facilitate the fabrication of tandem OPVs from P3HT:PCBM active layers.[444] Thus far, graphene has not acted as a primary light absorber in OPVs or DSSCs even though such cells have been hypothesized to have PCE values of 12% in a single cell geometry and 24% in a tandem geometry.[445] However, graphene quantum dots (GQDs) have been employed as light absorbers in TiO$_2$-based DSSCs as seen in Fig. 12g.[446]

Graphene-semiconductor Schottky junctions have also been considered for photovoltaic applications. The first graphene/p-Si Schottky junction was demonstrated by the Wu group using CVD graphene, leading to a PCE of 2.2%.[447] Subsequently, the same photovoltaic effect was realized using Si nanowires[448, 449] and cadmium chalcogenides.[450, 451] Since these devices increased the junction surface area, the PCE increased to 2.86%. Chemical doping of planar graphene/p-Si junctions using AuCl$_3$ has since been attempted to tune the work function, but the observed PCE values were less than 0.1%.[452] An alternative graphene-Si Schottky junction solar cells was achieved by chemical doping CVD graphene with bis(trifluoromethanesulfonyl)amide (TFSA).[453] In this study, the junction between n-doped silicon and p-doped graphene yielded more than a 4-fold improvement in PCE response (8.6%) compared to undoped graphene cells (1.9%). The current-voltage curves and cell schematic are shown in Fig. 12h.[453] The observed enhancement in PCE resulted from both reduced sheet resistance of the doped graphene as well as an increase in Schottky barrier height for more efficient charge separation.[453] Future efforts will likely focus on chemical methods to further improve the integration of graphene with complementary photovoltaic materials.

## 4. Carbon Nanomaterials for Sensing Applications

While the sensitivity of the electronic properties of carbon nanomaterials to the surrounding environment poses challenges in some applications, it offers a distinct advantage for sensors. For more than a decade, carbon nanomaterials have been used to sense a variety of analytes including gases, solvents, and biomolecules. While carbon nanomaterial-based electronic sensors have outperformed conventional technologies in terms of sensitivity, many challenges remain in terms of selectivity, reversibility, reusability, and scalability.[454-457] In this section, we review the achievements and challenges for carbon nanomaterials in chemical and biological sensors. Due to the presence of extensive reviews on carbon nanomaterial-based electrochemical sensors,[20,458,19,21,459-462] this section will focus primarily on electronic sensors.

## 4.1. Carbon nanotubes for sensing applications

One of the earliest reports of gas sensors based on single CNT FETs was made by Kong et al. in 2000.[463] Large shifts in threshold voltages were observed in the negative and positive directions upon exposure to ammonia ($NH_3$) and nitrogen dioxide ($NO_2$), respectively, as seen in Fig. 13a.[463] The sensitivity (defined as ratio of resistance before and after gas exposure) ranged from 100 to 1000 with response times of a few seconds to one minute. These performance metrics indicated that single CNT devices outperformed commercially available metal oxide or conducting polymer sensors.[464,465] The authors speculated that bulk doping of the CNTs by adsorbed gases was likely contributing to the observed resistance modulation. However, the lack of definitive proof regarding the sensing mechanism inspired a series of subsequent studies focusing on the mechanism of chemical sensing by CNT FETs.[466] Through these studies, the range of analytes was expanded to include alcohols,[467] oxygen,[468] benzene,[469] hydrogen,[470] and water.[471] While some papers confirmed bulk doping as the underlying mechanism,[466] others reported that the observed resistance change was a contact effect and not a bulk doping phenomenon.[123, 472, 473] The debate ultimately revealed that semiconducting CNTs can act as Schottky barrier transistors, and that the barrier height between the metal and nanotubes can be tuned by the metal work function. Since the work function of the metal can be modulated by local dipoles induced by adsorbed gas molecules, the threshold voltage and polarity of CNT FETs will correspondingly change as a function of the surrounding environment.[474]

In contrast to semiconducting CNTs, chemically induced Schottky barrier height modulation in metallic CNTs produces a lower variation in resistivity and thus reduced sensitivity. However, chemiresistive sensing is still possible using single metallic CNT devices. In this case, a stronger chemical interaction (e.g., covalent interaction) is needed to obtain large resistance changes. A significant advance in this direction was achieved by the Collins group who developed a point functionalization scheme wherein a controlled density of defects is created on a CNT (metallic or semiconducting) via an electrochemical redox reaction.[475] Selective functionalization of various chemical species is then achieved on these defect sites, thereby tuning the detection selectivity. This scheme has been used to detect single molecule reactions in CNT devices[476, 477] in addition to oxidization or reduction events based on the redox potential.[475] Selective electrodeposition of metals can also be performed at these defect sites, thus enabling hydrogen sensors for Pd-decorated CNT FETs.[478]

Although the aforementioned methods enable sensitivity down to the single molecule level, detection selectivity requires additional attention. To address this requirement, CNTs were non-covalently functionalized with moieties having selectivity to specific target analytes. For example, polyethyleneimine (PEI)-coated CNT FETs are selectively sensitive to $NO_2$, while Nafion-coated devices are selective towards $NH_3$.[479] In a unique approach to selective sensing, the Johnson group used semiconducting SWCNTs encapsulated with single-stranded DNA (ss-DNA) as shown in Fig. 13b. The DNA sequences were chosen to have specific binding affinity for a series of analytes including methanol, propionic acid, trimethylamine (TMA), dinitrotoluene (DNT), and dimethyl methylphosphonate (DMMP). The ss-DNA functionalized CNT devices showed a significant current modulation as high as 20-30% versus 0-1% in bare control CNT devices at the same exposure dose.[480, 481] Highly selective interactions between ss-DNA and CNTs of different chiralities may provide further tunability for this class of chemical sensors.[95]

In an effort to simplify fabrication, research has also been devoted to using CNT thin films and vertically grown CNT arrays for chemical sensing applications. Early reports used as-grown heterogeneous CNT thin films as chemiresistive sensors for $NO_2$,[482] $NH_3$,[483] and DMMP.[484] Specificity issues were addressed to some extent by implementing both covalent[483] and non-covalent[484] functionalizations of these CNT thin films. Since each device is effectively averaging the influence of the large number of CNTs in the thin film, device-to-device reproducibility with high sensitivity can be achieved over large areas.[485] In a related approach, CNTs were also functionalized with metal nanoparticles that improved selectivity for a variety of gases.[486] In this case, the sensing selectivity originates from the metal-CNT barrier height modulation following gas exposure.[487] Gas sensors based on vertical arrays of CNTs have also been reported.[488, 489] The conductance of vertically aligned conducting polymer-CNT films was found to be highly sensitive to vapors of small organic molecules such as THF, ethanol, and cyclohexane.[489]

Another approach for gas sensing using large-area as-grown CNTs is the chemosensitive capacitor. CNT films (see Fig. 13c) were used as one electrode of the capacitor, while a heavily doped silicon wafer was employed as the other electrode. The authors observed that the capacitance of this device was extremely sensitive (down to 0.5 ppb). The capacitance changes were also fast (~4 sec), reversible, and selective (using non-covalent functionalization) as seen in Fig. 13d, thus providing an advantage over conventional chemicapacitive sensors that take minutes to respond and refresh.[490] The sensing mechanism was attributed to both capacitance and conductance changes.[491, 492, 493]

A potential issue with sensors based on as-grown CNT films is the differential response of the constituent metallic and semiconducting CNTs. For example, although the channel length and network morphology can be tailored to achieve a net semiconducting character in the final device characteristics, the metallic CNTs in the channel will have a relatively weak response to adsorbates. It is also well known that metallic and semiconducting CNTs interact differently with adsorbates and dopants.[381, 494, 495] Monodisperse CNTs thus present an opportunity for improved sensing as explored in two recently published reports.[496, 497] In a paper by Nakano et al., monodisperse semiconducting CNTs showed an order of magnitude improvement in sensitivity compared to as-grown CNTs as shown in Fig. 13e.[496] The other study focused on arrays of semiconductor CNT (sorted via selective dispersion and DGU) devices, revealing that CNTs with narrow diameter distributions near 1.4 nm have the highest sensitivity to hydrogen (see Fig. 13f).[498]

Beyond gases and small organic molecules, the electrical response of CNTs can be modulated by biomolecular adsorbates, thus providing opportunities for biological sensors. An early study used as-grown CNT TFTs to detect proteins.[499] In these devices, non-specific binding was suppressed by passivating the CNTs with polyethylene oxide (PEO) based molecules. The PEO end of these molecules was then biotinylated (i.e., covalently attached to biotin), enabling selective sensing of streptavidin as shown in Fig. 14a.[499] Similar results were also obtained with staphylococcal protein A (SpA) and human immunoglobulin (IgG).[499,500] The biotin-steptavidin pair interaction was also detected in single CNT FETs in an independent study.[501] Similar studies were subsequently completed to sense more complex biomolecular systems including glucose oxidase-glucose,[502] thrombin aptamer-thrombin,[503] adenoviruses,[504] prostate specific antigen-antibody interactions,[505] and the enzymatic hydrolysis of starch.[506] Recently, single molecule level lysosome dynamics have also been detected in single CNT FETs.[507]

Additional research has focused on detecting DNA hybridization with an eye towards DNA sequencing. The first report of DNA hybridization on CNT TFTs showed specific detection of complementary DNA sequences via conductance modulation as shown in Fig. 14b.[508] Although the authors speculated that charge transfer doping was the underlying sensing mechanism, it was later shown by Tang et al.[509] that the sensing was contact dominated. In particular, it was observed that efficient DNA hybridization on gold electrodes reduces the work function of gold, thus changing the Schottky barrier height at the metal-CNT contact as depicted in Fig.14c.[509] While CNT TFT-based electronic biosensors are relatively unstudied, one example on unsorted CNT TFTs shows that the sensitivity in detecting streptavidin using biotinylated CNT networks decreases with CNT network density (see Fig. 14d), suggesting larger sensitivity for semiconducting CNTs versus metallic CNTs.[510] Consequently, monodisperse semiconducting CNTs may lead to further improvements in CNT TFT biosensors.

## 4.2. Graphene for sensing applications

Like SWCNTs, every carbon atom in single-layer graphene is on the surface, which implies that its electronic properties should be sensitive to the surrounding environment.

Consequently, the electrical conductivity of graphene is strongly modulated by adsorbates, providing clear opportunities for sensing. As was the case for CNTs discussed above, electrochemical sensing has been achieved with graphene; however, previous reviews have already covered this topic in detail,[21, 461, 511] so this section focuses on graphene-based electrical detection strategies.

The ultra-high sensitivity of graphene FET sensors was first demonstrated by the Manchester group.[512] Bottom-gate graphene FETs showed thermally reversible detection of common chemical vapors such as $NO_2$, $NH_3$, CO, and $H_2O$ (see Fig. 15a and 15b). The measurements reveal exceptionally high sensitivity down to the single molecule level through modulation of the transverse Hall resistance ($\rho_{xy}$) at high magnetic field (10 T). The Hall bar geometry of these graphene devices eliminated the contacts as sensing sites, thus avoiding the ambiguities introduced by the FET geometry discussed earlier for CNTs.[512] Later, theoretical investigations of charge transfer doping in graphene substantiated the proposed doping mechanism.[513]

In an effort to simplify device testing and fabrication, large-area solution-processed GO and r-GO films have also been investigated for chemical sensing (Fig. 15c).[514] A variety of chemical species including nerve agents and explosives have been detected down to the ppb level. A comparison with CNT TFT sensors in the same study revealed 1000x lower noise levels in r-GO sensors as seen in Fig. 15c.[514] Reductions in noise level and improved sensitivity have also been observed by suspending graphene[515] and/or removing resist residues.[516] A similar report by the Kaner group corroborated charge transfer doping as the sensing mechanism for r-GO chemical sensors.[517] An alternative method for realizing macroscopic chemical sensors using graphene-based materials was recently reported by Yavari et al.[518] Their approach uses three-dimensionally interconnected graphene foam (Fig. 15d) obtained by growing CVD graphene on a Ni foam[519] and then etching away the Ni. This strategy results in a robust foam structure of few layer graphene with a high specific surface area (850 m$^2$/g). These sensors showed sensitivities of about 20 ppm for NH$_3$ and NO$_2$, and were fully reversible by Joule heating.[518] While graphene sensors have demonstrated high sensitivity, selectivity is much less frequently addressed. An exception is a recent study that achieved a degree of sensing selectivity by identifying different gases via low frequency noise spectral density.[520] In another report, large-area CVD graphene was utilized as the base material for chemical sensors,[521] although the sensitivity was inferior (~100 ppb) to r-GO sensors (0.1-5 ppb).

Graphene has also been investigated for biosensing applications. Following the historical trajectory of CNT biosensors, graphene biosensors have been used for the label-free detection of proteins.[522] While these initial reports lacked sensing selectivity, subsequent studies have achieved selective detection of biomolecules using non-covalently functionalized graphene. For example, a Pt nanoparticle modified r-GO FET was used to specifically detect DNA,[523] and r-GO functionalized with a Au nanoparticle-antibody conjugate was used to detect a protein specific to the antibody.[524] Non-covalently functionalized CVD graphene similarly allows highly sensitive and specific detection of glucose and glutamate in solution.[525]

Graphene has been predicted and demonstrated to be a promising system for high fidelity DNA sequencing due to its single atom thick structure. The initial concept of using a nanogap in graphene for this purpose was proposed in a theoretical paper by Postma.[526] The graphene nanogap is meant to act as an electrode as well as a membrane pore for sequencing ss-DNA passing through the gap. Since each base has a unique electronic structure and unique density of states, the traversal of each base type is expected to give rise to a specific value of conductance across the gap. Repeated fingerprinting could then sequence DNA,[526] as was later verified independently using first principles calculations.[527] This concept has received much attention, leading to more theoretical studies with different pore geometries and pore edge functionalization.[528-530] Other papers emphasized the use of graphene nanoribbons either in suspended form[531, 532] or with a circular nanopore[529, 530] for sequencing.

In practice, DNA translocation is observed to induce ionic current blockades across graphene nanopores.[533] This concept is similar to earlier work on inorganic solid state nanopore systems.[534] These fluctuations in ionic current across graphene nanopores have been verified independently by three groups.[535-537] Fig. 15e shows the experimental setup.[537] In all cases, the nanopores (2-40 nm) were sculpted in graphene using a high energy electron beam. A scatter plot of conductance/current variations versus time for blockage events was included in all three reports (see Fig. 15f).[535] The differences between folding and unfolding events can be distinguished from the scatter plot and from the shape of the current versus time signal. While it was anticipated that graphene nanopores may outperform other existing solid state nanopore systems due to their single atom thickness, it was observed that graphene nanopores possess significantly higher noise spectral density compared to silicon nitride pores, possibly due to additional defects and thus leakage current pathways in graphene. The defect density issue has been partially addressed by growing a 5 nm thick $TiO_2$ film via atomic layer deposition over the graphene.[537] Nevertheless, significantly more research will be required before the theoretical vision of graphene-based DNA sequencing is experimentally realized.

## 5. Conclusions and Future Outlook

It is apparent from the extensive literature coverage in this review that all forms of graphitic nanocarbon have significant promise for electronic, optoelectronic, photovoltaic, and sensing applications. While each class of carbon nanomaterial may ultimately impact each of these application areas, the unique strengths of fullerenes, carbon nanotubes, and graphene imply that they will have disproportionate representation in devices that play to their individual advantages. For example, the solution processability of functionalized fullerenes has led to their dominance as electron acceptors in organic photovoltaics. In contrast, semiconducting CNTs are particularly well-suited for applications that require a finite band gap such as digital electronics, optoelectronics, and electrical sensors. Finally, the high carrier mobility and optical transparency of graphene make it a leading candidate for radio frequency analog circuits and transparent conductors.

While significant progress has been achieved, additional challenges must be addressed before the full commercial potential of carbon nanomaterials is realized. Although advances in growth and post-synthetic separation methods have dramatically improved the monodispersity of carbon nanomaterials, device-to-device variability remains an issue. For example, the threshold voltages of an array of field-effect transistors fabricated from semiconducting CNTs typically vary by several volts, which is at least an order of magnitude higher than acceptable in modern-day integrated circuits. These variations can be partially addressed by further improvements in the purity and monodispersity of the semiconducting CNT band gap, but it is likely that much of the device inhomogeneity can be attributed to extrinsic factors such as the effect of metal contacts, dielectric layers, underlying substrate, and surrounding environment. Similar extrinsic issues are also limiting the performance of graphene-based radio frequency analog circuits due to the sensitivity of the electronic properties of graphene to surface adsorbates. Chemical methods for controlling and passivating carbon nanomaterial surfaces and interfaces are thus of paramount importance. Similarly, strategies for chemically enhancing and tuning the properties of carbon nanomaterials (e.g., chemically functionalizing graphene to open a band gap) are critical to further improvements in device characteristics. For these reasons, we anticipate that carbon nanomaterials will continue to be a subject of intense and fruitful research even as the first applications reach the marketplace.

ACKNOWLEDGEMENTS: The authors acknowledge support from the National Science Foundation (DMR-1006391 and DMR-1121262) and the Nanoelectronics Research Initiative at the Materials Research Center of Northwestern University, as well as by AFOSR (FA9550-08-1-0331).

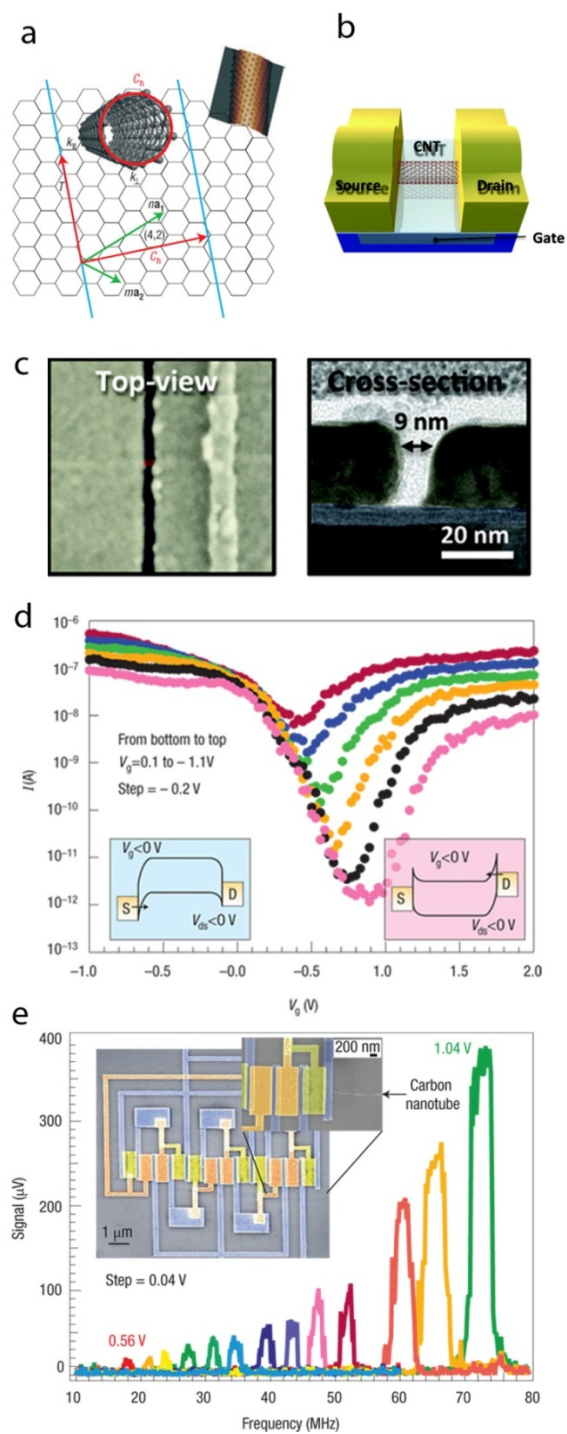

**Fig. 1.** Digital electronics based on individual CNT FETs. a) A schematic of the lattice structure of graphene. Wrapping a rectangular section of graphene along the chiral vector ($C_h$) conceptually produces a SWCNT. Reprinted with permission from ref.[6] © 2007 Macmillan Publishers Ltd. b) Schematic of a bottom-gate CNT FET. Reprinted with permission from ref.[112] © 2012, American Chemical Society. c) A top-view SEM image and cross-sectional TEM image

of a sub-10 nm channel bottom-gate CNT FET. Reprinted with permission from ref.[112] © 2012, American Chemical Society. d) Transfer characteristics of an ambipolar CNT FET. Insets show injection of electrons and holes for positive and negative $V_g$, respectively. e) Supply voltage dependent frequency spectra of a 5-stage ring oscillator fabricated from an individual CNT. The voltage is increased from 0.56 V to 1.04 V in steps of 0.04 V, from left to right. The inset shows a false-color scanning electron micrograph of the ring oscillator. Reprinted with permission from ref.[6] © 2007, Macmillan Publishers Ltd.

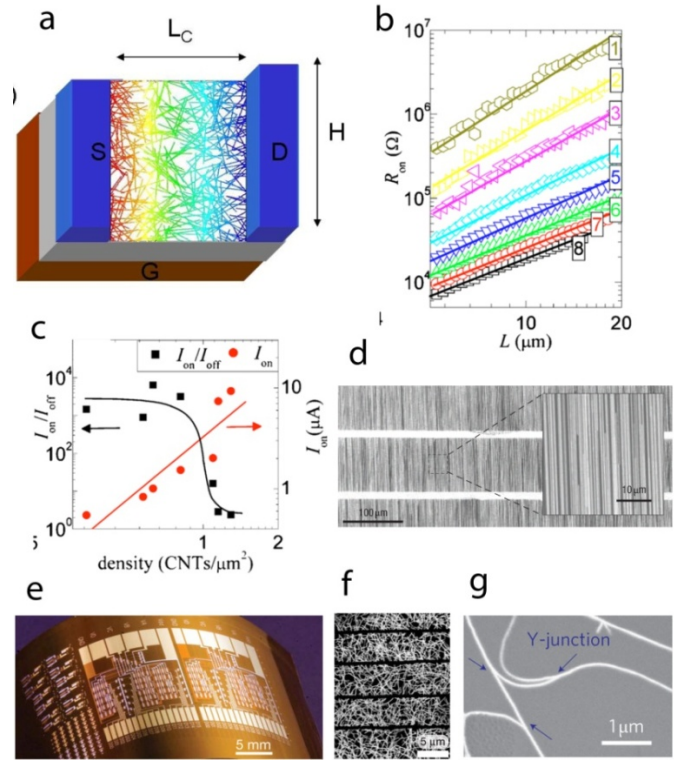

**Fig. 2.** Large-area digital electronics based on as-grown CNT TFTs. a) Schematic of a CNT TFT with a percolating network of semiconducting and metallic CNTs in the channel. The color variation from drain to source shows the simulated electrostatic potential within percolation theory. Reprinted with permission from ref.[164] © 2005 American Physical Society. b) Scaling of channel resistance with channel length ($L$) for CNT density increasing from top (1) to bottom (8). Reprinted with permission from ref.[167] © 2010 American Institute of Physics. c) A plot of on/off ratio and on-current as a function of CNT density shows the inherent tradeoff between these two metrics. Reprinted with permission from ref.[167] © 2010 American Institute of Physics. d) Array of aligned CNTs grown by CVD. Reprinted with permission from ref.[175], [167] © 2007 Macmillan Publishers Ltd. e) A medium-scale circuit consisting of ~100 CNT TFTs fabricated on a flexible substrate. Reprinted with permission from ref.[183] © 2008 Macmillan Publishers Ltd. f) Scanning electron micrograph of etched CNT strips to reduce the effect of metallic CNTs in the circuit shown in (e). Reprinted with permission from ref.[183] © 2008 Macmillan Publishers Ltd. g) Covalently bonded CNT-CNT junctions (indicated by arrows) for decreased contact resistance in high performance CNT TFTs. Reprinted with permission from ref.[184] © 2011 Macmillan Publishers Ltd.

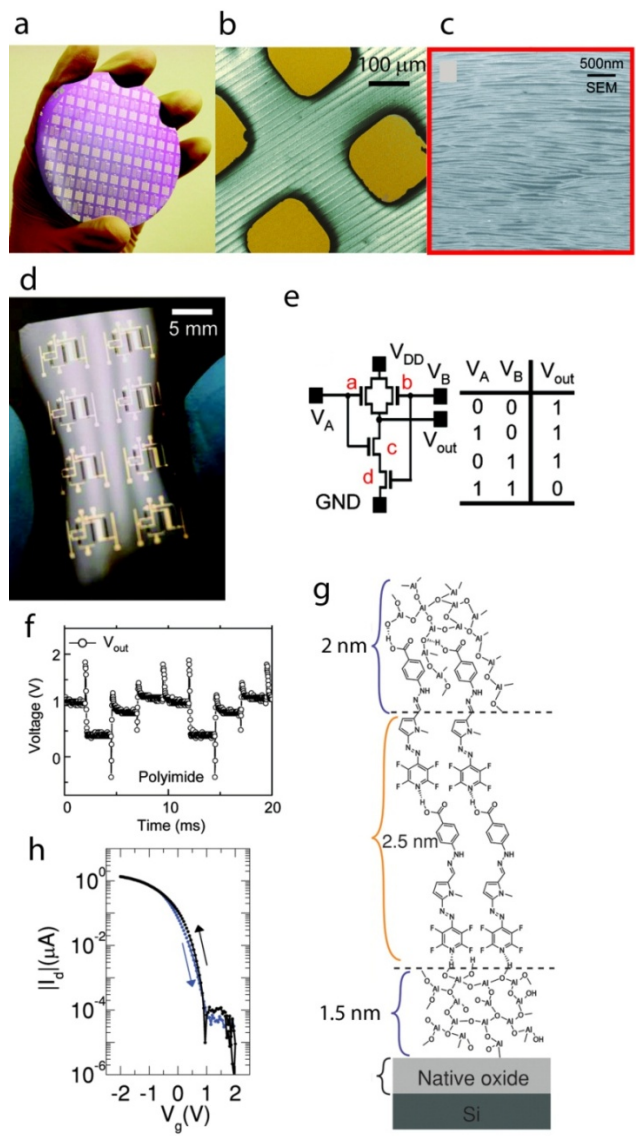

**Fig. 3.** Monodisperse semiconducting CNTs for digital electronics. a) An optical micrograph of a 3 inch wafer of CNT TFTs fabricated from 98% pure semiconducting CNTs. Reprinted with permission from ref.[188] © 2010, American Chemical Society. b) 99% pure semiconducting CNTs are aligned over large areas using evaporation-driven self-assembly. Reprinted with permission from ref.[195] © 2012 Wiley-VCH. c) An scanning electron micrograph of aligned semiconducting CNTs incorporated in top-gate CNT TFTs. Reprinted with permission from ref.[190] © 2008, American Chemical Society. d) An optical micrograph of a printed semiconducting CNT circuit on a flexible polyimide substrate. e) Circuit diagram and logic sequence of a NAND gate. f) Dynamic response of NAND gate at 100 Hz. Reprinted with permission from ref.[187] © 2010, American Chemical Society. g) Chemical structure of 6 nm thick hybrid inorganic-organic SAND. Reprinted with permission from ref.[192] © 2012, American Chemical Society. h) Hysteresis-free ambient transfer characteristics of a semiconducting CNT TFT using SAND as the gate dielectric.

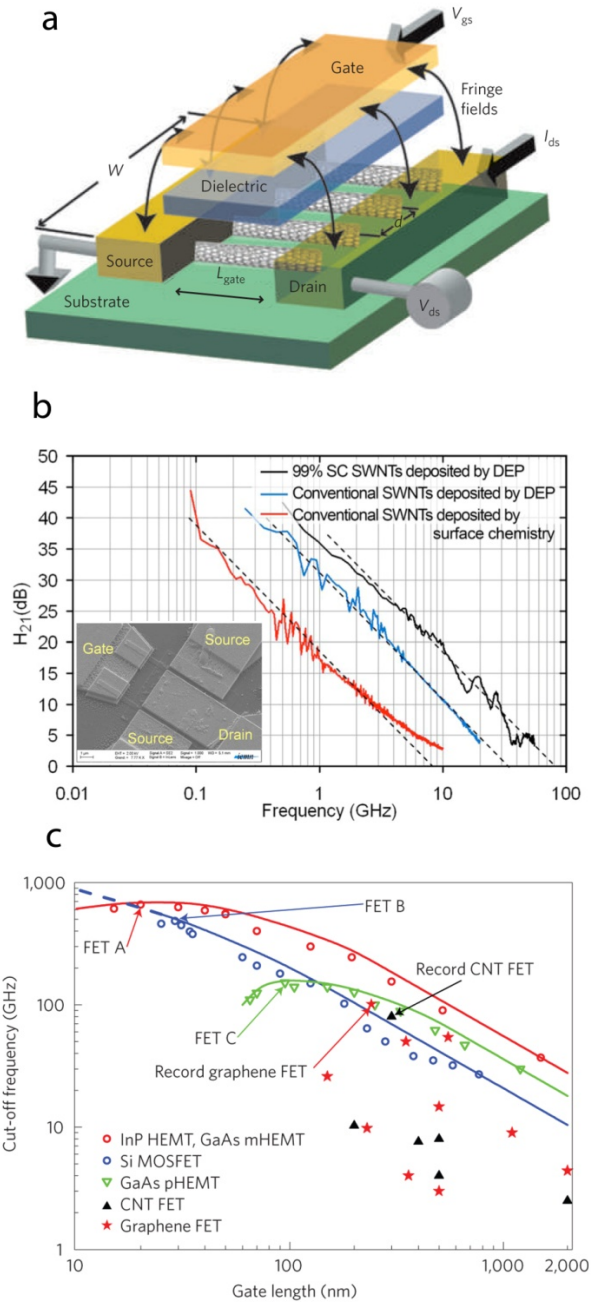

**Fig. 4**. CNTs for RF electronics. a) A schematic of a high performance RF device where parasitic capacitance due to fringe fields is minimized by optimized gate geometry and transconductance is increased through the use of an array of CNTs in the channel. Reprinted with permission from ref.[202] © 2009 Macmillan Publishers Ltd.  b) Current-gain versus frequency for the three CNT TFTs described in the legend.  Dielectrophoretic assembly of 99% semiconducting CNTs leads to unity gain at 80 GHz.  The inset shows a scanning electron micrograph of the CNT RF device (scale bar = 2 μm). Reprinted with permission from ref.[213] ©

2009 American Institute of Physics. c) Cut-off frequency versus gate length for CNT and graphene RF devices are compared with competing RF HEMTs and MOSFETs. The points are the experimental data, and the correspondingly colored lines are drawn to guide the eye. Reprinted with permission from ref.[214] © 2010 Macmillan Publishers Ltd.

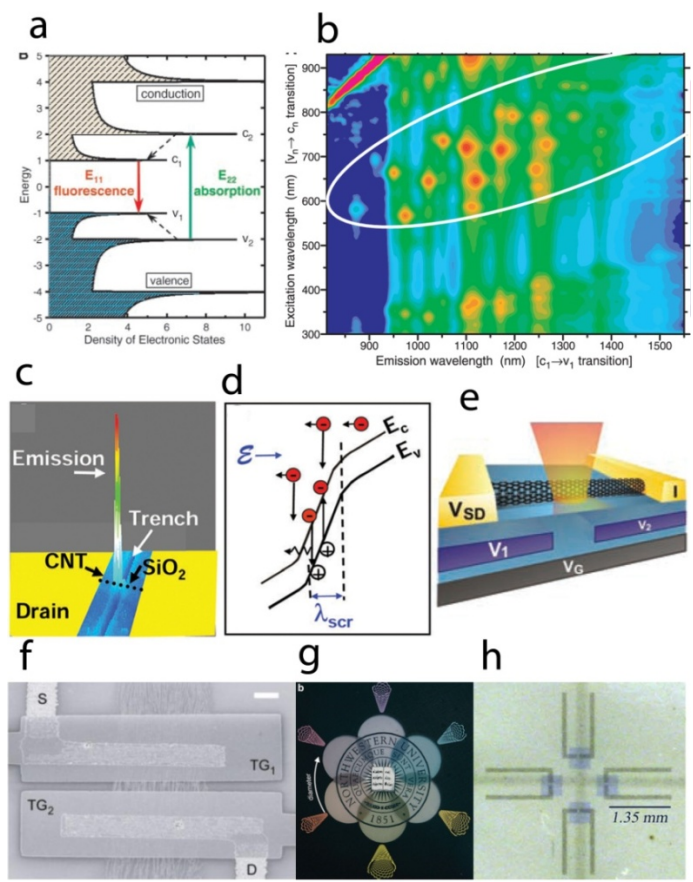

**Fig. 5.** CNTs for optoelectronic applications. a) A diagram of the van Hove singularities in the one-dimensional electronic density of states and typical optical transitions in CNTs.  b) CNT photoluminescence map that shows peaks at different excitation and emission energies corresponding to different chiralities of CNTs. Reprinted with permission from ref.[132] © 2002, American Association for the Advancement of Science.   c) Efficient electroluminescence from a p-type CNT FET where the CNT is partially suspended over a trench in the gate dielectric.  d) Electroluminescence peaks are observed at the dielectric discontinuity due to impact excitation processes. Reprinted with permission from ref.[138] © 2005, American Association for the Advancement of Science.  e) A schematic of the split-gate geometry device that generates multiple electron-hole pairs from a single photon excitation. Reprinted with permission from ref.[239] © 2009, American Association for the Advancement of Science.  f) A scanning electron micrograph of a polarized light-emitting diode realized with a split-gate geometry device on an aligned array of DGU-sorted semiconducting CNTs. The emitted light polarization is along the length of the CNTs. Reprinted with permission from ref.[233] © 2009, The Optical Society. g) Optical micrograph of conducting sheets of DGU-sorted metallic CNTs with different CNT diameters leading to the visible colors of the films. Reprinted with permission from ref.[185] © 2008, American Chemical Society.  h) An optical micrograph of a transparent, flexible printed circuit using CNTs as electrodes and pentacene as the active channel. Reprinted with permission from ref.[243] © 2009, Elsevier B.V.

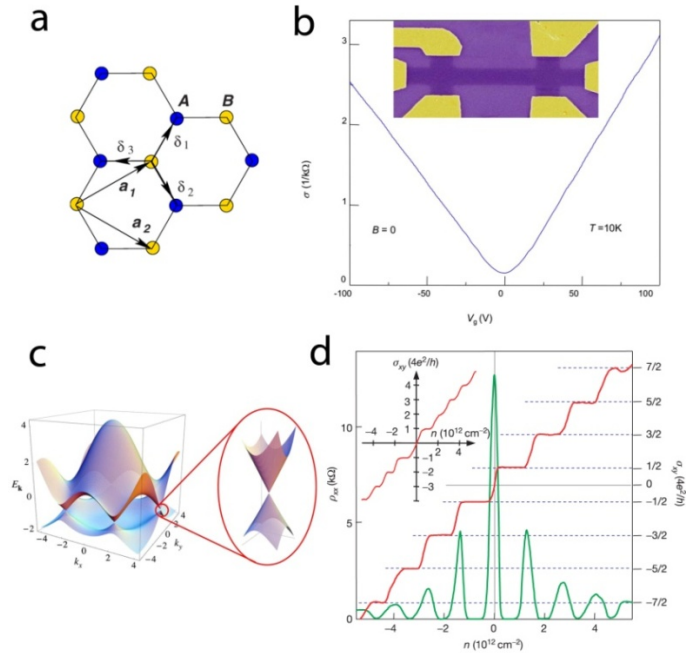

**Fig. 6.** Electronic structure of graphene. a) A schematic of the honeycomb lattice of graphene shows two carbon atoms in the unit cell. b) Conductivity of single layer graphene plotted as a function of gate bias at T = 10 K. The inset shows an optical micrograph of a Hall bar graphene device. c) Electronic dispersion of graphene. Right: zoomed in linear dispersion near one of the Dirac points. Reprinted with permission from ref.[12] © 2009, American Physical Society. d) Hall conductivity ($\sigma_{xy}$, right axis) and longitudinal resistivity ($\rho_{xx}$, left axis) showing anomalous ½-integer quantum hall effect in single layer graphene. Inset: integer quantum hall effect in bilayer graphene. Reprinted with permission from ref.[259] © 2005 Macmillan Publishers Ltd.

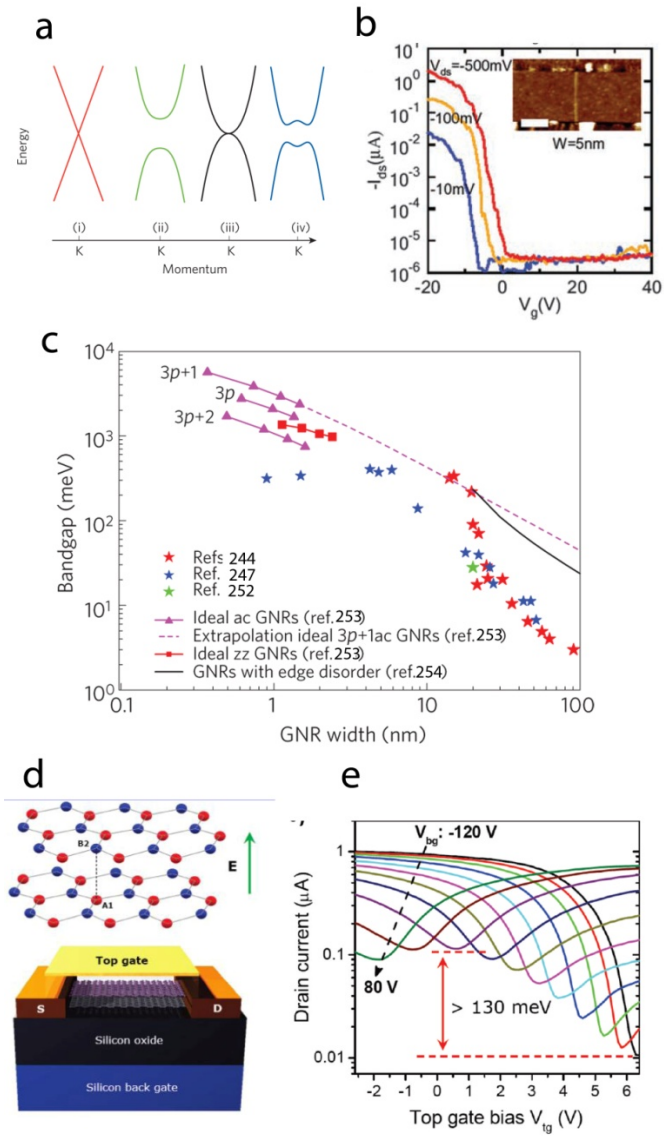

**Fig. 7.** Graphene for digital electronics. a) Band dispersion around the K point for single layer graphene (i), graphene nanoribbon (GNR) (ii), unbiased bilayer graphene (BLG) (iii), and BLG with vertically applied electric field (iv). Reprinted with permission from ref.[214] © 2010 Macmillan Publishers Ltd. b) Transfer characteristics of a GNR FET for different drain biases. Inset: AFM image of the GNR FET (scale bar = 100 nm). Reprinted with permission from ref.[277] © 2008, American Association for the Advancement of Science. c) Band gap versus GNR width obtained by different methods. Points are experimental data from the references shown in legend, and lines are calculated trends. Reprinted with permission from ref.[214] © 2010 Macmillan Publishers Ltd. d) Lattice structure of BLG with Bernal stacking; a schematic of a BLG FET with a top gate to enable the application of a vertical electric field to induce a band gap. e) Transfer characteristics of a BLG FET (drain current versus top-gate bias, $V_{TG}$) for different bottom-gate biases ($V_{BG}$). Highest on/off ratio (100) is obtained for the largest displacement field ($V_{BG}$ = -120 V, $V_{TG}$ = 6V ). Reprinted with permission from ref.[298] © 2010, American Chemical Society.

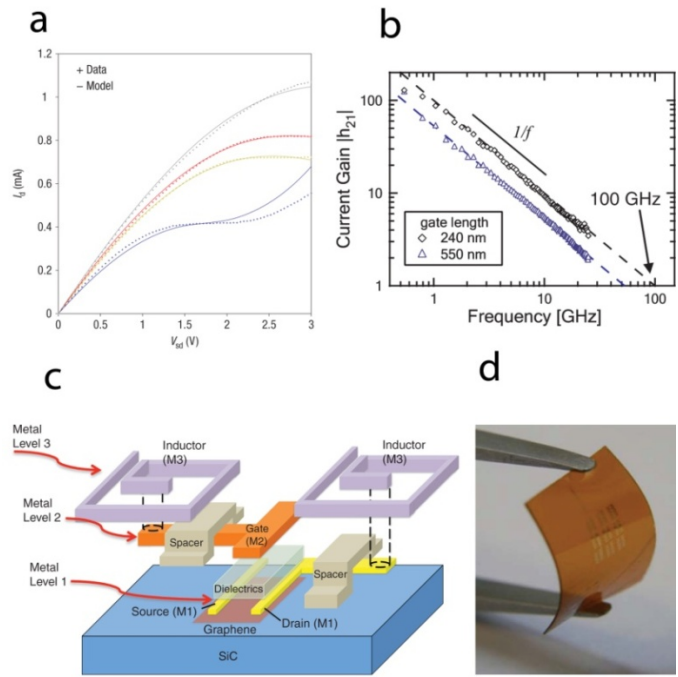

**Fig. 8.** Graphene for RF electronics. a) Output characteristics of graphene FETs show poor current saturation. Dotted lines are experimental data, and solid lines are model fits. Reprinted with permission from ref.[313] © 2008 Macmillan Publishers Ltd. b) Small-signal current gain of two graphene FETs with gate lengths of 240 nm (black) and 550 nm (blue). Reprinted with permission from ref.[322] © 2010, American Association for the Advancement of Science. c) Schematic of a graphene mixer circuit consisting of a top-gate graphene transistor and two metal inductors. The spacer is made of 120 nm thick $SiO_2$. Reprinted with permission from ref.[323] © 2011, American Association for the Advancement of Science. d) An optical image of a 2.2 GHz graphene device fabricated on a flexible polyimide substrate using DGU-sorted single layer graphene. Reprinted with permission from ref.[326] © 2012, American Chemical Society.

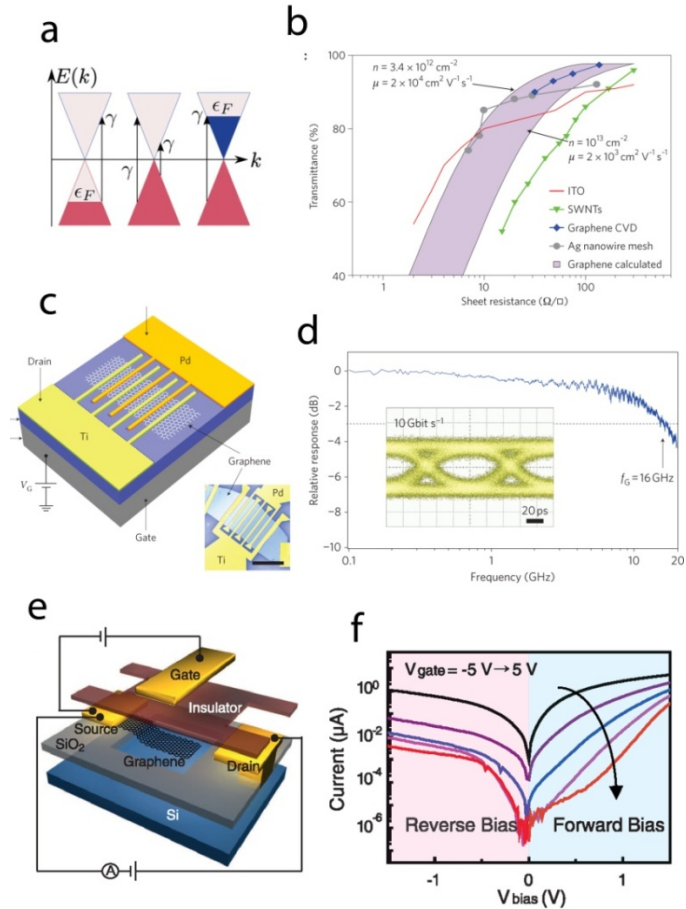

**Fig. 9.** Graphene for optoelectronics and novel electronic devices. a) Schematic of the linear dispersion of graphene near the K point shows the widely tunable photoresponse for unbiased graphene. Biased graphene has a photoexcitation range that is limited by the empty valence band or filled conduction band. Reprinted with permission from ref.[13] © 2012, American Physical Society. b) Transmittance versus sheet resistance of graphene-based transparent conductors as compared to CNT thin films and commercial ITO. The purple region denotes the theoretical bounds for graphene with the range of doping levels shown in the legend. Reprinted with permission from ref.[16] © 2010 Macmillan Publishers Ltd. c) Schematic (upper) and an optical image (lower) of a metal-graphene-metal photodetector with asymmetric metal electrodes (Ti and Pd) in an interdigitated geometry to obtain a large contact area. d) Relative photoresponse versus light intensity modulation frequency shows -3 dB photoresponse at 16 GHz for the photodetector in (c). Inset: corresponding receiver eye-diagram. The scale bar is 20 picoseconds. Reprinted with permission from ref.[336] © 2010 Macmillan Publishers Ltd. e) A schematic diagram of a graphene based barrister. f) The switching behavior of a graphene barrister is shown in forward and reverse bias. The gate bias is changed from -5 V to 5 V in steps of 2 V from top (black) to bottom (red) curves. Reprinted with permission from ref.[342] © 2010, American Association for the Advancement of Science.

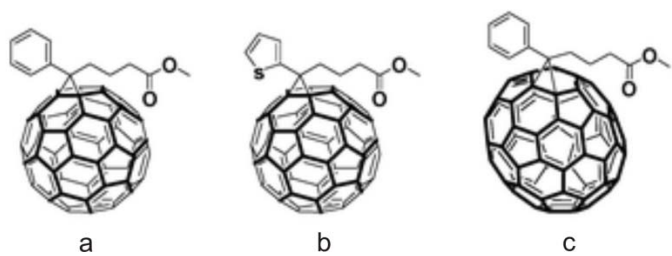

**Fig. 10.** Molecular structures of fullerene based acceptors. a) $PC_{61}BM$. b) $ThC_{61}BM$. c) $PC_{71}BM$. Reprinted with permission from ref.[350] © 2010, The Royal Society of Chemistry.

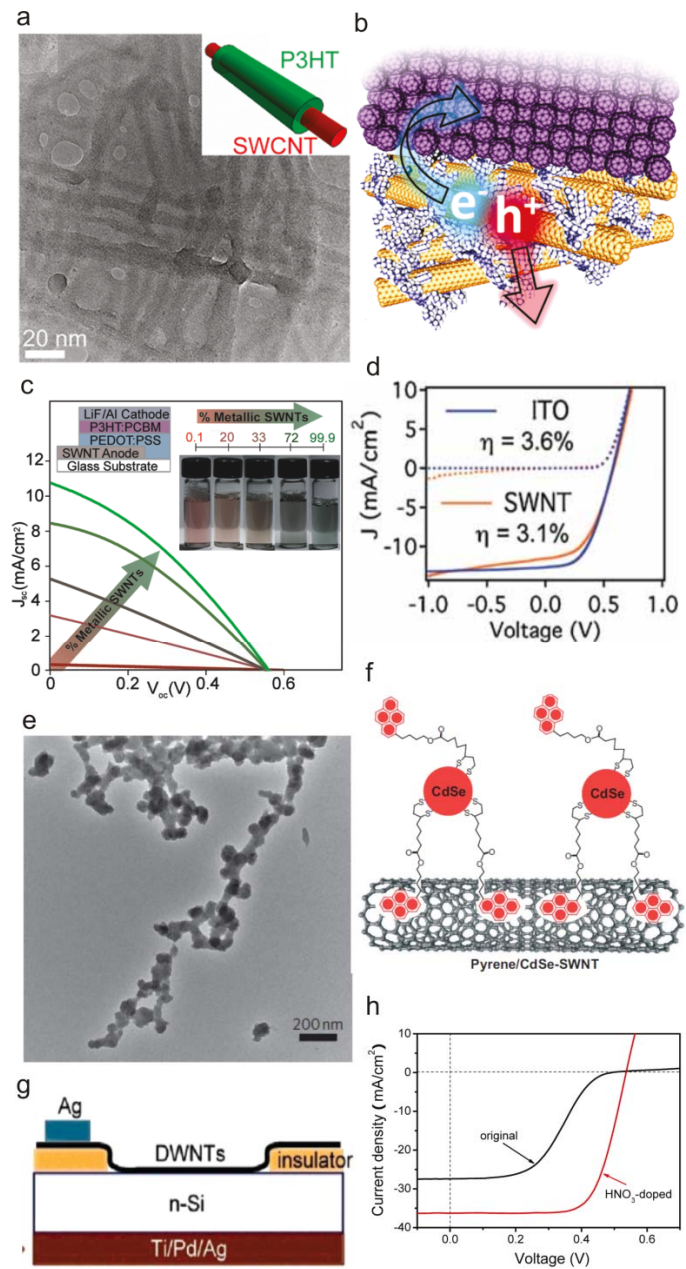

**Fig. 11.** CNTs in photovoltaics. a) TEM image of a semiconducting SWCNT-P3HT hybrid. Reprinted with permission from ref.[364] © 2011, American Chemical Society. b) Schematic of a bilayer semiconducting SWCNT-$C_{60}$ solar cell showing charge separation at the interface. Reprinted with permission from ref.[368] © 2010, American Chemical Society. c) Current-voltage curves of a P3HT-PCBM solar cell with transparent CNT anodes of varying metallic content. The increase in efficiency as a function of metallic content can be seen. Inset: device schematic and solutions of SWCNTs with increasing metallic content. Reprinted with permission from ref.[381] © 2011, Wiley-VCH. d) Comparison of current-voltage curves under dark (dotted) and illumination (solid) of a control device with an ITO electrode (blue) and a CNT thin film

electrode (red). Reprinted with permission from ref.[373] © 2009, Wiley-VCH. e) TEM micrograph of a virus-templated CNT-TiO$_2$ composite used for DSSCs. Reprinted with permission from ref.[394] © 1998, Macmillan Publishers Ltd. f) Schematic of a pyrene functionalized CdSe quantum dot-CNT composite used for quantum dot solar cells. Reprinted with permission from ref.[399] © 2008, Wiley-VCH. g) Schematic drawing of a DWNT:n-Si thin film solar cell. Reprinted with permission from ref.[402] © 2007, American Chemical Society. h) Current-voltage curves of a CVD grown CNT thin film:n-Si solar cell before (black) and after (red) nitric acid doping. Reprinted with permission from ref.[405] © 2011, American Chemical Society.

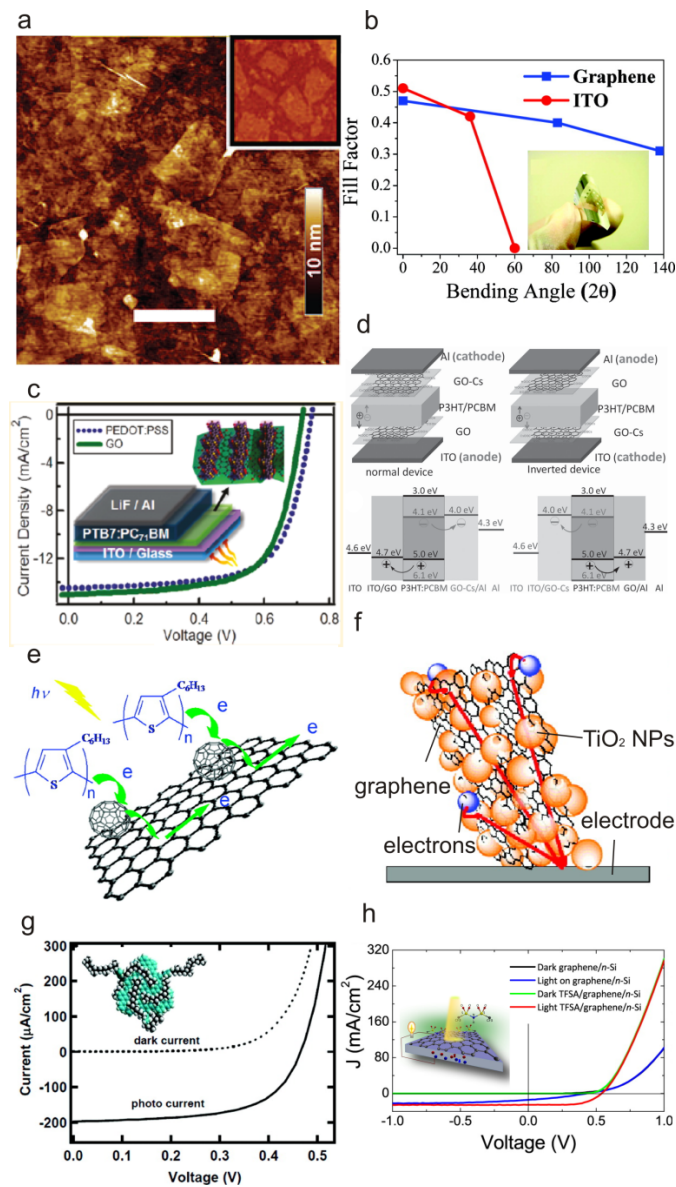

**Fig. 12.** Graphene in photovoltaics. a) AFM image of a thick r-GO film. Inset: AFM image of a sub-monolayer coverage r-GO film showing individual flake boundaries. Reprinted with permission from ref.[414] © 2008, American Chemical Society. b) Comparison of fill factor versus bending angle for cells with a control ITO anode (red) and a CVD graphene anode (blue). The improved mechanical robustness of CVD graphene is apparent. Reprinted with permission from ref.[417] © 2010, American Chemical Society. c) Comparison of current-voltage curves of PTB7:PCBM organic solar cells using PEDOT:PSS (blue dotted) and graphene oxide (green solid) as hole transport layers. Reprinted with permission from ref.[431] © 2011, American Chemical Society. d) Diagrams depicting normal and inverted cell geometries using GO and GO-Cs as hole transport and electron transport layers respectively. Reprinted with permission from ref.[434] © 2012, Wiley-VCH. e) Illustration of covalently grafted, fullerene-modified graphene with P3HT in the active layer. Reprinted with permission from ref.[440] © 2011, American Chemical Society.. f) TiO$_2$ nanoparticles with graphene additive for high efficiency DSSCs. The graphene additive provides conductive pathways while minimizing recombination. Reprinted with permission from ref.[442] © 2010, American Chemical Society. g) Current-voltage curves in the dark and light of a photoelectrochemical cell sensitized with GQDs. Reprinted with permission from ref.[446] © 2010, American Chemical Society. h) Current-voltage characteristics of control and TFSA-doped graphene:n-Si Schottky junction solar cells in the dark and light. Reprinted with permission from ref.[453] © 2012, American Chemical Society.

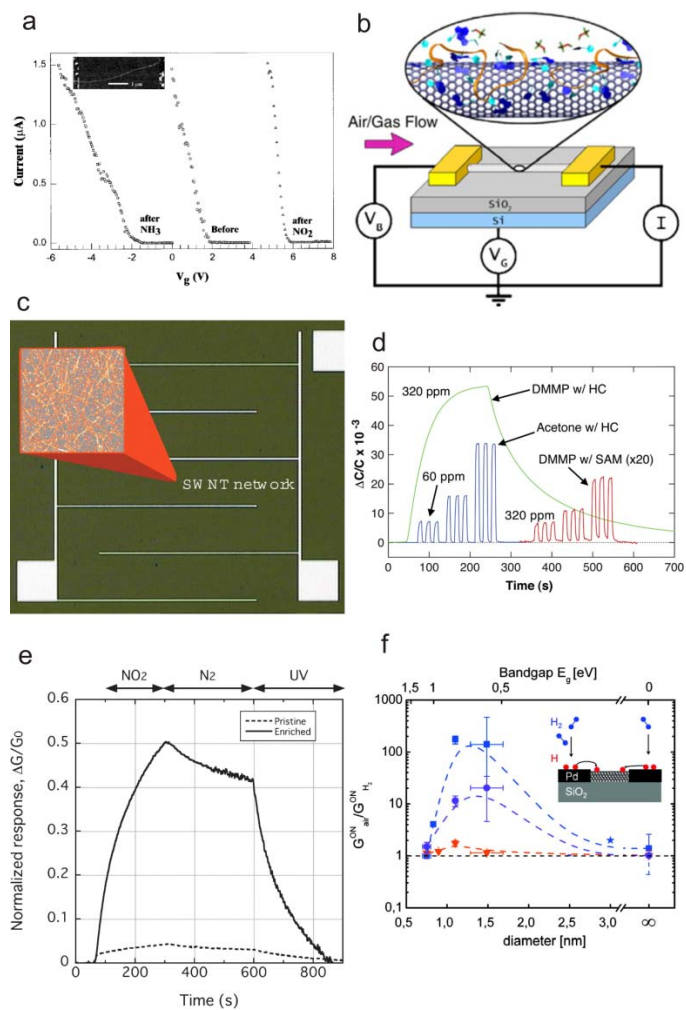

**Fig. 13.** CNTs for chemical sensing. a) Transfer characteristics of a single semiconducting SWCNT FET (AFM image in inset) before and after exposure to $NH_3$ and $NO_2$. Reprinted with permission from ref.[463] © 2000, American Association for the Advancement of Science. b) Schematic depiction of a single CNT FET non-covalently functionalized with ss-DNA molecules for selective detection of nerve agents such as DMMP. Reprinted with permission from ref.[481] © 2006, IOP Publishing Ltd. c) Optical micrograph of a chemicapacitive sensor fabricated from as-grown CNT networks. d) Change in capacitance versus time upon introduction of analyte species (e.g., DMMP and acetone) for a non-covalently functionalized CNT network chemicapacitve sensor. HC (polycarbosilane) and allyltrichlorosilane are used for non-covalent functionalization. Reprinted with permission from ref.[490] © 2005, American Association for the Advancement of Science. e) Sensitivity comparison for unsorted and sorted semiconducting SWCNT thin film sensors upon exposure to $NO_2$. The sensitivity of the sorted semiconducting SWCNT sensor is 10 times greater than the unsorted case. Reprinted with permission from ref.[496] © 2012, The Japan Society of Applied Physics. f) Sensitivity of semiconducting SWCNT hydrogen sensor versus SWCNT diameter. The highest sensitivity (~3 orders of magnitude change in conductance) is achieved for 1.4 nm diameter SWCNTs. Reprinted with permission from ref.[498] © 2011, American Chemical Society.

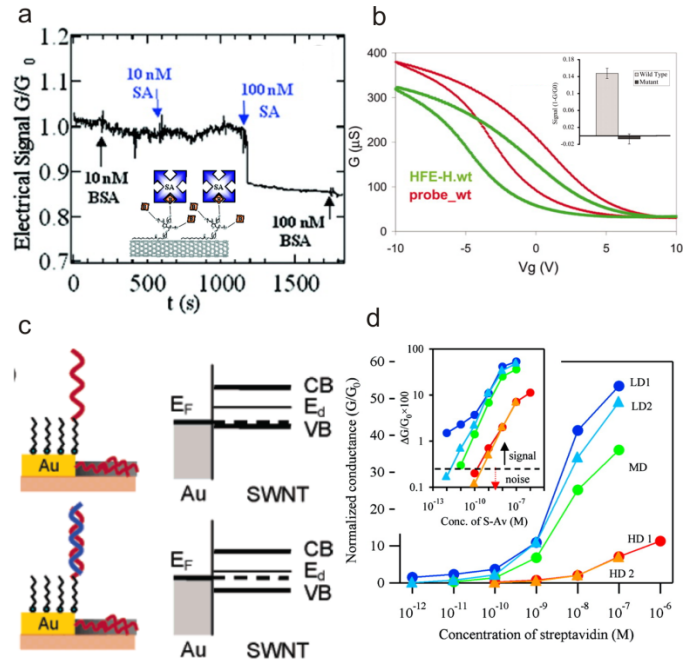

**Fig. 14.** Nanotubes for biosensing. a) Electrical response of a PEO-biotin (B) functionalized CNT network device, showing specific sensitivity to 100 nM streptavidin (SA) in contrast to no response to bovine serum albumin (BSA). Reprinted with permission from ref.[499] © 2003, National Academy of Sciences, U.S.A. b) Comparison of transfer characteristics of an as-grown CNT network FET coated with a specific probe DNA sequence before (red) and after (green) exposure to a target DNA sequence that is complementary to the probe. The inset shows a comparison of the magnitudes of the electrical signals upon exposure to the complementary target sequence and a mutated target with a single nucleotide polymorphism (SNP). Reprinted with permission from ref.[508] © 2006, National Academy of Sciences, U.S.A. c) Schematic depiction of the DNA sensing mechanism for CNT FETs. The rise in metal work function following hybridization of ss-DNA to its complementary sequence changes the Fermi level at the metal-semiconductor Schottky junction, thereby producing a conductance modulation. Reprinted with permission from ref.[509] © 2006, American Chemical Society. d) Conductance variation as function of streptavidin concentration for a non-covalently functionalized CNT network FET with low, medium, and high tube density (LD, MD, and HD) in the channel. The inset shows a semilog plot. Reprinted with permission from ref.[510] © 2010, American Chemical Society.

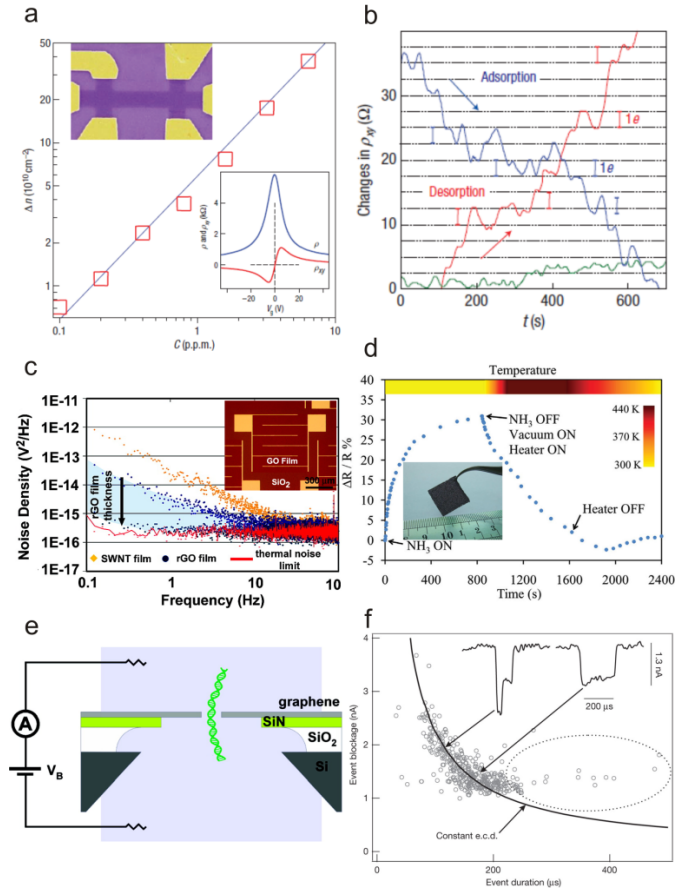

**Fig. 15**. Graphene for sensing applications. a) Change in carrier density versus concentration of $NO_2$ (in ppm) of a graphene FET sensor in a Hall bar geometry (top inset). The longitudinal and transverse resistances as a function of gate voltage are shown in the bottom inset. b) Hall resistance fluctuations in steps of +/- 1 e of a three layer graphene device at B = 10 T, indicating single molecule adsorption and desorption events. Reprinted with permission from ref.[512] © 2007, Macmillan Publishers Ltd. c) Comparison of the noise spectral density of a large r-GO film sensor versus a CNT thin-film sensor. The noise level in the r-GO sensor is lower by an order of magnitude compared to the CNT sensor and reduces further with decreasing r-GO film thickness. Reprinted with permission from ref.[514] © 2008, American Chemical Society. d) Change in resistance versus time for a graphene foam (inset) gas sensor upon exposure to $NH_3$. Complete reversibility is observed following heating at 440 K. Reprinted with permission from ref.[518] © 2011, Macmillan Publishers Ltd. e) Schematic representation of the experimental setup used to detect DNA translocation through individual graphene nanopores. The ionic current is measured as a function of time. Reprinted with permission from ref.[537] © 2010, American Chemical Society. f) The magnitude of current fluctuations versus time for a large number of events is plotted. The inset shows the ionic current versus time. Reprinted with permission from ref.[535] © 2010, Macmillan Publishers Ltd.

**References:**

1. P. M. Ajayan, *Chem. Rev.*, 1999, **99**, 1787-1800.
2. G. S. Hammond and V. J. Kuck, *Fullerenes: Synthesis, properties, and chemistry of large carbon clusters*, 1992.
3. M. J. Allen, V. C. Tung and R. B. Kaner, *Chem. Rev.*, 2009, **110**, 132-145.
4. M. C. Hersam, *Nat. Nano.*, 2008, **3**, 387-394.
5. D. Jariwala, A. Srivastava and P. M. Ajayan, *J. Nanosci. Nanotechnol.*, 2011, **11**, 6621-6641.
6. P. Avouris, Z. Chen and V. Perebeinos, *Nat. Nano.*, 2007, **2**, 605-615.
7. P. Avouris, *MRS Bulletin*, 2004, **29**, 403-410.
8. P. Avouris and R. Martel, *MRS Bulletin*, 2010, **35**, 306-313.
9. P. Avouris, M. Freitag and V. Perebeinos, *Nat. Photon.*, 2008, **2**, 341-350.
10. Q. Cao and J. A. Rogers, *Adv. Mater.*, 2009, **21**, 29-53.
11. N. Rouhi, D. Jain and P. J. Burke, *ACS Nano*, 2011, **5**, 8471-8487.
12. A. H. Castro Neto, F. Guinea, N. M. R. Peres, K. S. Novoselov and A. K. Geim, *Rev. Mod. Phys.*, 2009, **81**, 109-162.
13. N. M. R. Peres, *Rev. Mod. Phys.*, 2010, **82**, 2673-2700.
14. S. Das Sarma, S. Adam, E. H. Hwang and E. Rossi, *Rev. Mod. Phys.*, 2011, **83**, 407-470.
15. M. S. Fuhrer, C. N. Lau and A. H. MacDonald, *MRS Bulletin*, 2010, **35**, 289-295.
16. F. Bonaccorso, Z. Sun, T. Hasan and A. C. Ferrari, *Nat. Photon.*, 2010, **4**, 611-622.
17. C. J. Brabec, S. Gowrisanker, J. J. M. Halls, D. Laird, S. Jia and S. P. Williams, *Adv. Mater.*, 2010, **22**, 3839-3856.
18. B. C. Thompson and J. M. J. Fréchet, *Angew. Chem. Int. Ed.*, 2008, **47**, 58-77.
19. Y. X. Liu, X. C. Dong and P. Chen, *Chem. Soc. Rev.*, 2012, **41**, 2283-2307.
20. J. Wang, *Electroanalysis*, 2005, **17**, 7-14.
21. M. Pumera, A. Ambrosi, A. Bonanni, E. L. K. Chng and H. L. Poh, *Trac-Trends Anal. Chem.*, 2010, **29**, 954-965.
22. W. Yang, K. R. Ratinac, S. P. Ringer, P. Thordarson, J. J. Gooding and F. Braet, *Angew. Chem. Int. Ed.*, 2010, **49**, 2114-2138.
23. R. Saito, G. Dresselhaus and M. S. Dresselhaus, *Physical properties of carbon nanotubes*, Imperial College Press, London, 1998.
24. Stephanie Reich, Christian Thomsen and J. Maultzsch, *Carbon nanotubes: Basic concepts and physical properties,* Wiley-VCH, 2004.
25. P. J. F. Harris, *Carbon Nanotube Science: Synthesis, Properties and Applications*, Cambridge University Press, 2011.
26. H.-S. Philip Wong and D. Akinwande, *Carbon Nanotube and Graphene Device Physics*, Cambridge University Press, 2011.
27. H. W. Kroto, J. R. Heath, S. C. O'Brien, R. F. Curl and R. E. Smalley, *Nature*, 1985, **318**, 162-163.
28. E. A. Rohlfing, D. M. Cox and A. Kaldor, *J. Chem. Phys.*, 1984, **81**, 3322-3330.
29. G. Churilov, *Instruments and Experimental Techniques*, 2000, **43**, 1-10.
30. R. E. Smalley, *Acc. Chem. Res.*, 1992, **25**, 98-105.
31. S. Stevenson, G. Rice, T. Glass, K. Harich, F. Cromer, M. R. Jordan, J. Craft, E. Hadju, R. Bible, M. M. Olmstead, K. Maitra, A. J. Fisher, A. L. Balch and H. C. Dorn, *Nature*, 1999, **401**, 55-57.
32. D. S. Bethune, R. D. Johnson, J. R. Salem, M. S. de Vries and C. S. Yannoni, *Nature*, 1993, **366**, 123-128.
33. K.-H. Homann, *Angew. Chem. Int. Ed.*, 1998, **37**, 2434-2451.

34. H. Richter, A. J. Labrocca, W. J. Grieco, K. Taghizadeh, A. L. Lafleur and J. B. Howard, *J. Phys. Chem. B*, 1997, **101**, 1556-1560.
35. J. B. Howard, J. T. McKinnon, Y. Makarovsky, A. L. Lafleur and M. E. Johnson, *Nature*, 1991, **352**, 139-141.
36. L. T. Scott, *Angew. Chem. Int. Ed.*, 2004, **43**, 4994-5007.
37. D. S. Bethune, C. H. Klang, M. S. de Vries, G. Gorman, R. Savoy, J. Vazquez and R. Beyers, *Nature*, 1993, **363**, 605-607.
38. T. W. Ebbesen and P. M. Ajayan, *Nature*, 1992, **358**, 220-222.
39. W. Kratschmer, L. D. Lamb, K. Fostiropoulos and D. R. Huffman, *Nature*, 1990, **347**, 354-358.
40. S. Iijima, *Nature*, 1991, **354**, 56-58.
41. S. Iijima and T. Ichihashi, *Nature*, 1993, **363**, 603-605.
42. A. Thess, R. Lee, P. Nikolaev, H. J. Dai, P. Petit, J. Robert, C. H. Xu, Y. H. Lee, S. G. Kim, A. G. Rinzler, D. T. Colbert, G. E. Scuseria, D. Tomanek, J. E. Fischer and R. E. Smalley, *Science*, 1996, **273**, 483-487.
43. T. Guo, P. Nikolaev, A. G. Rinzler, D. Tomanek, D. T. Colbert and R. E. Smalley, *J. Phys. Chem.*, 1995, **99**, 10694-10697.
44. T. Guo, P. Nikolaev, A. Thess, D. T. Colbert and R. E. Smalley, *Chem. Phys. Lett.*, 1995, **243**, 49-54.
45. W. Z. Li, S. S. Xie, L. X. Qian, B. H. Chang, B. S. Zou, W. Y. Zhou, R. A. Zhao and G. Wang, *Science*, 1996, **274**, 1701-1703.
46. K. Hata, D. N. Futaba, K. Mizuno, T. Namai, M. Yumura and S. Iijima, *Science*, 2004, **306**, 1362-1364.
47. S. Talapatra, S. Kar, S. K. Pal, R. Vajtai, L. Ci, P. Victor, M. M. Shaijumon, S. Kaur, O. Nalamasu and P. M. Ajayan, *Nat. Nano.*, 2006, **1**, 112-116.
48. D. Jariwala, K. Chandra, A. Cao, S. Talapatra, M. Shima, D. Anuhya, V. S. S. S. Prasad, R. Ribeiro, P. C. Canfield, D. Wu, A. Srivastava, R. K. Mandal, A. K. Pramanick, R. Vajtai, P. M. Ajayan and G. V. S. Sastry, *The Banaras Metallurgist*, 2012, **17**, 57-67.
49. B. C. Satishkumar, A. Govindaraj, R. Sen and C. N. R. Rao, *Chem. Phys. Lett.*, 1998, **293**, 47-52.
50. R. Sen, A. Govindaraj and C. N. R. Rao, *Chem. Mat.*, 1997, **9**, 2078-2081.
51. P. Nikolaev, M. J. Bronikowski, R. K. Bradley, F. Rohmund, D. T. Colbert, K. A. Smith and R. E. Smalley, *Chem. Phys. Lett.*, 1999, **313**, 91-97.
52. E. T. Thostenson, Z. Ren and T.-W. Chou, *Composites Science and Technology*, 2001, **61**, 1899-1912.
53. P. M. Ajayan and T. W. Ebbesen, *Reports on Progress in Physics*, 1997, **60**, 1025.
54. M. Meyyappan, D. Lance, C. Alan and H. David, *Plasma Sources Science and Technology*, 2003, **12**, 205.
55. M. Terrones, *Annual Review of Materials Research*, 2003, **33**, 419-501.
56. K. S. Novoselov, A. K. Geim, S. V. Morozov, D. Jiang, Y. Zhang, S. V. Dubonos, I. V. Grigorieva and A. A. Firsov, *Science*, 2004, **306**, 666-669.
57. L. Xuekun, Y. Minfeng, H. Hui and S. R. Rodney, *Nanotechnology*, 1999, **10**, 269.
58. Y. Zhang, J. P. Small, W. V. Pontius and P. Kim, *Appl. Phys. Lett.*, 2005, **86**, 073104-073103.
59. C. Berger, Z. Song, T. Li, X. Li, A. Y. Ogbazghi, R. Feng, Z. Dai, A. N. Marchenkov, E. H. Conrad, P. N. First and W. A. de Heer, *J. Phys. Chem. B*, 2004, **108**, 19912-19916.
60. K. V. Emtsev, A. Bostwick, K. Horn, J. Jobst, G. L. Kellogg, L. Ley, J. L. McChesney, T. Ohta, S. A. Reshanov, J. Rohrl, E. Rotenberg, A. K. Schmid, D. Waldmann, H. B. Weber and T. Seyller, *Nat. Mater.*, 2009, **8**, 203-207.
61. D. V. Badami, *Nature*, 1962, **193**, 569-570.
62. O. Chuhei and N. Ayato, *J. Phys.: Condens. Matter*, 1997, **9**, 1.
63. W. S. Hummers and R. E. Offeman, *J. Am. Chem. Soc.*, 1958, **80**, 1339-1339.

64. B. C. Brodie, *Ann. Chim. Phys.*, 1860, **59**, 466.
65. L. Staudenmaier, *Berichte der deutschen chemischen Gesellschaft*, 1898, **31**, 1481-1487.
66. S. Stankovich, D. A. Dikin, R. D. Piner, K. A. Kohlhaas, A. Kleinhammes, Y. Jia, Y. Wu, S. T. Nguyen and R. S. Ruoff, *Carbon*, 2007, **45**, 1558-1565.
67. S. Stankovich, R. D. Piner, X. Chen, N. Wu, S. T. Nguyen and R. S. Ruoff, *J. Mater. Chem.*, 2006, **16**, 155-158.
68. V. C. Tung, M. J. Allen, Y. Yang and R. B. Kaner, *Nat. Nano.*, 2009, **4**, 25-29.
69. C. Gómez-Navarro, R. T. Weitz, A. M. Bittner, M. Scolari, A. Mews, M. Burghard and K. Kern, *Nano Lett.*, 2007, **7**, 3499-3503.
70. D. A. Sokolov, K. R. Shepperd and T. M. Orlando, *J. Phys. Chem. Lett.*, 2010, **1**, 2633-2636.
71. S. Guo and S. Dong, *Chem. Soc. Rev.*, 2011, **40**, 2644-2672.
72. S. Park and R. S. Ruoff, *Nat. Nano.*, 2009, **4**, 217-224.
73. D. R. Dreyer, S. Park, C. W. Bielawski and R. S. Ruoff, *Chem. Soc. Rev.*, 2010, **39**, 228-240.
74. V. Singh, D. Joung, L. Zhai, S. Das, S. I. Khondaker and S. Seal, *Progress in Materials Science*, 2011, **56**, 1178-1271.
75. M. Lotya, Y. Hernandez, P. J. King, R. J. Smith, V. Nicolosi, L. S. Karlsson, F. M. Blighe, S. De, Z. Wang, I. T. McGovern, G. S. Duesberg and J. N. Coleman, *J. Am. Chem. Soc.*, 2009, **131**, 3611-3620.
76. Y. Hernandez, V. Nicolosi, M. Lotya, F. M. Blighe, Z. Sun, S. De, I. T. McGovern, B. Holland, M. Byrne, Y. K. Gun'Ko, J. J. Boland, P. Niraj, G. Duesberg, S. Krishnamurthy, R. Goodhue, J. Hutchison, V. Scardaci, A. C. Ferrari and J. N. Coleman, *Nat. Nano.*, 2008, **3**, 563-568.
77. Y. T. Liang and M. C. Hersam, *J. Am. Chem. Soc.*, 2010, **132**, 17661-17663.
78. K. S. Kim, Y. Zhao, H. Jang, S. Y. Lee, J. M. Kim, K. S. Kim, J.-H. Ahn, P. Kim, J.-Y. Choi and B. H. Hong, *Nature*, 2009, **457**, 706-710.
79. A. Reina, X. Jia, J. Ho, D. Nezich, H. Son, V. Bulovic, M. S. Dresselhaus and J. Kong, *Nano Lett.*, 2008, **9**, 30-35.
80. X. Li, W. Cai, J. An, S. Kim, J. Nah, D. Yang, R. Piner, A. Velamakanni, I. Jung, E. Tutuc, S. K. Banerjee, L. Colombo and R. S. Ruoff, *Science*, 2009, **324**, 1312-1314.
81. A. Srivastava, C. Galande, L. Ci, L. Song, C. Rai, D. Jariwala, K. F. Kelly and P. M. Ajayan, *Chem. Mat.*, 2010, **22**, 3457-3461.
82. A. E. Karu and M. Beer, *J. Appl. Phys.*, 1966, **37**, 2179-2181.
83. M. Eizenberg and J. M. Blakely, *Surf. Sci.*, 1979, **82**, 228-236.
84. Y. Zhu, S. Murali, W. Cai, X. Li, J. W. Suk, J. R. Potts and R. S. Ruoff, *Adv. Mater.*, 2010, **22**, 3906-3924.
85. J. Xiao and M. E. Meyerhoff, *Journal of Chromatography A*, 1995, **715**, 19-29.
86. W. A. Scrivens, P. V. Bedworth and J. M. Tour, *J. Am. Chem. Soc.*, 1992, **114**, 7917-7919.
87. K. Nagata, E. Dejima, Y. Kikuchi and M. Hashiguchi, *Chem. Lett.*, 2005, **34**, 178-179.
88. Y. Liu, Y.-W. Yang and Y. Chen, *Chem. Commun.*, 2005, 4208-4210.
89. S. M. Bachilo, L. Balzano, J. E. Herrera, F. Pompeo, D. E. Resasco and R. B. Weisman, *J. Am. Chem. Soc.*, 2003, **125**, 11186-11187.
90. P. G. Collins, M. Hersam, M. Arnold, R. Martel and P. Avouris, *Phys. Rev. Lett.*, 2001, **86**, 3128-3131.
91. M. Kanungo, H. Lu, G. G. Malliaras and G. B. Blanchet, *Science*, 2009, **323**, 234-237.
92. M. C. LeMieux, M. Roberts, S. Barman, Y. W. Jin, J. M. Kim and Z. Bao, *Science*, 2008, **321**, 101-104.
93. T. Tanaka, H. H. Jin, Y. Miyata and H. Kataura, *Appl. Phys. Express*, 2008, **1,** 114001.
94. R. Krupke, F. Hennrich, H. v. Löhneysen and M. M. Kappes, *Science*, 2003, **301**, 344-347.
95. X. Tu, S. Manohar, A. Jagota and M. Zheng, *Nature*, 2009, **460**, 250-253.

96. H. Liu, D. Nishide, T. Tanaka and H. Kataura, *Nat Commun*, 2011, **2**, 309.
97. T. Tanaka, H. Jin, Y. Miyata, S. Fujii, H. Suga, Y. Naitoh, T. Minari, T. Miyadera, K. Tsukagoshi and H. Kataura, *Nano Lett.*, 2009, **9**, 1497-1500.
98. J. Liu and M. C. Hersam, *MRS Bulletin*, 2010, **35**, 315-321.
99. H. Zhang, B. Wu, W. Hu and Y. Liu, *Chem. Soc. Rev.*, 2011, **40**, 1324-1336.
100. M. S. Arnold, S. I. Stupp and M. C. Hersam, *Nano Lett.*, 2005, **5**, 713-718.
101. M. S. Arnold, A. A. Green, J. F. Hulvat, S. I. Stupp and M. C. Hersam, *Nat. Nano.*, 2006, **1**, 60-65.
102. A. A. Green and M. C. Hersam, *ACS Nano*, 2011, **5**, 1459-1467.
103. A. A. Green and M. C. Hersam, *Nat. Nano.*, 2009, **4**, 64-70.
104. A. A. Green and M. C. Hersam, *Adv. Mater.*, 2011, **23**, 2185-2190.
105. A. Green, M. Duch and M. Hersam, *Nano Research*, 2009, **2**, 69-77.
106. A. L. Antaris, J.-W. T. Seo, A. A. Green and M. C. Hersam, *ACS Nano*, 2010, **4**, 4725-4732.
107. A. A. Green and M. C. Hersam, *Nano Lett.*, 2009, **9**, 4031-4036.
108. A. A. Green and M. C. Hersam, *J. Phys. Chem. Lett.*, 2009, **1**, 544-549.
109. X. Sun, D. Luo, J. Liu and D. G. Evans, *ACS Nano*, 2010, **4**, 3381-3389.
110. X. Sun, Z. Liu, K. Welsher, J. Robinson, A. Goodwin, S. Zaric and H. Dai, *Nano Research*, 2008, **1**, 203-212.
111. L. Liu, H. Zhou, R. Cheng, W. J. Yu, Y. Liu, Y. Chen, J. Shaw, X. Zhong, Y. Huang and X. Duan, *ACS Nano*, 2012, **6**, 8241-8249.
112. A. D. Franklin, M. Luisier, S.-J. Han, G. Tulevski, C. M. Breslin, L. Gignac, M. S. Lundstrom and W. Haensch, *Nano Lett.*, 2012, **12**, 758-762.
113. T. Ando and T. Nakanishi, *J. Phys. Soc. Jpn.*, 1998, **67**, 1704-1713.
114. T. Durkop, S. A. Getty, E. Cobas and M. S. Fuhrer, *Nano Lett.*, 2004, **4**, 35-39.
115. V. Perebeinos, J. Tersoff and P. Avouris, *Phys. Rev. Lett.*, 2005, **94**, 086802.
116. X. Zhou, J.-Y. Park, S. Huang, J. Liu and P. L. McEuen, *Phys. Rev. Lett.*, 2005, **95**, 146805.
117. N. W. Ashcroft and N. D. Mermin, *Solid State Physics*, Saunders, 1976.
118. H. J. Hrostowski, F. J. Morin, T. H. Geballe and G. H. Wheatley, *Phys. Rev.*, 1955, **100**, 1672-1676.
119. Z. Yao, C. L. Kane and C. Dekker, *Phys. Rev. Lett.*, 2000, **84**, 2941-2944.
120. A. Javey, J. Guo, M. Paulsson, Q. Wang, D. Mann, M. Lundstrom and H. Dai, *Phys. Rev. Lett.*, 2004, **92**, 106804.
121. Y.-F. Chen and M. S. Fuhrer, *Phys. Rev. Lett.*, 2005, **95**, 236803.
122. S. Datta, *Electronic Transport in Mesoscopic Systems*, Cambridge University Press, 1997.
123. F. Léonard and J. Tersoff, *Phys. Rev. Lett.*, 2000, **84**, 4693-4696.
124. S. Heinze, J. Tersoff, R. Martel, V. Derycke, J. Appenzeller and P. Avouris, *Phys. Rev. Lett.*, 2002, **89**, 106801.
125. R. Martel, V. Derycke, C. Lavoie, J. Appenzeller, K. K. Chan, J. Tersoff and P. Avouris, *Phys. Rev. Lett.*, 2001, **87**, 256805.
126. Z. H. Chen, J. Appenzeller, J. Knoch, Y. M. Lin and P. Avouris, *Nano Lett.*, 2005, **5**, 1497-1502.
127. W. Kim, A. Javey, O. Vermesh, O. Wang, Y. M. Li and H. J. Dai, *Nano Lett.*, 2003, **3**, 193-198.
128. V. K. Sangwan, V. W. Ballarotto, M. S. Fuhrer and E. D. Williams, *Appl. Phys. Lett.*, 2008, **93**, 113112.
129. S. Ilani, L. A. K. Donev, M. Kindermann and P. L. McEuen, *Nat. Phys.*, 2006, **2**, 687-691.
130. F. Wang, G. Dukovic, L. E. Brus and T. F. Heinz, *Science*, 2005, **308**, 838-841.
131. M. J. O'Connell, S. M. Bachilo, C. B. Huffman, V. C. Moore, M. S. Strano, E. H. Haroz, K. L. Rialon, P. J. Boul, W. H. Noon, C. Kittrell, J. Ma, R. H. Hauge, R. B. Weisman and R. E. Smalley, *Science*, 2002, **297**, 593-596.
132. S. M. Bachilo, M. S. Strano, C. Kittrell, R. H. Hauge, R. E. Smalley and R. B. Weisman, *Science*, 2002, **298**, 2361-2366.

133. GrossoG, GravesJ, A. T. Hammack, A. A. High, L. V. Butov, HansonM and A. C. Gossard, *Nat. Photon.*, 2009, **3**, 577-580.
134. F. Wang, G. Dukovic, L. E. Brus and T. F. Heinz, *Phys. Rev. Lett.*, 2004, **92**, 177401.
135. L. Huang, H. N. Pedrosa and T. D. Krauss, *Phys. Rev. Lett.*, 2004, **93**, 017403.
136. A. A. High, E. E. Novitskaya, L. V. Butov, M. Hanson and A. C. Gossard, *Science*, 2008, **321**, 229-231.
137. J. A. Misewich, R. Martel, P. Avouris, J. C. Tsang, S. Heinze and J. Tersoff, *Science*, 2003, **300**, 783-786.
138. J. Chen, V. Perebeinos, M. Freitag, J. Tsang, Q. Fu, J. Liu and P. Avouris, *Science*, 2005, **310**, 1171-1174.
139. S. J. Tans, M. H. Devoret, H. Dai, A. Thess, R. E. Smalley, L. J. Geerligs and C. Dekker, *Nature*, 1997, **386**, 474-477.
140. S. J. Tans, A. R. M. Verschueren and C. Dekker, *Nature*, 1998, **393**, 49-52.
141. M. Bockrath, D. H. Cobden, J. Lu, A. G. Rinzler, R. E. Smalley, L. Balents and P. L. McEuen, *Nature*, 1999, **397**, 598-601.
142. H. W. C. Postma, T. Teepen, Z. Yao, M. Grifoni and C. Dekker, *Science*, 2001, **293**, 76-79.
143. M. Bockrath, D. H. Cobden, P. L. McEuen, N. G. Chopra, A. Zettl, A. Thess and R. E. Smalley, *Science*, 1997, **275**, 1922-1925.
144. A. Javey, J. Guo, Q. Wang, M. Lundstrom and H. J. Dai, *Nature*, 2003, **424**, 654-657.
145. A. Javey, J. Guo, D. B. Farmer, Q. Wang, D. Wang, R. G. Gordon, M. Lundstrom and H. Dai, *Nano Lett.*, 2004, **4**, 447-450.
146. J. Appenzeller, Y. M. Lin, J. Knoch and P. Avouris, *Phys. Rev. Lett.*, 2004, **93**, 196805.
147. A. Javey, H. Kim, M. Brink, Q. Wang, A. Ural, J. Guo, P. McIntyre, P. McEuen, M. Lundstrom and H. Dai, *Nat. Mater.*, 2002, **1**, 241-246.
148. A. S. Sedra and K. C. Smith, *Microelectronic Circuits*, Oxford University Press, USA, 2009.
149. A. Bachtold, P. Hadley, T. Nakanishi and C. Dekker, *Science*, 2001, **294**, 1317-1320.
150. V. Derycke, R. Martel, J. Appenzeller and P. Avouris, *Nano Lett.*, 2001, **1**, 453-456.
151. A. Javey, Q. Wang, A. Ural, Y. Li and H. Dai, *Nano Lett.*, 2002, **2**, 929-932.
152. Z. H. Chen, J. Appenzeller, Y. M. Lin, J. Sippel-Oakley, A. G. Rinzler, J. Y. Tang, S. J. Wind, P. M. Solomon and P. Avouris, *Science*, 2006, **311**, 1735-1735.
153. L. Ding, Z. Zhang, S. Liang, T. Pei, S. Wang, Y. Li, W. Zhou, J. Liu and L.-M. Peng, *Nat Commun*, 2012, **3**, 677.
154. A. Vijayaraghavan, S. Blatt, D. Weissenberger, M. Oron-Carl, F. Hennrich, D. Gerthsen, H. Hahn and R. Krupke, *Nano Lett.*, 2007, **7**, 1556-1560.
155. M. Ganzhorn, A. Vijayaraghavan, A. A. Green, S. Dehm, A. Voigt, M. Rapp, M. C. Hersam and R. Krupke, *Adv. Mater.*, 2011, **23**, 1734-1738.
156. M. R. Islam, K. J. Kormondy, E. Silbar and S. I. Khondaker, *Nanotechnology*, 2012, **23**, 125201.
157. K. J. Kormondy, P. Stokes and S. I. Khondaker, *Nanotechnology*, 2011, **22**, 415201.
158. P. Stokes and S. I. Khondaker, *Nanotechnology*, 2008, **19**, 175202.
159. P. Stokes, E. Silbar, Y. M. Zayas and S. I. Khondaker, *Appl. Phys. Lett.*, 2009, **94**, 113104.
160. P. Stokes and S. I. Khondaker, *Appl. Phys. Lett.*, 2010, **96**, 083110.
161. S. Shekhar, P. Stokes and S. I. Khondaker, *ACS Nano*, 2011, **5**, 1739-1746.
162. S. R. Forrest, *Nature*, 2004, **428**, 911.
163. T. J. Marks, *MRS Bulletin*, 2010, **35**, 1018-1027.
164. S. Kumar, J. Y. Murthy and M. A. Alam, *Phys. Rev. Lett.*, 2005, **95**, 066802.
165. E. S. Snow, J. P. Novak, P. M. Campbell and D. Park, *Appl. Phys. Lett.*, 2003, **82**, 2145-2147.
166. A. Behnam, J. Guo and A. Ural, *J. Appl. Phys.*, 2007, **102**, 044313.

167. V. K. Sangwan, A. Behnam, V. W. Ballarotto, M. S. Fuhrer, A. Ural and E. D. Williams, *Appl. Phys. Lett.*, 2010, **97**, 043111.
168. C. Kocabas, N. Pimparkar, O. Yesilyurt, S. J. Kang, M. A. Alam and J. A. Rogers, *Nano Lett.*, 2007, **7**, 1195-1202.
169. A. Behnam, L. Noriega, Y. Choi, Z. C. Wu, A. G. Rinzler and A. Ural, *Appl. Phys. Lett.*, 2006, **89**, 093107.
170. Q. Cao, M. Xia, C. Kocabas, M. Shim, J. A. Rogers and S. V. Rotkin, *Appl. Phys. Lett.*, 2007, **90**, 023516.
171. M. S. Fuhrer, J. Nygard, L. Shih, M. Forero, Y. G. Yoon, M. S. C. Mazzoni, H. J. Choi, J. Ihm, S. G. Louie, A. Zettl and P. L. McEuen, *Science*, 2000, **288**, 494-497.
172. M. A. Topinka, M. W. Rowell, D. Goldhaber-Gordon, M. D. McGehee, D. S. Hecht and G. Gruner, *Nano Lett.*, 2009, **9**, 1866-1871.
173. V. K. Sangwan, V. W. Ballarotto, D. R. Hines, M. S. Fuhrer and E. D. Williams, *Solid-State Electron.*, 2010, **54**, 1204-1210.
174. S. J. Kang, C. Kocabas, H. S. Kim, Q. Cao, M. A. Meitl, D. Y. Khang and J. A. Rogers, *Nano Lett.*, 2007, **7**, 3343-3348.
175. S. J. Kang, C. Kocabas, T. Ozel, M. Shim, N. Pimparkar, M. A. Alam, S. V. Rotkin and J. A. Rogers, *Nat. Nano.*, 2007, **2**, 230-236.
176. C. Kocabas, S. H. Hur, A. Gaur, M. A. Meitl, M. Shim and J. A. Rogers, *Small*, 2005, **1**, 1110-1116.
177. P. C. Collins, M. S. Arnold and P. Avouris, *Science*, 2001, **292**, 706-709.
178. K. Ryu, A. Badmaev, C. Wang, A. Lin, N. Patil, L. Gomez, A. Kumar, S. Mitra, H. S. P. Wong and C. W. Zhou, *Nano Lett.*, 2009, **9**, 189-197.
179. C. Wang, K. Ryu, L. De Arco, A. Badmaev, J. Zhang, X. Lin, Y. Che and C. Zhou, *Nano Research*, 2010, **3**, 831-842.
180. S. Shekhar, M. Erementchouk, M. N. Leuenberger and S. I. Khondaker, *Appl. Phys. Lett.*, 2011, **98**, 243121.
181. S. Shekhar, H. Heinrich and S. I. Khondaker, *Carbon*, 2012, **50**, 1635-1643.
182. G. Zhang, P. Qi, X. Wang, Y. Lu, X. Li, R. Tu, S. Bangsaruntip, D. Mann, L. Zhang and H. Dai, *Science*, 2006, **314**, 974-977.
183. Q. Cao, H. S. Kim, N. Pimparkar, J. P. Kulkarni, C. J. Wang, M. Shim, K. Roy, M. A. Alam and J. A. Rogers, *Nature*, 2008, **454**, 495.
184. D.-m. Sun, M. Y. Timmermans, Y. Tian, A. G. Nasibulin, E. I. Kauppinen, S. Kishimoto, T. Mizutani and Y. Ohno, *Nat. Nano.*, 2011, **6**, 156-161.
185. A. A. Green and M. C. Hersam, *Nano Lett.*, 2008, **8**, 1417-1422.
186. T. P. Tyler, T. A. Shastry, B. J. Leever and M. C. Hersam, *Adv. Mater.*, 2012, **24**, 4765-4768.
187. M. Ha, Y. Xia, A. A. Green, W. Zhang, M. J. Renn, C. H. Kim, M. C. Hersam and C. D. Frisbie, *ACS Nano*, 2010, **4**, 4388-4395.
188. C. Wang, J. Zhang and C. Zhou, *ACS Nano*, 2010, **4**, 7123-7132.
189. C. Wang, J. Zhang, K. Ryu, A. Badmaev, L. G. De Arco and C. Zhou, *Nano Lett.*, 2009, **9**, 4285-4291.
190. M. Engel, J. P. Small, M. Steiner, M. Freitag, A. A. Green, M. C. Hersam and P. Avouris, *ACS Nano*, 2008, **2**, 2445-2452.
191. B. K. Sarker, S. Shekhar and S. I. Khondaker, *ACS Nano*, 2011, **5**, 6297-6305.
192. V. K. Sangwan, R. Ponce Ortiz, J. M. P. Alaboson, J. D. Emery, M. J. Bedzyk, L. J. Lauhon, T. J. Marks and M. C. Hersam, *ACS Nano*, 2012, **6**, 7480–7488.
193. S. Fujii, T. Tanaka, Y. Miyata, H. Suga, Y. Naitoh, T. Minari, T. Miyadera, K. Tsukagoshi and H. Kataura, *Appl. Phys. Express*, 2009, **2,** 071601.
194. M. E. Roberts, M. C. LeMieux, A. N. Sokolov and Z. Bao, *Nano Lett.*, 2009, **9**, 2526-2531.

195. T. A. Shastry, J.-W. T. Seo, J. J. Lopez, H. N. Arnold, J. Z. Kelter, V. K. Sangwan, L. J. Lauhon, T. J. Marks and M. C. Hersam, *Small*, 2012, 10.1002/smll.201201398.
196. S.-H. Hur, M.-H. Yoon, A. Gaur, M. Shim, A. Facchetti, T. J. Marks and J. A. Rogers, *J. Am. Chem. Soc.*, 2005, **127**, 13808-13809.
197. R. o. P. Ortiz, A. Facchetti and T. J. Marks, *Chem. Rev.*, 2009, **110**, 205-239.
198. S. A. DiBenedetto, A. Facchetti, M. A. Ratner and T. J. Marks, *Adv. Mater.*, 2009, **21**, 1407-1433.
199. M.-H. Yoon, A. Facchetti and T. J. Marks, *Proc. Natl. Acad. Sci. USA*, 2005, **102**, 4678-4682.
200. G. N. Ostojic and M. C. Hersam, *Small*, 2012, **8**, 1840-1845.
201. S.-p. Han, H. T. Maune, R. D. Barish, M. Bockrath and W. A. Goddard, *Nano Lett.*, 2012, **12**, 1129-1135.
202. C. Rutherglen, D. Jain and P. Burke, *Nat. Nano.*, 2009, **4**, 811-819.
203. S. M. Sze, *Physics of Semiconductor Devices*, Wiley-Interscience, 1981.
204. D. V. Singh, K. A. Jenkins, J. Appenzeller, D. Neumayer, A. Grill and H. S. P. Wong, *Nanotechnology, IEEE Transactions on*, 2004, **3**, 383-387.
205. S. D. Li, Z. Yu, S. F. Yen, W. C. Tang and P. J. Burke, *Nano Lett.*, 2004, **4**, 753-756.
206. S. Rosenblatt, H. Lin, V. Sazonova, S. Tiwari and P. L. McEuen, *Appl. Phys. Lett.*, 2005, **87**, 153111.
207. J. Appenzeller and D. J. Frank, *Appl. Phys. Lett.*, 2004, **84**, 1771-1773.
208. E. Cobas and M. S. Fuhrer, *Appl. Phys. Lett.*, 2008, **93**, 043120.
209. C. Kocabas, H. S. Kim, T. Banks, J. A. Rogers, A. A. Pesetski, J. E. Baumgardner, S. V. Krishnaswamy and H. Zhang, *Proc. Natl. Acad. Sci. USA*, 2008, **105**, 1405-1409.
210. C. Kocabas, S. Dunham, Q. Cao, K. Cimino, X. Ho, H.-S. Kim, D. Dawson, J. Payne, M. Stuenkel, H. Zhang, T. Banks, M. Feng, S. V. Rotkin and J. A. Rogers, *Nano Lett.*, 2009, **9**, 1937-1943.
211. C. Rutherglen, D. Jain and P. Burke, *Appl. Phys. Lett.*, 2008, **93**, 083119-083113.
212. C. Wang, A. Badmaev, A. Jooyaie, M. Bao, K. L. Wang, K. Galatsis and C. Zhou, *ACS Nano*, 2011, **5**, 4169-4176.
213. L. Nougaret, H. Happy, G. Dambrine, V. Derycke, J. P. Bourgoin, A. A. Green and M. C. Hersam, *Appl. Phys. Lett.*, 2009, **94**, 243505-243503.
214. F. Schwierz, *Nat. Nano.*, 2010, **5**, 487-496.
215. C. Rutherglen and P. Burke, *Nano Lett.*, 2007, **7**, 3296-3299.
216. J. Minhun, K. Jaeyoung, N. Jinsoo, L. Namsoo, L. Chaemin, L. Gwangyong, K. Junseok, K. Hwiwon, J. Kyunghwan, A. D. Leonard, J. M. Tour and C. Gyoujin, *Electron Devices, IEEE Transactions on*, 2010, **57**, 571-580.
217. H. Liangbing, L. Jianfeng, L. Jun, G. George and M. Tobin, *Nanotechnology*, 2010, **21**, 155202.
218. G. Dukovic, F. Wang, D. Song, M. Y. Sfeir, T. F. Heinz and L. E. Brus, *Nano Lett.*, 2005, **5**, 2314-2318.
219. V. Perebeinos, J. Tersoff and P. Avouris, *Phys. Rev. Lett.*, 2004, **92**, 257402.
220. S. Lebedkin, F. Hennrich, T. Skipa and M. M. Kappes, *J. Phys. Chem. B*, 2003, **107**, 1949-1956.
221. J. Lefebvre, D. G. Austing, J. Bond and P. Finnie, *Nano Lett.*, 2006, **6**, 1603-1608.
222. V. Perebeinos and P. Avouris, *Phys. Rev. Lett.*, 2008, **101**, 057401.
223. M. Freitag, J. Chen, J. Tersoff, J. C. Tsang, Q. Fu, J. Liu and P. Avouris, *Phys. Rev. Lett.*, 2004, **93**, 076803.
224. M. Freitag, V. Perebeinos, J. Chen, A. Stein, J. C. Tsang, J. A. Misewich, R. Martel and P. Avouris, *Nano Lett.*, 2004, **4**, 1063-1066.
225. J. Zaumseil, X. Ho, J. R. Guest, G. P. Wiederrecht and J. A. Rogers, *ACS Nano*, 2009, **3**, 2225-2234.
226. E. Adam, C. M. Aguirre, L. Marty, B. C. St-Antoine, F. Meunier, P. Desjardins, D. Ménard and R. Martel, *Nano Lett.*, 2008, **8**, 2351-2355.
227. H. Qian, C. Georgi, N. Anderson, A. A. Green, M. C. Hersam, L. Novotny and A. Hartschuh, *Nano Lett.*, 2008, **8**, 1363-1367.

228. V. Perebeinos and P. Avouris, *Phys. Rev. B*, 2006, **74**, 121410.
229. S. O. Koswatta, V. Perebeinos, M. S. Lundstrom and P. Avouris, *Nano Lett.*, 2008, **8**, 1596-1601.
230. V. Perebeinos and P. Avouris, *Nano Lett.*, 2007, **7**, 609-613.
231. M. Freitag, J. C. Tsang, J. Kirtley, A. Carlsen, J. Chen, A. Troeman, H. Hilgenkamp and P. Avouris, *Nano Lett.*, 2006, **6**, 1425-1433.
232. T. Mueller, M. Kinoshita, M. Steiner, V. Perebeinos, A. A. Bol, D. B. Farmer and P. Avouris, *Nat. Nano.*, 2010, **5**, 27-31.
233. M. Kinoshita, M. Steiner, M. Engel, J. P. Small, A. A. Green, M. C. Hersam, R. Krupke, E. E. Mendez and P. Avouris, *Opt. Express*, 2010, **18**, 25738-25745.
234. D. Mann, Y. K. Kato, A. Kinkhabwala, E. Pop, J. Cao, X. Wang, L. Zhang, Q. Wang, J. Guo and H. Dai, *Nat. Nano.*, 2007, **2**, 33-38.
235. M. Freitag, Y. Martin, J. A. Misewich, R. Martel and P. Avouris, *Nano Lett.*, 2003, **3**, 1067-1071.
236. J. U. Lee, *Appl. Phys. Lett.*, 2005, **87**, 073101.
237. Y. H. Ahn, A. W. Tsen, B. Kim, Y. W. Park and J. Park, *Nano Lett.*, 2007, **7**, 3320-3323.
238. K. Balasubramanian, M. Burghard, K. Kern, M. Scolari and A. Mews, *Nano Lett.*, 2005, **5**, 507-510.
239. N. M. Gabor, Z. Zhong, K. Bosnick, J. Park and P. L. McEuen, *Science*, 2009, **325**, 1367-1371.
240. M. E. Itkis, F. Borondics, A. Yu and R. C. Haddon, *Science*, 2006, **312**, 413-416.
241. K. W. Lee, S. P. Lee, H. Choi, K. H. Mo, J. W. Jang, H. Kweon and C. E. Lee, *Appl. Phys. Lett.*, 2007, **91**, 023110.
242. A. Kumar and C. Zhou, *ACS Nano*, 2010, **4**, 11-14.
243. A. Southard, V. Sangwan, J. Cheng, E. D. Williams and M. S. Fuhrer, *Org. Electron.*, 2009, **10**, 1556-1561.
244. V. K. Sangwan, A. Southard, T. L. Moore, V. W. Ballarotto, D. R. Hines, M. S. Fuhrer and E. D. Williams, *Microelectronic Engineering*, 2011, **88**, 3150-3154.
245. X. Liu, J. Si, B. Chang, G. Xu, Q. Yang, Z. Pan, S. Xie, P. Ye, J. Fan and M. Wan, *Appl. Phys. Lett.*, 1999, **74**, 164-166.
246. Y. Sakakibara, S. Tatsuura, H. Kataura, M. Tokumoto and Y. Achiba, *Jpn. J. Appl. Phys.*, 2003, **42**, L494.
247. P. R. Wallace, *Phys. Rev.*, 1947, **71**, 622-634.
248. R. Peierls, *Ann Inst Henri Poincare*, 1935, **5**, 45.
249. L. D. Landau, *Phys. Z. Sowjetunion*, 1937, **11**, 9.
250. J. A. Venables, G. D. T. Spiller and M. Hanbucken, *Reports on Progress in Physics*, 1984, **47**, 399.
251. T. Ando, *J. Phys. Soc. Jpn.*, 2005, **74**, 777-817.
252. T. Ando, *J. Phys. Soc. Jpn.*, 2006, **75**, 074716.
253. K. Nomura and A. H. MacDonald, *Phys. Rev. Lett.*, 2007, **98**, 076602.
254. V. V. Cheianov and V. I. Fal'ko, *Phys. Rev. B*, 2006, **74**, 041403.
255. S. Adam, E. H. Hwang, V. M. Galitski and S. Das Sarma, *Proc. Natl. Acad. Sci. USA*, 2007, **104**, 18392-18397.
256. M. I. Katsnelson, K. S. Novoselov and A. K. Geim, *Nat. Phys.*, 2006, **2**, 620-625.
257. K. S. Novoselov, E. McCann, S. V. Morozov, V. I. Fal'ko, M. I. Katsnelson, U. Zeitler, D. Jiang, F. Schedin and A. K. Geim, *Nat. Phys.*, 2006, **2**, 177-180.
258. Y. Zhang, Y.-W. Tan, H. L. Stormer and P. Kim, *Nature*, 2005, **438**, 201-204.
259. K. S. Novoselov, A. K. Geim, S. V. Morozov, D. Jiang, M. I. Katsnelson, I. V. Grigorieva, S. V. Dubonos and A. A. Firsov, *Nature*, 2005, **438**, 197-200.
260. K. I. Bolotin, K. J. Sikes, Z. Jiang, M. Klima, G. Fudenberg, J. Hone, P. Kim and H. L. Stormer, *Solid State Commun.*, 2008, **146**, 351-355.
261. D. K. Efetov and P. Kim, *Phys. Rev. Lett.*, 2010, **105**, 256805.

262. J. H. Chen, C. Jang, S. Adam, M. S. Fuhrer, E. D. Williams and M. Ishigami, *Nat. Phys.*, 2008, **4**, 377-381.
263. J.-H. Chen, C. Jang, S. Xiao, M. Ishigami and M. S. Fuhrer, *Nat. Nano.*, 2008, **3**, 206-209.
264. E. H. Hwang and S. Das Sarma, *Phys. Rev. B*, 2008, **77**, 115449.
265. C. R. Dean, A. F. Young, I. Meric, C. Lee, L. Wang, S. Sorgenfrei, K. Watanabe, T. Taniguchi, P. Kim, K. L. Shepard and J. Hone, *Nat. Nano.*, 2010, **5**, 722-726.
266. A. S. Mayorov, R. V. Gorbachev, S. V. Morozov, L. Britnell, R. Jalil, L. A. Ponomarenko, P. Blake, K. S. Novoselov, K. Watanabe, T. Taniguchi and A. K. Geim, *Nano Lett.*, 2011, **11**, 2396-2399.
267. C. Berger, Z. Song, X. Li, X. Wu, N. Brown, D. Maud, C. Naud and W. A. de Heer, *physica status solidi (a)*, 2007, **204**, 1746-1750.
268. W. Gannett, W. Regan, K. Watanabe, T. Taniguchi, M. F. Crommie and A. Zettl, *Appl. Phys. Lett.*, 2011, **98**, 242105-242103.
269. K. Nakada, M. Fujita, G. Dresselhaus and M. S. Dresselhaus, *Phys. Rev. B*, 1996, **54**, 17954-17961.
270. Y.-W. Son, M. L. Cohen and S. G. Louie, *Phys. Rev. Lett.*, 2006, **97**, 216803.
271. A. K. Geim and K. S. Novoselov, *Nat. Mater.*, 2007, **6**, 183-191.
272. L. Brey and H. A. Fertig, *Phys. Rev. B*, 2006, **73**, 235411.
273. F. Sols, F. Guinea and A. H. C. Neto, *Phys. Rev. Lett.*, 2007, **99**, 166803.
274. M. Y. Han, B. Özyilmaz, Y. Zhang and P. Kim, *Phys. Rev. Lett.*, 2007, **98**, 206805.
275. L. Jiao, L. Zhang, X. Wang, G. Diankov and H. Dai, *Nature*, 2009, **458**, 877-880.
276. D. V. Kosynkin, A. L. Higginbotham, A. Sinitskii, J. R. Lomeda, A. Dimiev, B. K. Price and J. M. Tour, *Nature*, 2009, **458**, 872-876.
277. X. Li, X. Wang, L. Zhang, S. Lee and H. Dai, *Science*, 2008, **319**, 1229-1232.
278. L. Ci, L. Song, D. Jariwala, A. L. Elías, W. Gao, M. Terrones and P. M. Ajayan, *Adv. Mater.*, 2009, **21**, 4487-4491.
279. U. K. Parashar, S. Bhandari, R. K. Srivastava, D. Jariwala and A. Srivastava, *Nanoscale*, 2011, **3**, 3876-3882.
280. M. Sprinkle, M. Ruan, Y. Hu, J. Hankinson, M. Rubio Roy, B. Zhang, X. Wu, C. Berger and W. A. de Heer, *Nat. Nano.*, 2010, **5**, 727-731.
281. J. Cai, P. Ruffieux, R. Jaafar, M. Bieri, T. Braun, S. Blankenburg, M. Muoth, A. P. Seitsonen, M. Saleh, X. Feng, K. Mullen and R. Fasel, *Nature*, 2010, **466**, 470-473.
282. X. Wang, X. Li, L. Zhang, Y. Yoon, P. K. Weber, H. Wang, J. Guo and H. Dai, *Science*, 2009, **324**, 768-771.
283. Z. Chen, Y.-M. Lin, M. J. Rooks and P. Avouris, *Physica E: Low-dimensional Systems and Nanostructures*, 2007, **40**, 228-232.
284. L. Yang, C.-H. Park, Y.-W. Son, M. L. Cohen and S. G. Louie, *Phys. Rev. Lett.*, 2007, **99**, 186801.
285. M. Evaldsson, I. V. Zozoulenko, H. Xu and T. Heinzel, *Phys. Rev. B*, 2008, **78**, 161407.
286. J. Bai, X. Zhong, S. Jiang, Y. Huang and X. Duan, *Nat. Nano.*, 2010, **5**, 190-194.
287. M. Z. Hossain, M. A. Walsh and M. C. Hersam, *J. Am. Chem. Soc.*, 2010, **132**, 15399-15403.
288. Q. H. Wang and M. C. Hersam, *Nano Lett.*, 2010, **11**, 589-593.
289. D. C. Elias, R. R. Nair, T. M. G. Mohiuddin, S. V. Morozov, P. Blake, M. P. Halsall, A. C. Ferrari, D. W. Boukhvalov, M. I. Katsnelson, A. K. Geim and K. S. Novoselov, *Science*, 2009, **323**, 610-613.
290. R. R. Nair, W. Ren, R. Jalil, I. Riaz, V. G. Kravets, L. Britnell, P. Blake, F. Schedin, A. S. Mayorov, S. Yuan, M. I. Katsnelson, H.-M. Cheng, W. Strupinski, L. G. Bulusheva, A. V. Okotrub, I. V. Grigorieva, A. N. Grigorenko, K. S. Novoselov and A. K. Geim, *Small*, 2010, **6**, 2877-2884.
291. M. Z. Hossain, J. E. Johns, K. H. Bevan, H. J. Karmel, Y. T. Liang, S. Yoshimoto, K. Mukai, T. Koitaya, J. Yoshinobu, M. Kawai, A. M. Lear, L. L. Kesmodel, S. L. Tait and M. C. Hersam, *Nat. Chem.*, 2012, **4**, 305-309.
292. G. Eda and M. Chhowalla, *Adv. Mater.*, 2010, **22**, 2392-2415.

293. G. Giovannetti, P. A. Khomyakov, G. Brocks, P. J. Kelly and J. van den Brink, *Phys. Rev. B*, 2007, **76**, 073103.
294. R. Quhe, J. Zheng, G. Luo, Q. Liu, R. Qin, J. Zhou, D. Yu, S. Nagase, W.-N. Mei, Z. Gao and J. Lu, *NPG Asia Mater*, 2012, **4**, e6.
295. A. Ramasubramaniam, D. Naveh and E. Towe, *Nano Lett.*, 2011, **11**, 1070-1075.
296. I. Forbeaux, J. M. Themlin and J. M. Debever, *Phys. Rev. B*, 1998, **58**, 16396-16406.
297. C. Berger, Z. Song, X. Li, X. Wu, N. Brown, C. Naud, D. Mayou, T. Li, J. Hass, A. N. Marchenkov, E. H. Conrad, P. N. First and W. A. de Heer, *Science*, 2006, **312**, 1191-1196.
298. F. Xia, D. B. Farmer, Y.-m. Lin and P. Avouris, *Nano Lett.*, 2010, **10**, 715-718.
299. E. V. Castro, K. S. Novoselov, S. V. Morozov, N. M. R. Peres, J. M. B. L. dos Santos, J. Nilsson, F. Guinea, A. K. Geim and A. H. C. Neto, *Phys. Rev. Lett.*, 2007, **99**, 216802.
300. E. McCann, *Phys. Rev. B*, 2006, **74**, 161403.
301. Y. Zhang, T.-T. Tang, C. Girit, Z. Hao, M. C. Martin, A. Zettl, M. F. Crommie, Y. R. Shen and F. Wang, *Nature*, 2009, **459**, 820-823.
302. K. F. Mak, C. H. Lui, J. Shan and T. F. Heinz, *Phys. Rev. Lett.*, 2009, **102**, 256405.
303. C. H. Lui, Z. Li, K. F. Mak, E. Cappelluti and T. F. Heinz, *Nat. Phys.*, 2011, **7**, 944-947.
304. W. J. Yu, L. Liao, S. H. Chae, Y. H. Lee and X. Duan, *Nano Lett.*, 2011, **11**, 4759-4763.
305. A. Lherbier, X. Blase, Y.-M. Niquet, F. Triozon and S. Roche, *Phys. Rev. Lett.*, 2008, **101**, 036808.
306. D. Wei, Y. Liu, Y. Wang, H. Zhang, L. Huang and G. Yu, *Nano Lett.*, 2009, **9**, 1752-1758.
307. L. S. Panchakarla, K. S. Subrahmanyam, S. K. Saha, A. Govindaraj, H. R. Krishnamurthy, U. V. Waghmare and C. N. R. Rao, *Adv. Mater.*, 2009, **21**, 4726-4730.
308. L. Ci, L. Song, C. Jin, D. Jariwala, D. Wu, Y. Li, A. Srivastava, Z. F. Wang, K. Storr, L. Balicas, F. Liu and P. M. Ajayan, *Nat. Mater.*, 2010, **9**, 430-435.
309. L. Song, L. Balicas, D. J. Mowbray, R. B. Capaz, K. Storr, L. Ci, D. Jariwala, S. Kurth, S. G. Louie, A. Rubio and P. M. Ajayan, *Phys. Rev. B*, 2012, **86**, 075429.
310. H. Xu, Z. Zhang, H. Xu, Z. Wang, S. Wang and L.-M. Peng, *ACS Nano*, 2011, **5**, 5031-5037.
311. S. Russo, M. F. Craciun, M. Yamamoto, A. F. Morpurgo and S. Tarucha, *Physica E: Low-dimensional Systems and Nanostructures*, 2010, **42**, 677-679.
312. R. S. Shishir and D. K. Ferry, *J. Phys.: Condens. Matter*, 2009, **21**, 344201.
313. I. Meric, M. Y. Han, A. F. Young, B. Ozyilmaz, P. Kim and K. L. Shepard, *Nat. Nano.*, 2008, **3**, 654-659.
314. M. C. Lemme, T. J. Echtermeyer, M. Baus and H. Kurz, *Electron Device Letters, IEEE*, 2007, **28**, 282-284.
315. S. Kim, J. Nah, I. Jo, D. Shahrjerdi, L. Colombo, Z. Yao, E. Tutuc and S. K. Banerjee, *Appl. Phys. Lett.*, 2009, **94**, 062107-062103.
316. A. Pirkle, R. M. Wallace and L. Colombo, *Appl. Phys. Lett.*, 2009, **95**, 133106-133103.
317. D. B. Farmer, H.-Y. Chiu, Y.-M. Lin, K. A. Jenkins, F. Xia and P. Avouris, *Nano Lett.*, 2009, **9**, 4474-4478.
318. B. Lee, G. Mordi, M. J. Kim, Y. J. Chabal, E. M. Vogel, R. M. Wallace, K. J. Cho, L. Colombo and J. Kim, *Appl. Phys. Lett.*, 2010, **97**, 043107-043103.
319. J. M. P. Alaboson, Q. H. Wang, J. D. Emery, A. L. Lipson, M. J. Bedzyk, J. W. Elam, M. J. Pellin and M. C. Hersam, *ACS Nano*, 2011, **5**, 5223-5232.
320. Q. H. Wang and M. C. Hersam, *Nat. Chem.*, 2009, **1**, 206-211.
321. Y.-M. Lin, K. A. Jenkins, A. Valdes-Garcia, J. P. Small, D. B. Farmer and P. Avouris, *Nano Lett.*, 2008, **9**, 422-426.
322. Y.-M. Lin, C. Dimitrakopoulos, K. A. Jenkins, D. B. Farmer, H.-Y. Chiu, A. Grill and P. Avouris, *Science*, 2010, **327**, 662.

323. Y.-M. Lin, A. Valdes-Garcia, S.-J. Han, D. B. Farmer, I. Meric, Y. Sun, Y. Wu, C. Dimitrakopoulos, A. Grill, P. Avouris and K. A. Jenkins, *Science*, 2011, **332**, 1294-1297.
324. Y. Wu, Y.-m. Lin, A. A. Bol, K. A. Jenkins, F. Xia, D. B. Farmer, Y. Zhu and P. Avouris, *Nature*, 2011, **472**, 74-78.
325. L. Liao, Y.-C. Lin, M. Bao, R. Cheng, J. Bai, Y. Liu, Y. Qu, K. L. Wang, Y. Huang and X. Duan, *Nature*, 2010, **467**, 305-308.
326. C. Sire, F. Ardiaca, S. Lepilliet, J.-W. T. Seo, M. C. Hersam, G. Dambrine, H. Happy and V. Derycke, *Nano Lett.*, 2012, **12**, 1184-1188.
327. Z. Wang, Z. Zhang, H. Xu, L. Ding, S. Wang and L.-M. Peng, *Appl. Phys. Lett.*, 2010, **96**, 173104-173103.
328. H.-Y. Chen and J. Appenzeller, *Nano Lett.*, 2012, **12**, 2067-2070.
329. S. Bae, H. Kim, Y. Lee, X. Xu, J.-S. Park, Y. Zheng, J. Balakrishnan, T. Lei, H. Ri Kim, Y. I. Song, Y.-J. Kim, K. S. Kim, B. Ozyilmaz, J.-H. Ahn, B. H. Hong and S. Iijima, *Nat. Nano.*, 2010, **5**, 574-578.
330. R. R. Nair, P. Blake, A. N. Grigorenko, K. S. Novoselov, T. J. Booth, T. Stauber, N. M. R. Peres and A. K. Geim, *Science*, 2008, **320**, 1308.
331. K. F. Mak, J. Shan and T. F. Heinz, *Phys. Rev. Lett.*, 2011, **106**, 046401.
332. G. Eda, Y.-Y. Lin, C. Mattevi, H. Yamaguchi, H.-A. Chen, I. S. Chen, C.-W. Chen and M. Chhowalla, *Adv. Mater.*, 2010, **22**, 505-509.
333. T. Gokus, R. R. Nair, A. Bonetti, M. Böhmler, A. Lombardo, K. S. Novoselov, A. K. Geim, A. C. Ferrari and A. Hartschuh, *ACS Nano*, 2009, **3**, 3963-3968.
334. S. Essig, C. W. Marquardt, A. Vijayaraghavan, M. Ganzhorn, S. Dehm, F. Hennrich, F. Ou, A. A. Green, C. Sciascia, F. Bonaccorso, K. P. Bohnen, H. v. Löhneysen, M. M. Kappes, P. M. Ajayan, M. C. Hersam, A. C. Ferrari and R. Krupke, *Nano Lett.*, 2010, **10**, 1589-1594.
335. F. Xia, T. Mueller, Y.-m. Lin, A. Valdes-Garcia and P. Avouris, *Nat. Nano.*, 2009, **4**, 839-843.
336. T. Mueller, F. Xia and P. Avouris, *Nat. Photon.*, 2010, **4**, 297-301.
337. N. M. Gabor, J. C. W. Song, Q. Ma, N. L. Nair, T. Taychatanapat, K. Watanabe, T. Taniguchi, L. S. Levitov and P. Jarillo-Herrero, *Science*, 2011, **334**, 648-652.
338. D. Sun, G. Aivazian, A. M. Jones, J. S. Ross, W. Yao, D. Cobden and X. Xu, *Nat. Nano.*, 2012, **7**, 114-118.
339. M. C. Duch, G. R. S. Budinger, Y. T. Liang, S. Soberanes, D. Urich, S. E. Chiarella, L. A. Campochiaro, A. Gonzalez, N. S. Chandel, M. C. Hersam and G. M. Mutlu, *Nano Lett.*, 2011, **11**, 5201-5207.
340. S. K. Banerjee, L. F. Register, E. Tutuc, D. Reddy and A. H. MacDonald, *Electron Device Letters, IEEE*, 2009, **30**, 158-160.
341. L. Britnell, R. V. Gorbachev, R. Jalil, B. D. Belle, F. Schedin, A. Mishchenko, T. Georgiou, M. I. Katsnelson, L. Eaves, S. V. Morozov, N. M. R. Peres, J. Leist, A. K. Geim, K. S. Novoselov and L. A. Ponomarenko, *Science*, 2012, **335**, 947-950.
342. H. Yang, J. Heo, S. Park, H. J. Song, D. H. Seo, K.-E. Byun, P. Kim, I. Yoo, H.-J. Chung and K. Kim, *Science*, 2012, **336**, 1140-1143.
343. N. S. Sariciftci, L. Smilowitz, A. J. Heeger and F. Wudl, *Science*, 1992, **258**, 1474-1476.
344. N. S. Sariciftci, D. Braun, C. Zhang, V. I. Srdanov, A. J. Heeger, G. Stucky and F. Wudl, *Appl. Phys. Lett.*, 1993, **62**, 585-587.
345. N. S. Sariciftci, L. Smilowitz, A. J. Heeger and F. Wudl, *Synthetic Metals*, 1993, **59**, 333-352.
346. J. C. Hummelen, B. W. Knight, F. LePeq, F. Wudl, J. Yao and C. L. Wilkins, *J. Org. Chem.*, 1995, **60**, 532-538.
347. G. Yu, J. Gao, J. C. Hummelen, F. Wudl and A. J. Heeger, *Science*, 1995, **270**, 1789-1791.
348. L. M. Popescu, P. vant Hof, A. B. Sieval, H. T. Jonkman and J. C. Hummelen, *Appl. Phys. Lett.*, 2006, **89**, 213507.

349. M. Lenes, S. W. Shelton, A. B. Sieval, D. F. Kronholm, J. C. Hummelen and P. W. M. Blom, *Adv. Funct. Mater.*, 2009, **19**, 3002-3007.
350. J. L. Delgado, P.-A. Bouit, S. Filippone, M. A. Herranz and N. Martin, *Chem. Commun.*, 2010, **46**, 4853-4865.
351. M. M. Wienk, J. M. Kroon, W. J. H. Verhees, J. Knol, J. C. Hummelen, P. A. van Hal and R. A. J. Janssen, *Angewandte Chemie*, 2003, **115**, 3493-3497.
352. R. B. Ross, C. M. Cardona, D. M. Guldi, S. G. Sankaranarayanan, M. O. Reese, N. Kopidakis, J. Peet, B. Walker, G. C. Bazan, E. Van Keuren, B. C. Holloway and M. Drees, *Nat. Mater.*, 2009, **8**, 208-212.
353. Y. He and Y. Li, *Physical Chemistry Chemical Physics*, 2011, **13**, 1970-1983.
354. G. D. Scholes and G. Rumbles, *Nat. Mater.*, 2006, **5**, 683-696.
355. A. C. Dillon, *Chem. Rev.*, 2010, **110**, 6856-6872.
356. T. Umeyama and H. Imahori, *Energy Environ. Sci.*, 2008, **1**, 120-133.
357. D. M. Guldi, A. Rahman, V. Sgobba and C. Ehli, *Chem. Soc. Rev.*, 2006, **35**, 471-487.
358. D. D. Tune, B. S. Flavel, R. Krupke and J. G. Shapter, *Adv. Energy Mater.*, 2012, **2**, 1043-1055.
359. E. Kymakis, I. Alexandrou and G. A. J. Amaratunga, *J. Appl. Phys.*, 2003, **93**, 1764-1768.
360. E. Kymakis, E. Koudoumas, I. Franghiadakis and G. A. J. Amaratunga, *Journal of Physics D: Applied Physics*, 2006, **39**, 1058.
361. S. Berson, R. de Bettignies, S. Bailly, S. Guillerez and B. Jousselme, *Adv. Funct. Mater.*, 2007, **17**, 3363-3370.
362. M. Bernardi, M. Giulianini and J. C. Grossman, *ACS Nano*, 2010, **4**, 6599-6606.
363. J. Geng and T. Zeng, *J. Am. Chem. Soc.*, 2006, **128**, 16827-16833.
364. S. Ren, M. Bernardi, R. R. Lunt, V. Bulovic, J. C. Grossman and S. Gradečak, *Nano Lett.*, 2011, **11**, 5316-5321.
365. S. Chaudhary, H. Lu, A. M. Müller, C. J. Bardeen and M. Ozkan, *Nano Lett.*, 2007, **7**, 1973-1979.
366. J. M. Holt, A. J. Ferguson, N. Kopidakis, B. A. Larsen, J. Bult, G. Rumbles and J. L. Blackburn, *Nano Lett.*, 2010, **10**, 4627-4633.
367. D. J. Bindl, N. S. Safron and M. S. Arnold, *ACS Nano*, 2010, **4**, 5657-5664.
368. D. J. Bindl, M.-Y. Wu, F. C. Prehn and M. S. Arnold, *Nano Lett.*, 2010, **11**, 455-460.
369. M.-H. Ham, G. L. C. Paulus, C. Y. Lee, C. Song, K. Kalantar-zadeh, W. Choi, J.-H. Han and M. S. Strano, *ACS Nano*, 2010, **4**, 6251-6259.
370. A. J. Ferguson, J. L. Blackburn, J. M. Holt, N. Kopidakis, R. C. Tenent, T. M. Barnes, M. J. Heben and G. Rumbles, *J. Phys. Chem. Lett.*, 2010, **1**, 2406-2411.
371. B. G. Lewis and D. C. Paine, *MRS Bulletin*, 2000, **25**, 22-27.
372. B. S. Shim, Z. Tang, M. P. Morabito, A. Agarwal, H. Hong and N. A. Kotov, *Chem. Mat.*, 2007, **19**, 5467-5474.
373. R. C. Tenent, T. M. Barnes, J. D. Bergeson, A. J. Ferguson, B. To, L. M. Gedvilas, M. J. Heben and J. L. Blackburn, *Adv. Mater.*, 2009, **21**, 3210-3216.
374. A. J. Miller, R. A. Hatton and S. R. P. Silva, *Appl. Phys. Lett.*, 2006, **89**, 133117-133113.
375. H. Ago, K. Petritsch, M. S. P. Shaffer, A. H. Windle and R. H. Friend, *Adv. Mater.*, 1999, **11**, 1281-1285.
376. M. W. Rowell, M. A. Topinka, M. D. McGehee, H.-J. Prall, G. Dennler, N. S. Sariciftci, L. Hu and G. Gruner, *Appl. Phys. Lett.*, 2006, **88**, 233506-233503.
377. A. D. Pasquier, H. E. Unalan, A. Kanwal, S. Miller and M. Chhowalla, *Appl. Phys. Lett.*, 2005, **87**, 203511-203513.
378. J. van de Lagemaat, T. M. Barnes, G. Rumbles, S. E. Shaheen, T. J. Coutts, C. Weeks, I. Levitsky, J. Peltola and P. Glatkowski, *Appl. Phys. Lett.*, 2006, **88**, 233503-233503.
379. G. Fanchini, S. Miller, B. B. Parekh and M. Chhowalla, *Nano Lett.*, 2008, **8**, 2176-2179.

380. B. Gao, Y. F. Chen, M. S. Fuhrer, D. C. Glattli and A. Bachtold, *Phys. Rev. Lett.*, 2005, **95**, 196802.
381. T. P. Tyler, R. E. Brock, H. J. Karmel, T. J. Marks and M. C. Hersam, *Adv. Energy Mater.*, 2011, **1**, 785-791.
382. J. L. Blackburn, T. M. Barnes, M. C. Beard, Y.-H. Kim, R. C. Tenent, T. J. McDonald, B. To, T. J. Coutts and M. J. Heben, *ACS Nano*, 2008, **2**, 1266-1274.
383. K. S. Mistry, B. A. Larsen, J. D. Bergeson, T. M. Barnes, G. Teeter, C. Engtrakul and J. L. Blackburn, *ACS Nano*, 2011, **5**, 3714-3723.
384. B. O'Regan and M. Gratzel, *Nature*, 1991, **353**, 737-740.
385. K. H. Jung, J. S. Hong, R. Vittal and K. J. Kim, *Chem. Lett.*, 2002, 864-865.
386. K. Suzuki, M. Yamaguchi, M. Kumagai and S. Yanagida, *Chem. Lett.*, 2003, **32**, 28-29.
387. S.-R. Jang, R. Vittal and K.-J. Kim, *Langmuir*, 2004, **20**, 9807-9810.
388. S. L. Kim, S. R. Jang, R. Vittal, J. Lee and K. J. Kim, *J. Appl. Electrochem.*, 2006, **36**, 1433-1439.
389. T. Y. Lee, P. S. Alegaonkar and J. B. Yoo, *Thin Solid Films*, 2007, **515**, 5131-5135.
390. H. Yu, X. Quan, S. Chen and H. Zhao, *J. Phys. Chem. C*, 2007, **111**, 12987-12991.
391. A. Kongkanand, R. Martínez Domínguez and P. V. Kamat, *Nano Lett.*, 2007, **7**, 676-680.
392. P. Brown, K. Takechi and P. V. Kamat, *J. Phys. Chem. C*, 2008, **112**, 4776-4782.
393. F. Hao, P. Dong, J. Zhang, Y. Zhang, P. E. Loya, R. H. Hauge, J. Li, J. Lou and H. Lin, *Sci. Rep.*, 2012, **2**, 368.
394. X. Dang, H. Yi, M.-H. Ham, J. Qi, D. S. Yun, R. Ladewski, M. S. Strano, P. T. Hammond and A. M. Belcher, *Nat. Nano.*, 2011, **6**, 377-384.
395. P. V. Kamat, *J. Phys. Chem. C*, 2008, **112**, 18737-18753.
396. D. M. Guldi, G. M. A. Rahman, M. Prato, N. Jux, S. Qin and W. Ford, *Angew. Chem. Int. Ed.*, 2005, **44**, 2015-2018.
397. L. Sheeney-Haj-Ichia, B. Basnar and I. Willner, *Angew. Chem. Int. Ed.*, 2005, **44**, 78-83.
398. D. M. Guldi, G. M. A. Rahman, V. Sgobba, N. A. Kotov, D. Bonifazi and M. Prato, *J. Am. Chem. Soc.*, 2006, **128**, 2315-2323.
399. L. Hu, Y. L. Zhao, K. Ryu, C. Zhou, J. F. Stoddart and G. Grüner, *Adv. Mater.*, 2008, **20**, 939-946.
400. B. Farrow and P. V. Kamat, *J. Am. Chem. Soc.*, 2009, **131**, 11124-11131.
401. L. Yang, S. Wang, Q. Zeng, Z. Zhang, T. Pei, Y. Li and L.-M. Peng, *Nat. Photon.*, 2011, **5**, 672-676.
402. J. Wei, Y. Jia, Q. Shu, Z. Gu, K. Wang, D. Zhuang, G. Zhang, Z. Wang, J. Luo, A. Cao and D. Wu, *Nano Lett.*, 2007, **7**, 2317-2321.
403. Y. Jia, J. Wei, K. Wang, A. Cao, Q. Shu, X. Gui, Y. Zhu, D. Zhuang, G. Zhang, B. Ma, L. Wang, W. Liu, Z. Wang, J. Luo and D. Wu, *Adv. Mater.*, 2008, **20**, 4594-4598.
404. Z. Li, V. P. Kunets, V. Saini, Y. Xu, E. Dervishi, G. J. Salamo, A. R. Biris and A. S. Biris, *ACS Nano*, 2009, **3**, 1407-1414.
405. Y. Jia, A. Cao, X. Bai, Z. Li, L. Zhang, N. Guo, J. Wei, K. Wang, H. Zhu, D. Wu and P. M. Ajayan, *Nano Lett.*, 2011, **11**, 1901-1905.
406. L. Zhang, Y. Jia, S. Wang, Z. Li, C. Ji, J. Wei, H. Zhu, K. Wang, D. Wu, E. Shi, Y. Fang and A. Cao, *Nano Lett.*, 2010, **10**, 3583-3589.
407. Q. Shu, J. Wei, K. Wang, H. Zhu, Z. Li, Y. Jia, X. Gui, N. Guo, X. Li, C. Ma and D. Wu, *Nano Lett.*, 2009, **9**, 4338-4342.
408. C. X. Guo, G. H. Guai and C. M. Li, *Adv. Energy Mater.*, 2011, **1**, 448-452.
409. X. Wan, G. Long, L. Huang and Y. Chen, *Adv. Mater.*, 2011, **23**, 5342-5358.
410. X. Wang, L. Zhi and K. Mullen, *Nano Lett.*, 2007, **8**, 323-327.
411. J. Wu, H. A. Becerril, Z. Bao, Z. Liu, Y. Chen and P. Peumans, *Appl. Phys. Lett.*, 2008, **92**, 263302-263303.
412. Z. Yin, S. Sun, T. Salim, S. Wu, X. Huang, Q. He, Y. M. Lam and H. Zhang, *ACS Nano*, 2010, **4**, 5263-5268.

413. Z. Yin, S. Wu, X. Zhou, X. Huang, Q. Zhang, F. Boey and H. Zhang, *Small*, 2010, **6**, 307-312.
414. H. A. Becerril, J. Mao, Z. Liu, R. M. Stoltenberg, Z. Bao and Y. Chen, *ACS Nano*, 2008, **2**, 463-470.
415. V. C. Tung, L.-M. Chen, M. J. Allen, J. K. Wassei, K. Nelson, R. B. Kaner and Y. Yang, *Nano Lett.*, 2009, **9**, 1949-1955.
416. Y. Wang, X. Chen, Y. Zhong, F. Zhu and K. P. Loh, *Appl. Phys. Lett.*, 2009, **95**, 063302-063303.
417. L. Gomez De Arco, Y. Zhang, C. W. Schlenker, K. Ryu, M. E. Thompson and C. Zhou, *ACS Nano*, 2010, **4**, 2865-2873.
418. H. Bi, F. Huang, J. Liang, X. Xie and M. Jiang, *Adv. Mater.*, 2011, **23**, 3202-3206.
419. A. Kasry, M. A. Kuroda, G. J. Martyna, G. S. Tulevski and A. A. Bol, *ACS Nano*, 2010, **4**, 3839-3844.
420. Y. Wang, S. W. Tong, X. F. Xu, B. Özyilmaz and K. P. Loh, *Adv. Mater.*, 2011, **23**, 1514-1518.
421. F. Güneş, H.-J. Shin, C. Biswas, G. H. Han, E. S. Kim, S. J. Chae, J.-Y. Choi and Y. H. Lee, *ACS Nano*, 2010, **4**, 4595-4600.
422. K. Ki Kang, R. Alfonso, S. Yumeng, P. Hyesung, L. Lain-Jong, L. Young Hee and K. Jing, *Nanotechnology*, 2010, **21**, 285205.
423. S. De and J. N. Coleman, *ACS Nano*, 2010, **4**, 2713-2720.
424. M. Girtan and M. Rusu, *Sol. Energ. Mat. Sol. Cells*, 2010, **94**, 446-450.
425. M. O. Reese, A. J. Morfa, M. S. White, N. Kopidakis, S. E. Shaheen, G. Rumbles and D. S. Ginley, *Sol. Energ. Mat. Sol. Cells*, 2008, **92**, 746-752.
426. M. Jørgensen, K. Norrman and F. C. Krebs, *Sol. Energ. Mat. Sol. Cells*, 2008, **92**, 686-714.
427. J.-A. Yan, L. Xian and M. Y. Chou, *Phys. Rev. Lett.*, 2009, **103**, 086802.
428. K. P. Loh, Q. Bao, G. Eda and M. Chhowalla, *Nat. Chem.*, 2010, **2**, 1015-1024.
429. S.-S. Li, K.-H. Tu, C.-C. Lin, C.-W. Chen and M. Chhowalla, *ACS Nano*, 2010, **4**, 3169-3174.
430. J.-M. Yun, J.-S. Yeo, J. Kim, H.-G. Jeong, D.-Y. Kim, Y.-J. Noh, S.-S. Kim, B.-C. Ku and S.-I. Na, *Adv. Mater.*, 2011, **23**, 4923-4928.
431. I. P. Murray, S. J. Lou, L. J. Cote, S. Loser, C. J. Kadleck, T. Xu, J. M. Szarko, B. S. Rolczynski, J. E. Johns, J. Huang, L. Yu, L. X. Chen, T. J. Marks and M. C. Hersam, *J. Phys. Chem. Lett.*, 2011, **2**, 3006-3012.
432. Y. Liang, D. Feng, Y. Wu, S.-T. Tsai, G. Li, C. Ray and L. Yu, *J. Am. Chem. Soc.*, 2009, **131**, 7792-7799.
433. Y. Liang, Z. Xu, J. Xia, S.-T. Tsai, Y. Wu, G. Li, C. Ray and L. Yu, *Adv. Mater.*, 2010, **22**, E135-E138.
434. J. Liu, Y. Xue, Y. Gao, D. Yu, M. Durstock and L. Dai, *Adv. Mater.*, 2012, **24**, 2227.
435. J. Kim, V. C. Tung and J. Huang, *Adv. Energy Mater.*, 2011, **1**, 1052-1057.
436. V. C. Tung, J. Kim and J. Huang, *Adv. Energy Mater.*, 2012, **2**, 299-303.
437. J. T. Robinson, J. S. Burgess, C. E. Junkermeier, S. C. Badescu, T. L. Reinecke, F. K. Perkins, M. K. Zalalutdniov, J. W. Baldwin, J. C. Culbertson, P. E. Sheehan and E. S. Snow, *Nano Lett.*, 2010, **10**, 3001-3005.
438. Q. Liu, Z. Liu, X. Zhang, L. Yang, N. Zhang, G. Pan, S. Yin, Y. Chen and J. Wei, *Adv. Funct. Mater.*, 2009, **19**, 894-904.
439. Z. Liu, Q. Liu, Y. Huang, Y. Ma, S. Yin, X. Zhang, W. Sun and Y. Chen, *Adv. Mater.*, 2008, **20**, 3924-3930.
440. D. Yu, K. Park, M. Durstock and L. Dai, *J. Phys. Chem. Lett.*, 2011, **2**, 1113-1118.
441. D. Yu, Y. Yang, M. Durstock, J.-B. Baek and L. Dai, *ACS Nano*, 2010, **4**, 5633-5640.
442. N. Yang, J. Zhai, D. Wang, Y. Chen and L. Jiang, *ACS Nano*, 2010, **4**, 887-894.
443. V. C. Tung, J.-H. Huang, I. Tevis, F. Kim, J. Kim, C.-W. Chu, S. I. Stupp and J. Huang, *J. Am. Chem. Soc.*, 2011, **133**, 4940-4947.
444. V. C. Tung, J. Kim, L. J. Cote and J. Huang, *J. Am. Chem. Soc.*, 2011, **133**, 9262-9265.
445. V. Yong and J. M. Tour, *Small*, 2010, **6**, 313-318.
446. X. Yan, X. Cui, B. Li and L.-s. Li, *Nano Lett.*, 2010, **10**, 1869-1873.

447. X. Li, H. Zhu, K. Wang, A. Cao, J. Wei, C. Li, Y. Jia, Z. Li, X. Li and D. Wu, *Adv. Mater.*, 2010, **22**, 2743-2748.
448. G. Fan, H. Zhu, K. Wang, J. Wei, X. Li, Q. Shu, N. Guo and D. Wu, *ACS Appl. Mater. Interfaces*, 2011, **3**, 721-725.
449. C. Xie, P. Lv, B. Nie, J. Jie, X. Zhang, Z. Wang, P. Jiang, Z. Hu, L. Luo, Z. Zhu, L. Wang and C. Wu, *Appl. Phys. Lett.*, 2011, **99**, 133113.
450. L. Zhang, L. Fan, Z. Li, E. Shi, X. Li, H. Li, C. Ji, Y. Jia, J. Wei, K. Wang, H. Zhu, D. Wu and A. Cao, *Nano Research*, 2011, **4**, 891-900.
451. Y. Ye, Y. Dai, L. Dai, Z. Shi, N. Liu, F. Wang, L. Fu, R. Peng, X. Wen, Z. Chen, Z. Liu and G. Qin, *ACS Appl. Mater. Interfaces*, 2010, **2**, 3406-3410.
452. Y. Shi, K. K. Kim, A. Reina, M. Hofmann, L.-J. Li and J. Kong, *ACS Nano*, 2010, **4**, 2689-2694.
453. X. Miao, S. Tongay, M. K. Petterson, K. Berke, A. G. Rinzler, B. R. Appleton and A. F. Hebard, *Nano Lett.*, 2012, **12**, 2745-2750.
454. Z. Ting, M. Syed, V. M. Nosang and A. D. Marc, *Nanotechnology*, 2008, **19**, 332001.
455. P. Bondavalli, P. Legagneux and D. Pribat, *Sensors and Actuators B: Chemical*, 2009, **140**, 304-318.
456. B. L. Allen, P. D. Kichambare and A. Star, *Adv. Mater.*, 2007, **19**, 1439-1451.
457. D. R. Kauffman and A. Star, *Chem. Soc. Rev.*, 2008, **37**, 1197-1206.
458. Q. Zhao, Z. H. Gan and Q. K. Zhuang, *Electroanalysis*, 2002, **14**, 1609-1613.
459. K. Balasubramanian and M. Burghard, *Analytical and Bioanalytical Chemistry*, 2006, **385**, 452-468.
460. Y. Y. Shao, J. Wang, H. Wu, J. Liu, I. A. Aksay and Y. H. Lin, *Electroanalysis*, 2010, **22**, 1027-1036.
461. D. A. C. Brownson and C. E. Banks, *Analyst*, 2010, **135**, 2768-2778.
462. K. R. Ratinac, W. R. Yang, J. J. Gooding, P. Thordarson and F. Braet, *Electroanalysis*, 2011, **23**, 803-826.
463. J. Kong, N. R. Franklin, C. Zhou, M. G. Chapline, S. Peng, K. Cho and H. Dai, *Science*, 2000, **287**, 622-625.
464. Y. Shimizu and M. Egashira, *MRS Bulletin*, 1999, **24**, 18-24.
465. J. J. Miasik, A. Hooper and B. C. Tofield, *Journal of the Chemical Society-Faraday Transactions I*, 1986, **82**, 1117-&.
466. X. Liu, Z. Luo, S. Han, T. Tang, D. Zhang and C. Zhou, *Appl. Phys. Lett.*, 2005, **86**, 243501-243503.
467. T. Someya, J. Small, P. Kim, C. Nuckolls and J. T. Yardley, *Nano Lett.*, 2003, **3**, 877-881.
468. C. P. W. Paul, M. Natacha, T. Zhenni, M. Yoji, J. D. Carey and S. R. P. Silva, *Nanotechnology*, 2007, **18**, 175701.
469. C. Young Wook, O. Je Seung, Y. Seung Hwan, C. Hyang Hee and Y. Kyung-Hwa, *Nanotechnology*, 2007, **18**, 435504.
470. J. Kong, M. G. Chapline and H. Dai, *Adv. Mater.*, 2001, **13**, 1384-1386.
471. W. Kim, A. Javey, O. Vermesh, Q. Wang, Y. Li and H. Dai, *Nano Lett.*, 2003, **3**, 193-198.
472. S. Heinze, J. Tersoff and P. Avouris, *Appl. Phys. Lett.*, 2003, **83**, 5038-5040.
473. X. Cui, M. Freitag, R. Martel, L. Brus and P. Avouris, *Nano Lett.*, 2003, **3**, 783-787.
474. S. Auvray, J. Borghetti, M. F. Goffman, A. Filoramo, V. Derycke, J. P. Bourgoin and O. Jost, *Appl. Phys. Lett.*, 2004, **84**, 5106-5108.
475. B. R. Goldsmith, J. G. Coroneus, V. R. Khalap, A. A. Kane, G. A. Weiss and P. G. Collins, *Science*, 2007, **315**, 77-81.
476. B. R. Goldsmith, J. G. Coroneus, A. A. Kane, G. A. Weiss and P. G. Collins, *Nano Lett.*, 2007, **8**, 189-194.
477. J. Mannik, B. R. Goldsmith, A. Kane and P. G. Collins, *Phys. Rev. Lett.*, 2006, **97**, 016601.
478. V. R. Khalap, T. Sheps, A. A. Kane and P. G. Collins, *Nano Lett.*, 2010, **10**, 896-901.

479. P. Qi, O. Vermesh, M. Grecu, A. Javey, Q. Wang, H. Dai, S. Peng and K. J. Cho, *Nano Lett.*, 2003, **3**, 347-351.
480. C. Staii, A. T. Johnson, M. Chen and A. Gelperin, *Nano Lett.*, 2005, **5**, 1774-1778.
481. A. T. C. Johnson, S. Cristian, C. Michelle, K. Sam, J. Robert, M. L. Klein and A. Gelperin, *Semiconductor Science and Technology*, 2006, **21**, S17.
482. L. Valentini, I. Armentano, J. M. Kenny, C. Cantalini, L. Lozzi and S. Santucci, *Appl. Phys. Lett.*, 2003, **82**, 961-963.
483. E. Bekyarova, M. Davis, T. Burch, M. E. Itkis, B. Zhao, S. Sunshine and R. C. Haddon, *J. Phys. Chem. B*, 2004, **108**, 19717-19720.
484. J. P. Novak, E. S. Snow, E. J. Houser, D. Park, J. L. Stepnowski and R. A. McGill, *Appl. Phys. Lett.*, 2003, **83**, 4026-4028.
485. J. Li, Y. Lu, Q. Ye, L. Delzeit and M. Meyyappan, *Electrochemical and Solid-State Letters*, 2005, **8**, H100-H102.
486. A. Star, V. Joshi, S. Skarupo, D. Thomas and J.-C. P. Gabriel, *J. Phys. Chem. B*, 2006, **110**, 21014-21020.
487. D. R. Kauffman and A. Star, *Nano Lett.*, 2007, **7**, 1863-1868.
488. A. Modi, N. Koratkar, E. Lass, B. Wei and P. M. Ajayan, *Nature*, 2003, **424**, 171-174.
489. C. Wei, L. Dai, A. Roy and T. B. Tolle, *J. Am. Chem. Soc.*, 2006, **128**, 1412-1413.
490. E. S. Snow, F. K. Perkins, E. J. Houser, S. C. Badescu and T. L. Reinecke, *Science*, 2005, **307**, 1942-1945.
491. G. Esen, M. S. Fuhrer, M. Ishigami and E. D. Williams, *Appl. Phys. Lett.*, 2007, **90**, 123510-123513.
492. E. S. Snow and F. K. Perkins, *Nano Lett.*, 2005, **5**, 2414-2417.
493. E. S. Snow, F. K. Perkins and J. A. Robinson, *Chem. Soc. Rev.*, 2006, **35**, 790-798.
494. J. Lu, L. Lai, G. Luo, J. Zhou, R. Qin, D. Wang, L. Wang, W. N. Mei, G. Li, Z. Gao, S. Nagase, Y. Maeda, T. Akasaka and D. Yu, *Small*, 2007, **3**, 1566-1576.
495. Z. Chen, X. Du, M.-H. Du, C. D. Rancken, H.-P. Cheng and A. G. Rinzler, *Nano Lett.*, 2003, **3**, 1245-1249.
496. M. Nakano, M. Fujioka, K. Mai, H. Watanabe, Y. Martin and J. Suehiro, *Jpn. J. Appl. Phys.*, 2012, **51**, 045102.
497. Y. Battie, O. Ducloux, P. Thobois, N. Dorval, J. S. Lauret, B. Attal-Trétout and A. Loiseau, *Carbon*, 2011, **49**, 3544-3552.
498. M. Ganzhorn, A. Vijayaraghavan, S. Dehm, F. Hennrich, A. A. Green, M. Fichtner, A. Voigt, M. Rapp, H. von Löhneysen, M. C. Hersam, M. M. Kappes and R. Krupke, *ACS Nano*, 2011, **5**, 1670-1676.
499. R. J. Chen, S. Bangsaruntip, K. A. Drouvalakis, N. Wong Shi Kam, M. Shim, Y. Li, W. Kim, P. J. Utz and H. Dai, *Proc. Natl. Acad. Sci. USA*, 2003, **100**, 4984-4989.
500. R. J. Chen, H. C. Choi, S. Bangsaruntip, E. Yenilmez, X. Tang, Q. Wang, Y.-L. Chang and H. Dai, *J. Am. Chem. Soc.*, 2004, **126**, 1563-1568.
501. A. Star, J.-C. P. Gabriel, K. Bradley and G. Grüner, *Nano Lett.*, 2003, **3**, 459-463.
502. K. Besteman, J.-O. Lee, F. G. M. Wiertz, H. A. Heering and C. Dekker, *Nano Lett.*, 2003, **3**, 727-730.
503. H.-M. So, K. Won, Y. H. Kim, B.-K. Kim, B. H. Ryu, P. S. Na, H. Kim and J.-O. Lee, *J. Am. Chem. Soc.*, 2005, **127**, 11906-11907.
504. Y.-B. Zhang, M. Kanungo, A. J. Ho, P. Freimuth, D. van der Lelie, M. Chen, S. M. Khamis, S. S. Datta, A. T. C. Johnson, J. A. Misewich and S. S. Wong, *Nano Lett.*, 2007, **7**, 3086-3091.
505. C. Li, M. Curreli, H. Lin, B. Lei, F. N. Ishikawa, R. Datar, R. J. Cote, M. E. Thompson and C. Zhou, *J. Am. Chem. Soc.*, 2005, **127**, 12484-12485.

506. A. Star, V. Joshi, T.-R. Han, M. V. P. Altoé, G. Grüner and J. F. Stoddart, *Org. Lett.*, 2004, **6**, 2089-2092.
507. Y. Choi, I. S. Moody, P. C. Sims, S. R. Hunt, B. L. Corso, I. Perez, G. A. Weiss and P. G. Collins, *Science*, 2012, **335**, 319-324.
508. A. Star, E. Tu, J. Niemann, J.-C. P. Gabriel, C. S. Joiner and C. Valcke, *Proc. Natl. Acad. Sci. USA*, 2006, **103**, 921-926.
509. X. Tang, S. Bansaruntip, N. Nakayama, E. Yenilmez, Y.-l. Chang and Q. Wang, *Nano Lett.*, 2006, **6**, 1632-1636.
510. F. N. Ishikawa, M. Curreli, C. A. Olson, H.-I. Liao, R. Sun, R. W. Roberts, R. J. Cote, M. E. Thompson and C. Zhou, *ACS Nano*, 2010, **4**, 6914-6922.
511. Y. Shao, J. Wang, H. Wu, J. Liu, I. A. Aksay and Y. Lin, *Electroanalysis*, 2010, **22**, 1027-1036.
512. F. Schedin, A. K. Geim, S. V. Morozov, E. W. Hill, P. Blake, M. I. Katsnelson and K. S. Novoselov, *Nat. Mater.*, 2007, **6**, 652-655.
513. T. O. Wehling, K. S. Novoselov, S. V. Morozov, E. E. Vdovin, M. I. Katsnelson, A. K. Geim and A. I. Lichtenstein, *Nano Lett.*, 2007, **8**, 173-177.
514. J. T. Robinson, F. K. Perkins, E. S. Snow, Z. Wei and P. E. Sheehan, *Nano Lett.*, 2008, **8**, 3137-3140.
515. Z. Cheng, Q. Li, Z. Li, Q. Zhou and Y. Fang, *Nano Lett.*, 2010, **10**, 1864-1868.
516. Y. Dan, Y. Lu, N. J. Kybert, Z. Luo and A. T. C. Johnson, *Nano Lett.*, 2009, **9**, 1472-1475.
517. J. D. Fowler, M. J. Allen, V. C. Tung, Y. Yang, R. B. Kaner and B. H. Weiller, *ACS Nano*, 2009, **3**, 301-306.
518. F. Yavari, Z. Chen, A. V. Thomas, W. Ren, H.-M. Cheng and N. Koratkar, *Sci. Rep.*, 2011, **1**, 166.
519. Z. Chen, W. Ren, L. Gao, B. Liu, S. Pei and H.-M. Cheng, *Nat. Mater.*, 2011, **10**, 424-428.
520. S. Rumyantsev, G. Liu, M. S. Shur, R. A. Potyrailo and A. A. Balandin, *Nano Lett.*, 2012, **12**, 2294-2298.
521. F. Yavari, E. Castillo, H. Gullapalli, P. M. Ajayan and N. Koratkar, *Appl. Phys. Lett.*, 2012, **100**, 203120-203124.
522. Y. Ohno, K. Maehashi, Y. Yamashiro and K. Matsumoto, *Nano Lett.*, 2009, **9**, 3318-3322.
523. Z. Yin, Q. He, X. Huang, J. Zhang, S. Wu, P. Chen, G. Lu, Q. Zhang, Q. Yan and H. Zhang, *Nanoscale*, 2012, **4**, 293-297.
524. S. Mao, G. Lu, K. Yu, Z. Bo and J. Chen, *Adv. Mater.*, 2010, **22**, 3521-3526.
525. Y. Huang, X. Dong, Y. Shi, C. M. Li, L.-J. Li and P. Chen, *Nanoscale*, 2010, **2**, 1485-1488.
526. H. W. C. Postma, *Nano Lett.*, 2010, **10**, 420-425.
527. J. Prasongkit, A. Grigoriev, B. Pathak, R. Ahuja and R. H. Scheicher, *Nano Lett.*, 2011, **11**, 1941-1945.
528. Y. He, R. H. Scheicher, A. Grigoriev, R. Ahuja, S. Long, Z. Huo and M. Liu, *Adv. Funct. Mater.*, 2011, **21**, 2602-2602.
529. T. Nelson, B. Zhang and O. V. Prezhdo, *Nano Lett.*, 2010, **10**, 3237-3242.
530. K. K. Saha, M. Drndić and B. K. Nikolić, *Nano Lett.*, 2011, **12**, 50-55.
531. S. K. Min, W. Y. Kim, Y. Cho and K. S. Kim, *Nat. Nano.*, 2011, **6**, 162-165.
532. Y. Cho, S. K. Min, W. Y. Kim and K. S. Kim, *Phys. Chem. Chem. Phys.*, 2011, **13**, 14293-14296.
533. C. Sathe, X. Zou, J.-P. Leburton and K. Schulten, *ACS Nano*, 2011, **5**, 8842-8851.
534. B. M. Venkatesan and R. Bashir, *Nat. Nano.*, 2011, **6**, 615-624.
535. S. Garaj, W. Hubbard, A. Reina, J. Kong, D. Branton and J. A. Golovchenko, *Nature*, 2010, **467**, 190-193.
536. G. g. F. Schneider, S. W. Kowalczyk, V. E. Calado, G. g. Pandraud, H. W. Zandbergen, L. M. K. Vandersypen and C. Dekker, *Nano Lett.*, 2010, **10**, 3163-3167.

537. C. A. Merchant, K. Healy, M. Wanunu, V. Ray, N. Peterman, J. Bartel, M. D. Fischbein, K. Venta, Z. Luo, A. T. C. Johnson and M. Drndić, *Nano Lett.*, 2010, **10**, 2915-2921.